\documentclass[usenatbib]{mn2e}
\usepackage{natbibmnfix, graphicx, times}
\pdfoutput=1
\usepackage{epstopdf}
\usepackage{float}

\newcommand{\rev}{}

\newcommand{\msun}{M$_{\odot}$}
\newcommand{\kms}{km s$^{-1}$}
\newcommand{\vk}{$v_{\rm k}$}
\newcommand{\vesc}{$v_{\rm esc}$}
\newcommand{\tmrg}{$t_{\rm mrg}$}
\newcommand{\tcoal}{$t_{\rm coal}$}
\newcommand{\fgas}{$f_{\rm gas}$}
\newcommand{\tagn}{$t_{\rm AGN}$}

\title[Recoiling Black Holes in Merging Galaxies]{Recoiling Black Holes in Merging Galaxies: Relationship to AGN Lifetimes, Starbursts, and the M$_{\rm BH}-\sigma_*$ Relation}

\author[Blecha et al.]{Laura Blecha$^1$\thanks{Email: lblecha@cfa.harvard.edu}, Thomas J. Cox$^2$, Abraham Loeb$^1$, \& Lars Hernquist$^1$ \\ $^1$ Harvard University, Department of Astronomy, 60 Garden St., Cambridge, MA 02138, USA \\ $^2$ Carnegie Observatories, 813 Santa Barbara Street, Pasadena, CA 9110}

\begin{document}
\maketitle

\begin{abstract}
Gravitational-wave (GW) recoil of merging supermassive black holes (SMBHs) may influence the co-evolution of SMBHs and their host galaxies.  We examine this possibility using SPH/N-body simulations of gaseous galaxy mergers, including SMBH accretion, in which the merged BH receives a recoil kick. This enables us to follow the trajectories and accretion of recoiling BHs in self-consistent, evolving merger remnants.  In contrast to recent studies on similar topics, we conduct a large parameter study, generating a suite of over 200 simulations with more than 60 merger models and a range of recoil velocities (\vk).  With this, we can identify systematic trends in the behavior of recoiling BHs.  Our main results are as follows.  
~(1)\, BHs kicked at nearly the central escape speed (\vesc) may oscillate on large orbits for up to a Hubble time, but in gas-rich mergers, BHs kicked with up to $\sim 0.7$~\vesc~may be confined to the central few kpc of the galaxy, owing to gas drag and steep central potentials.
~(2)\,~\vesc~in gas-rich mergers may increase rapidly during final coalescence, in which case recoil trajectories may depend sensitively on the {\em timing} of the BH merger relative to the formation of the central potential well.  Delays of even a few $\times \, 10^7$ yr in the BH merger time may substantially suppress recoiling BH motion for a fixed kick speed.  
~(3)\, Recoil events generally reduce the lifetimes of bright active galactic nuclei (AGN), but short-period recoil oscillations may actually {\em extend} AGN lifetimes at lower luminosities.  
~(4)\, Bondi-Hoyle accretion dominates recoiling BH growth for~\vk/\vesc~$\la 0.6-0.8$; for higher~\vk, the BH accretes primarily from its ejected gas disk, which we model as a time-dependent accretion disk.  
~(5)\, Kinematically-offset AGN may be observable either immediately after a high-velocity recoil or during pericentric passages through a gas-rich remnant.  In either case, AGN lifetimes may be up to $\sim 10$ Myr (for $\Delta v>800$~\kms), though the latter generally have higher luminosities.
~(6)\, Spatially-offset AGN can occur for~\vk~$\ga 0.5-0.7$~\vesc~(for $\Delta R >1$ kpc); these generally have low luminosities and lifetimes of $\sim 1 - 100$ Myr.  
~(7)\, Rapidly-recoiling BHs may be up to $\sim 5$ times less massive than their stationary counterparts.  These mass deficits lower the normalization of the $M_{\rm BH}-\sigma_*$ relation and contribute to both intrinsic and overall scatter. 
~(8)\, Finally, the displacement of AGN feedback after a recoil event enhances central star formation rates in the merger remnant, thereby extending the starburst phase of the merger and creating a denser, more massive stellar cusp.

\end{abstract}


\section{Introduction}
\label{sec:intro}

Within the hierarchical structure formation paradigm, a significant fraction of galaxy growth occurs via successive mergers.  Given the ample evidence that most, if not all, local galaxies host central supermassive black holes \citep[SMBHs; e.g.,][]{korric95,richst98,ferfor05}, major galaxy mergers (those with mass ratios $q \ga 0.25$) should inevitably result in the formation of SMBH binaries.  The fate of these binaries is somewhat uncertain and likely depends on the conditions within their host galaxies.  In highly spheroidal, gas-poor galaxies, these binaries may ``stall" at separations of about a parsec for up to a Hubble time \citep[e.g.,][]{begelm80, milmer01, yu02}.  In gas-rich galaxy mergers, however, gas is driven to the central region of the merging galaxies simultaneously with the formation of the SMBH binary.  Numerical simulations indicate that BH merger timescales may be much shorter than galaxy merger timescales in this case \citep[$\sim 10^6 - 10^7$ yr from the hard binary stage; e.g.,][]{escala05,dotti07}.  This implies that the SMBH binaries most able to accrete gas and produce electromagnetic signatures are likely to be short-lived.  

To date, observations seem to confirm this scenario.  Several quasar {\em pairs} with kpc-scale separations have been found in merging galaxies \citep{komoss03, bianchi08, comerf09b, green10}.  Recently, spectroscopic and photometric observations have found that between $10^{-3} - 10^{-2}$ of active galactic nuclei (AGN) at $z\, \la\, 0.7$ may in fact contain dual BHs with separations of $\sim$ kpc \citep{comerf09a, smith10, liu10a, liu10b, smith10, fu10}.  However, only one unambiguous case of a SMBH {\em binary} has been found, with a separation of $\sim 7$ pc \citep{rodrig06}.  Several additional objects have been identified as candidate SMBH binaries but are still unconfirmed \citep{sillan88, komoss08, dotti09, bogdan09, borlau09}.  Thus, it is clear that the large majority of AGN do not contain binary SMBHs, which suggests that any substantial population of long-lived SMBH binaries must exist in gas-poor environments where they are quiescent.  

If most binaries are therefore assumed to merge on reasonably short timescales, then gravitational-wave (GW) recoil must be a common phenomenon throughout the merger history of galaxies.  GW recoil is a natural consequence of gravitational radiation from BH binary mergers \citep{peres62, bekens73, fitdet84}.  If the binary system has any asymmetry -- unequal masses, spins or spin orientations -- then gravitational waves are radiated asymmetrically, resulting in a net linear momentum flux from the final BH at the time of merger.  This causes the BH to recoil in the opposite direction.  Whether these recoil kicks were large enough to be astrophysically interesting was uncertain until a few years ago, because accurate calculations of the recoil velocity require BH merger simulations using full general relativity.  These simulations have revealed that GW recoils may be quite large.  Recoil velocities up to $4000$~\kms~are possible for special configurations, which is far greater than galactic escape speeds \citep{campan07a, campan07b}.  The implication that SMBHs may spend substantial time in motion and off-center has opened a new line of inquiry into the ramifications for SMBHs and their host galaxies.  

A useful starting point for such inquiries would be the distribution of recoil kick velocities, but this is quite difficult to ascertain in practice.  The final velocity depends sensitively on not just the mass ratio of the progenitor BHs, but on their spin magnitudes and orientations as well.  The distributions of SMBH binary mass ratios and spins at various redshifts have been estimated using halo merger trees and semi-analytical models of SMBH growth \citep[e.g.,][]{volont03,volont05,king08,berti08}.  These distributions depend on a number of model assumptions, however, and the BH spin {\em orientations} prior to merger are far more uncertain still.  Thus, based on recent results from BH merger simulations using full numerical relativity (NR), several groups have calculated kick probability distributions as a function of BH mass ratio for either fixed or random values of BH spin, with the assumption that the spins are randomly oriented \citep{schbuo07, campan07a, baker08, lousto10b, lousto10a, vanmet10}.  Their results are in good agreement with each other and imply that the fraction of high-velocity GW recoils is substantial.  This underscores the potential for recoil events to be an important component of galaxy mergers.  

It is quite possible, of course, that the spins of SMBH binaries are not randomly oriented, but are preferentially aligned in some way.  \citet{bogdan07} have suggested that torques in a circumbinary gas disk may align the BH spins with the orbital axis of the disk.  In this case, the resulting in-plane kicks would have a maximum recoil velocity of $< 200$~\kms, although spins that became anti-aligned by the same mechanism could result in recoil velocities up to 500~\kms~\citep[e.g.,][]{gonzal07a,campan07a,baker08}.  Additionally, simulations by \citet{dotti10} demonstrate efficient spin alignment of merging BHs in the presence of a highly coherent accretion flow, although it is unclear how efficient this process might be in a gaseous environment that includes, e.g., star formation.  \citet{kesden10} have recently demonstrated that BH spin alignment may instead occur via relativistic spin precession, regardless of whether a gas disk is present.  The aforementioned recoil kick probability distributions are therefore upper limits on the actual distributions.  

Numerous possible consequences of GW recoil events have been discussed in the literature.  GW recoils may have a large effect at high redshifts, where escape velocities of galaxies are smaller \citep[e.g.,][]{merrit04,madqua04,volont07}.  This is a concern for attempts to understand, for example, the origin of the $z=6$ SDSS quasars \citep[e.g.,][]{fan01,fan03}.  \citet{volree06} suggest that growth of SMBHs at high $z$ must occur only in highly-biased halos.  Using cosmological hydrodynamic simulations, \citet{sijack09} investigate BH growth in massive high-$z$ halos, including the ejection of BHs with recoil velocities above~\vesc.  Their findings are consistent with the observed populations of bright quasars at $z=6$, despite the effects of GW recoil.  

At lower redshifts, recoiling BHs may produce electromagnetic signatures.  The main signatures we will focus on here are recoiling AGN that are either spatially or kinematically offset from their host galaxies \citep{madqua04, loeb07, bleloe08, kommer08b, guedes09}.   Other possible signatures include flares from shocks induced by fallback of gas marginally bound to the ejected BH \citep{lippai08,shibon08,schkro08}, enhanced rates of stellar tidal disruptions \citep{kommer08a,stoloe10}, and compact stellar clusters around ejected BHs \citep{oleloe09,merrit09}.

Thus far, no confirmed GW recoil events have been observed.  An inherent challenge in observing offset quasars is that larger spatial or kinematic offsets are easier to resolve, but less gas will be bound to the recoiling BH at higher recoil velocities, so its AGN lifetime will be shorter.  \citet{bonshi07} conducted a search for kinematic offsets in SDSS quasar spectra and found a null result at a limit of 800~\kms.  Several recoil candidates have been proposed, but their extreme inferred velocities should be exceedingly rare.  Indeed, the recoil candidate with a 2600~\kms~offset, proposed by \citet{komoss08}, may in fact be a superposition of two galaxies \citep{heckma09,shield09a} or a binary SMBH system \citep{dotti09,bogdan09}.  Another candidate with an even higher (3500~\kms) offset is most likely a double-peaked emitter \citep{shield09b}.  Recently, \citet{civano10} have suggested that an unusual galaxy discovered in the COSMOS survey by \citet{comerf09b}, which those authors proposed to be a dual SMBH system, may in fact be a recoiling BH, as new spectra indicate a kinematic offset of 1200~\kms.  This candidate has a less extreme velocity than the others, but further observations are needed to confirm the nature of this object.  Additionally, the SMBH in M87 has recently been observed to be spatially offset by $\sim 7$ pc, which could possibly be explained by a past recoil event \citep{batche10}.

In addition to producing direct observational signatures, GW recoil may play a role in the co-evolution of SMBHs and their host galaxies.  Strong empirical correlations exist between SMBH mass and properties of the host galaxy bulge, including the bulge luminosity, mass, and stellar velocity dispersion \citep[e.g.,][]{korric95, magorr98, gebhar00,fermer00, merfer01,tremai02, marhun03}.  These correlations are well-reproduced by galaxy formation models in which much of BH and galaxy bulge growth occurs via successive mergers, and in which merger-triggered BH fueling is self-regulated via AGN feedback \citep[e.g.,][]{silree98, wyiloe03, dimatt05, hopkin06a}.  However, because GW recoil events may occur simultaneously with this rapid BH accretion phase, recoils could significantly disrupt the coordinated growth of BHs and galaxy bulges.  In particular, GW recoil may contribute to scatter in the $M_{\rm BH} - \sigma_*$ relation caused by ejected \citep{volont07} or bound recoiling \citep{bleloe08,sijack10} BHs; we will examine the latter possibility in greater detail.  It is also unclear {\em a priori} what effects, if any, GW recoil may have on the host galaxies themselves.  In purely collisionless galaxies, \citet{boylan04} and \citet{guamer08} have shown that bound recoiling BHs may scour out a stellar core.  In gas-rich galaxy mergers, copious of amounts of cold gas are driven to the central galactic region during coalescence, triggering a luminous starburst such that the galaxy may appear as a ULIRG \citep{sander88a,sanmir96}.  Feedback from a central AGN may terminate the starburst phase by expelling gas and dust from the central region \citep[e.g.,][]{hopkin06a,hopkin08c,somerv08}.  We will investigate whether observable starburst properties or gas and stellar dynamics may be affected by the sudden displacement of this central AGN via recoils \citep[see also recent work by][]{sijack10}.

Our current work was initiated as a follow-up of \citet{bleloe08}, hereafter BL08.  BL08 explored the trajectories and accretion of BHs on bound orbits in a static potential with stellar bulge and gaseous disk components.  They found that recoil kicks $\la$~\vesc~could produce long-lived ($\ga$ Gyr) oscillations of the BH; similar results were found by \citet{guamer08}, \citet{korlov08} \& \citet{guedes09}.  BL08 also demonstrated that GW recoil can affect SMBH growth even when the BH is not ejected entirely from the host galaxy.  By removing accreting BHs from the central dense region and thereby limiting the BH's fuel supply, BHs receiving recoil kicks $\ga 0.5$~\vesc~may be less massive than stationary BHs.   

In the present study, we use hydrodynamic simulations to self-consistently calculate the dynamics and accretion of recoiling BHs in realistic merger remnant potentials.  We simulate galaxy mergers using {\footnotesize GADGET-3} \citep{spring05a}, an SPH/N-body code, and apply a recoil kick to the BH at the time of BH merger.  Due to the detailed initial conditions and many free parameters involved in such merger simulations, we conduct a large parameter study with dozens of galaxy merger models and a wide range of kick velocities.  Each merger model is simulated with at least one kick velocity and also with no recoil kick, for comparison.  With this suite of recoil simulations, we are able to observe trends in the behavior of recoiling BHs in different environments.  When paired with our time-dependent, sub-resolution models for recoiling BH accretion, this approach allows us to, for example, determine velocity-dependent recoiling AGN lifetimes and estimate the effect of GW recoil on scatter in the $M_{\rm BH}-\sigma_*$ relation.

As this paper was in the final stages of preparation, two papers appeared that also involve hydrodynamic simulations of recoiling BHs in galaxy merger remnants \citep{guedes10, sijack10}.  However, each study used only a few simulations, probing only a small part of parameter space.  \citet{guedes10} used results of three merger simulations as initial conditions, and ran hydrodynamic simulations of recoiling BHs in these merger remnants for a short time.  These initial trajectories were used to calibrate semi-analytic calculations of the recoil trajectory in the remnant potential.  A range of kick velocities was tested in each of the three merger models used.  \citet{sijack10} simulate recoils in an isolated, stable disk galaxy; they also use this galaxy model as the initial condition for a full merger simulation with GW recoils.  Both studies also include prescriptions for accretion onto recoiling BHs.  We provide a comparison of our results to the findings of these papers in \S~\ref{sec:disc}.

Our simulation methods, GW recoil treatment, and galaxy merger models are outlined in \S~\ref{sec:simulations}. ~\S~\ref{sec:premrg} -~\ref{sec:accr_fb} contain our results, organized as follows.  In \S~\ref{sec:premrg}, the general characteristics of our galaxy merger simulations with stationary central BHs (i.e., no recoil kicks).  We discuss the variation in merger dynamics and remnant morphologies between models.  In \S~\ref{sec:bh_dynamics}, the dynamics of recoiling BHs are discussed.  We describe the general characteristics of recoil trajectories and trends between models in \S~\ref{ssec:bh_traj}.  In \S~\ref{ssec:tmrg}, we analyze the sensitivity of recoil trajectories to the BH merger time, and in \S~\ref{ssec:kick_orient} we investigate the dependence of recoil trajectories on the direction of the kick.  \S~\ref{sec:accr_fb} contains our results for the accretion and feedback of recoiling BHs.  In \S~\ref{ssec:bondi_only}, we examine accretion and feedback of recoiling BHs, compare to the fueling of stationary BHs, and calculate AGN lifetimes from Bondi-Hoyle accretion.  In \S~\ref{ssec:ejdisk}, we describe our time-dependent analytic model for calculating accretion onto a recoiling BH from a disk of gas ejected with the BH.  Using this accretion model, along with accretion from ambient gas, we present in \S~\ref{ssec:offsetagn} the velocity-dependent AGN luminosities and active lifetimes for recoiling BHs.  In particular, we calculate lifetimes for spatially-offset and kinematically-offset recoiling AGN.  In \S~\ref{ssec:msigma}, we compare the $M_{\rm BH}-\sigma_*$ relations derived from our set of no-recoil simulations and our set of high-velocity recoil simulations, and discuss implications for the observed $M_{\rm BH}-\sigma_*$ relation.  Finally, in \S~\ref{ssec:centralevol}, we explore the effects of GW recoil on star formation rates and the host galaxy structure.  We summarize and discuss our results in \S~\ref{sec:disc}.

\begin{table*}
\begin{center}
\begin{tabular}{l r r r r r r r r r r r r r r }

\multicolumn{10}{c}{Initial model parameters} & & \multicolumn{4}{c}{Resulting merger quantities}\\ 
\cline{1-10} \cline{12-15} 
Model & $q$ & $M_{\rm tot}$& $M_{\rm disk}$ & $f_{\rm gas}$ & $R_{\rm peri}$  & $\theta_1$ & $\phi_1$& $\theta_2$ & $\phi_2$  & & \tmrg & \vesc(\tmrg) & $v_{\rm esc,max}$ & \% diff. \\
 & &  [$10^{10}$~\msun] & [$10^{10}$~\msun] &  & [kpc] & [deg] & [deg] &  [deg] & [deg]  & & [Gyr] & [\kms] & [\kms]  & \\
\cline{1-10} \cline{12-15} 

{\bf q1fg0.6a} & {\bf 1.0} & {\bf 272.1} & {\bf 11.16} & {\bf 0.6} & {\bf 7.1} & {\bf 30} & {\bf 60} & {\bf -30} & {\bf 45}  & & {\bf 1.55} & {\bf 1258} & {\bf 1494}  & {\bf 18.7}  \\
q1fg0.5a & 1.0 & 272.1 & 11.16 & 0.5 &  7.1 & 30 & 60 & -30 &  45  & &			1.58 &	1157 &		1568  &		35.6  \\
q1fg0.4a & 1.0 & 272.1 & 11.16 & 0.4 & 7.1 & 30 & 60 & -30 & 45  && 			1.60 & 	1102 & 		1472 & 		33.6  \\
{\bf q1fg0.3a} & {\bf 1.0} & {\bf 272.1} & {\bf 11.16} & {\bf 0.3} & {\bf 7.1} & {\bf 30} & {\bf 60} & {\bf -30} & {\bf 45} & & {\bf 1.63} & {\bf 1118} &	{\bf 1337} & {\bf 19.6}  \\
q1fg0.3b & 1.0 & 272.1 & 11.16 & 0.3 & 14.3 & 30 & 60 & -30 & 45 & & 			2.01 &	1165 &		1256  &		7.8  \\
q1fg0.3c & 1.0 & 272.1 & 11.16 & 0.3 & 7.1 & 150 & 60 & -30 & 45 & & 			1.70 &	1113 &		1264  &		13.6  \\
q1fg0.3d & 1.0 & 272.1 & 11.16 & 0.3 & 7.1 & 150 & 0 & -30 & 45 & &			1.69 &	1152 &		1271  &		10.4  \\
q1fg0.3e & 1.0 & 272.1 & 11.16 & 0.3 & 7.1 & 90 & 60 & -30 & 45 & &			1.76 &	970 &		1194  &		23.1  \\
q1fg0.3M20w & 1.0 & 5442 & 223.1 & 0.3 & 19.4 & 30 & 60 & -30 & 45  && 		1.54	 & 	2555	 & 		2727 & 		6.7  \\
q1fg0.3M10x & 1.0 & 2721 & 111.6 & 0.3 & 15.4 & 30 & 60 & -30 & 45  && 		1.69	 & 	2067 & 		2290 & 		10.8  \\
q1fg0.3M0.1y & 1.0 & 27.21 & 1.116 & 0.3 & 3.3 & 30 & 60 & -30 & 45  && 		1.60	 & 	457 & 		641 & 		40.3  \\
q1fg0.3M0.05z & 1.0 & 13.61 & 0.5578 & 0.3 & 2.7 & 30 & 60 & -30 & 45  && 		2.00	 & 	338 & 		497 & 		47.0  \\
q1fg0.2a & 1.0 & 272.1 & 11.16 & 0.2 & 7.1 & 30 & 60 & -30 & 45 & &			1.67 &	1049 &		1182  &		12.6  \\
{\bf q1fg0.1a} & {\bf  1.0} & {\bf  272.1} & {\bf  11.16} & {\bf  0.1} & {\bf  7.1} & {\bf  30} & {\bf  60} & {\bf  -30} & {\bf  45} & & {\bf 1.72} & {\bf 876} & {\bf 931}  & {\bf 6.3}  \\
q1fg0.1b &  1.0 &  272.1 &  11.16 &  0.1 &  14.3 &  30 &  60 &  -30 &  45 & &		2.09 &	921 &		923  &		0.2  \\
q1fg0.1c &  1.0 &  272.1 &  11.16 &  0.1 &  7.1 &  150 &  60 &  -30 &  45 & &		1.80 &	891 &		956  &		7.3  \\
q1fg0.1d &  1.0 &  272.1 &  11.16 &  0.1 &  7.1 &  150 &  0 &  -30 &  45 & &		1.82 &	863 &		906  &		5.0  \\
q1fg0.1e &  1.0 &  272.1 &  11.16 &  0.1 &  7.1 &  90 &  60 &  -30 &  45 & &		1.84 &	860 &		918  &		6.7  \\
q1fg0.1M10x & 1.0 & 2721 & 111.6 & 0.1 & 15.4 & 30 & 60 & -30 & 45  && 		1.74	 & 	1924 &	 	2033	 & 		5.7  \\
q1fg0.04a & 1.0 & 272.1 & 11.16 & 0.04 & 7.1 & 30 & 60 & -30 & 45 & &			1.76 &	823 &		848  &		3.1  \\
{\bf q1fg0a} & {\bf 1.0} & {\bf 272.1} & {\bf 11.16} & {\bf 0.0} & {\bf 7.1} & {\bf 30} & {\bf 60} & {\bf -30} & {\bf 45} && {\bf 1.84} & {\bf 772} & {\bf 790}  & {\bf 2.3}  \\
q0.9fg0.4a & 0.9 & 258.5 & 10.60 & 0.4 & 7.1 & 30 & 60 & -30 & 45 & &			1.61 &	1104 &		1455  &		31.8  \\
q0.9fg0.3a & 0.9 & 258.5 & 10.60 & 0.3 & 7.1 & 30 & 60 & -30 & 45 & &			1.65 &	1154 &		1337  &		15.9  \\
q0.9fg0.1a & 0.9 & 258.5 & 10.60 & 0.1 & 7.1 & 30 & 60 & -30 & 45 & &			1.74 &	866 &		918  &		6.1  \\
q0.75fg0.4a & 0.75 & 238.1 & 9.762 & 0.4 & 7.1 & 30 & 60 & -30 & 45 & &		1.69 &	1113 &		1302  &		17.1  \\
q0.75fg0.3a & 0.75 & 238.1 & 9.762 & 0.3 & 7.1 & 30 & 60 & -30 & 45 & &		1.73 &	1090 &		1225  &		12.3  \\
q0.75fg0.1a & 0.75 & 238.1& 9.762  & 0.1 & 7.1 & 30 & 60 & -30 & 45 & &		1.83 &	810 &		867  &		7.1  \\
q0.667fg0.4a & 0.667 & 226.8 & 9.297 & 0.4 & 7.1 & 30 & 60 & -30 & 45 & & 		1.68 &	1086 &		1219  &		12.3  \\
q0.667fg0.3a & 0.667 & 226.8 & 9.297 & 0.3 & 7.1 & 30 & 60 & -30 & 45 & &		1.73 &	883 &		1166  &		32.1  \\
q0.667fg0.1a & 0.667 & 226.8 & 9.297 & 0.1 & 7.1 & 30 & 60 & -30 & 45 & &  		1.85 &	846 &		847  &		0.1  \\
{\bf q0.5fg0.6a}  & {\bf 0.5} & {\bf 204.1} & {\bf 8.368} & {\bf 0.6} & {\bf 7.1} & {\bf 30} & {\bf 60} & {\bf -30} & {\bf 45} &&	{\bf 1.88} & {\bf 914} & {\bf 1116}  & {\bf 22.1}  \\
q0.5fg0.5a  & 0.5 & 204.1 & 8.368 & 0.5 & 7.1 & 30 & 60 & -30 & 45 & & 			1.89 &	963 &		1076  &		11.7  \\
q0.5fg0.4a & 0.5 & 204.1 & 8.368 & 0.4 & 7.1 & 30 & 60 & -30 & 45 && 			1.95 & 	893 & 		983 & 		10.1  \\
q0.5fg0.4b & 0.5 & 204.1 & 8.368 & 0.4 & 14.3 & 30 & 60 & -30 & 45 & &			2.46 &	965 &		1060  &		9.8  \\
q0.5fg0.4c & 0.5 & 204.1 & 8.368 & 0.4 & 7.1 & 150 & 60 & -30 & 45  &&			1.90 &	1024 &		1112  &		8.6  \\
q0.5fg0.4d & 0.5 & 204.1 & 8.368 & 0.4 & 7.1 & 150 & 0 & -30 & 45  & &			1.90 &	1061 &		1117  &		5.2  \\
{\bf q0.5fg0.3a} & {\bf 0.5} & {\bf 204.1} & {\bf 8.368} & {\bf 0.3} &  {\bf 7.1} & {\bf 30} &  {\bf 60} &  {\bf -30} & {\bf 45}  && {\bf 1.97} & {\bf 993} & {\bf 994}  & {\bf 0.1}  \\
q0.5fg0.3b &  0.5 & 204.1 & 8.368 & 0.3 &  14.3 & 30 &  60 &  -30 &  45  & &		2.45 &	920 &		1003 &		9.0  \\
q0.5fg0.3c &  0.5 & 204.1 & 8.368 & 0.3 &  7.1 & 150 &  60 &  -30 &  45  & &		1.97 &	1012 &		1017  &		0.5  \\
q0.5fg0.3d &  0.5 & 204.1 & 8.368 & 0.3 &  7.1 & 150 &  0 &  -30 &  45  & &		1.92 &	1014 &		1069  &		5.4  \\
q0.5fg0.3e &  0.5 & 204.1 & 8.368 & 0.3 &  7.1 & 90 &  60 &  -30 &  45  & &		2.05 &	915 &		1126  &		23.1  \\
q0.5fg0.3f  &  0.5 & 204.1 & 8.368 & 0.3 & 7.1 & 90 & 0 & 0 & 0 & &				1.94 &	978 &		1029  &		5.2  \\
q0.5fg0.3g  &  0.5 & 204.1 & 8.368 & 0.3 & 7.1 & 60 & 60 & 150 & 0 & &			2.23 &	805 &		820  &		1.9  \\
q0.5fg0.3h  &  0.5 & 204.1 & 8.368 & 0.3 & 7.1 & 0 & 0 & 0 & 0 & &				1.84 &	927 &		1111  &		19.8  \\
q0.5fg0.3i  &  0.5 & 204.1 & 8.368 & 0.3 & 7.1 & 180 & 0 & 0 & 0 & & 			1.84 &	1028 &		1146  &		11.4  \\
q0.5fg0.3j  & 0.5 & 204.1 & 8.368 & 0.3 & 7.1 & 180 & 0 & 180 & 0 & &			2.09 &	823 &		930  &		13.0  \\
q0.5fg0.3k  &  0.5 & 204.1 & 8.368 & 0.3 & 7.1 & 10 & 0 & -10 & 0 & &			1.84 &	1046 &		1087  &		4.0  \\
q0.5fg0.3M10x & 0.5 & 2041 & 83.68 & 0.3 & 15.4 & 30 & 60 & -30 & 45  && 		2.02	& 	2081	 &	 	2070	 & 		-0.5 \\
q0.5fg0.2a &  0.5 & 204.1 & 8.368 &  0.2 &  7.1 &  30 &  60 &  -30 &  45  & &		2.02 &	830 &		862  &		3.8  \\
{\bf q0.5fg0.1a} & {\bf  0.5} & {\bf  204.1} & {\bf  8.368}  & {\bf  0.1} & {\bf  7.1} & {\bf  30} & {\bf  60} & {\bf  -30} & {\bf  45} && {\bf 2.12} & {\bf 777} & {\bf 798} & {\bf 2.7}  \\
q0.5fg0.1b &  0.5 &  204.1 &  8.368  &  0.1 &  14.3 &  30 &  60 &  -30 &  45  & &		2.56 &	812 &		813  &		0.1  \\
q0.5fg0.1c &  0.5 &  204.1 &  8.368  &  0.1 &  7.1 &  150 &  60 &  -30 &  45  & &		2.12 &	811 &		813  &		0.2  \\
q0.5fg0.1d &  0.5 &  204.1 &  8.368  &  0.1 &  7.1 &  150 & 0 &  -30 &  45  & &		2.12 &	814 &		810  &		-0.5  \\
q0.5fg0.1M10x & 0.5 & 2041 & 83.68 & 0.1 & 15.4 & 30 & 60 & -30 & 45  && 		2.07	& 	1737 & 		1737 & 		0.0  \\
q0.5fg0.04a &  0.5 & 204.1 & 8.368 &  0.04 &  7.1 &  30 &  60 &  -30 &  45  &&		2.18 &	740 &		748  &		1.0  \\
{\bf q0.5fg0a} & {\bf  0.5} & {\bf  204.1} & {\bf  8.368}  & {\bf  0} & {\bf  7.1} & {\bf  30} & {\bf  60} & {\bf  -30} & {\bf  45}  && {\bf 2.30} & {\bf 689} & {\bf 706}  & {\bf 2.4}  \\
q0.333fg0.4a & 0.333 & 181.4 & 7.438 &  0.4 &  7.1 &  30 &  60 &  -30 &  45 & &	2.28 &	810 &		816  &		0.7  \\
q0.333fg0.3a & 0.333 & 181.4 & 7.438 &  0.3 &  7.1 &  30 &  60 &  -30 &  45 & &	2.38 &	856 &		858  &		0.3  \\
q0.333fg0.1a & 0.333 & 181.4 & 7.438 &  0.1 &  7.1 &  30 &  60 &  -30 &  45 & &	2.57 &	741 &		744  &		0.3  \\
q0.25fg0.4a & 0.25 & 170.1 & 6.974 &  0.4 &  7.1 &  30 &  60 &  -30 &  45 & &		2.94 &	829 &		830  &		0.1  \\
q0.25fg0.3a & 0.25 & 170.1 & 6.974 &  0.3 &  7.1 &  30 &  60 &  -30 &  45 & &		2.90 &	852 &		852  &		-0.1  \\
q0.25fg0.1a & 0.25 & 170.1 & 6.974 &  0.1 &  7.1 &  30 &  60 &  -30 &  45 & &	3.18 &	722 &		725  &		0.3   
\end{tabular}
\end{center}
\caption{Galaxy merger models.  Boldface entries denote those for which we have varied~\vk.  Fiducial-mass merger models are labeled as q[{\em value}]fg[{\em value}][{\em orb}], where ``q" is the galaxy mass ratio, ``fg" is the initial gas fraction, and each letter {\em orb} corresponds to a specific orbit (``a" being our ``fiducial" configuration).  High- and low-mass models are denoted by q[{\em value}]fg[{\em value}]M[{\em factor}][{\em orb}], where ``M" is the ratio of the primary galaxy mass to the fiducial mass.  \label{table:models}}
\end{table*}

\section{Simulations}
\label{sec:simulations}

For our simulations of galaxy mergers, we employ the smoothed particle dynamics (SPH) code {\footnotesize GADGET-3} \citep{spring05a}, which conserves both energy and entropy \citep{sprher02}.  The code includes radiative cooling as well as a subresolution model for a multiphase interstellar medium (ISM) \citep{sprher03} that accounts for star formation and supernova feedback.  In addition, the code models BHs as gravitational ``sink" particles that contain a BH seed and a gas reservoir.  The reservoir is replenished by stochastic accretion of neighboring gas particles, but the actual accretion rate onto the BH is calculated smoothly using the Bondi-Hoyle-Lyttleton formula with locally-averaged values for the density and sound speed.  Because the gas around the BH cannot be resolved at higher densities below the spatial resolution limit, the accretion rate calculated from these values is multiplied by a constant factor.  Following other authors, we assume a value of 100 for this factor \citep[e.g.][]{spring05b,hopkin06c,johans09}.  Angular momentum is conserved during accretion of gas particles, but because this is a stochastic process owing to the finite mass resolution, we also introduce an analytic accretion drag force ($\propto \dot M \, v$) calculated from the Bondi accretion rate at each timestep.  These prescriptions are described in more detail in \citet{spring05b}.  

We note that, as in all hydrodynamic simulations involving BH accretion, the exact nature of the accretion on sub-resolution scales is unknown, so it is necessary to assume a sub-resolution model.  While the validity of the Bondi-Hoyle-Lyttleton formula (with a multiplicative factor) in our framework is indeed an assumption, few constraints exist on the gas flow from large scales down to the BH.  The Bondi-Hoyle-Lyttleton model provides a physically-motivated approximation of the gas captured by a BH, including a BH in motion, and its extensive use in the literature helps to place our work in a larger context of galaxy evolution studies.  We further assume that the accretion factor remains constant even after a GW recoil event.  This latter assumption is justified in that the spatial resolution limit, which motivates the use of this factor, always remains constant.  Furthermore, owing to the sink-particle treatment of BHs in the code, the accretion factor does not determine the amount of gas bound to the BH at each timestep; it influences only how this gas is accreted onto the BH.  From a practical standpoint, using the same accretion prescription throughout enables a more direct comparison between recoiling and stationary BHs -- one of the main goals of this study.

Because we are presently interested in GW recoil, we allow for an arbitrary velocity to be added to the remnant BH at the time of the BH merger,~\tmrg.  To more easily compare recoil events in different merger models, we scale this velocity to the central escape speed at the time of merger,~\vesc(\tmrg).  We define~\vesc~$= \sqrt{-2\Phi({\bf x}_{\rm BH})}$ at each timestep, i.e., the escape speed at the position of the BH, which is close to the position of the potential minimum in the absence of a recoil kick. \citet{campan07a}, \citet{baker08}, \& \citet{vanmet10} have all fit distributions to the data from full NR simulations of BH mergers, with results in good agreement.  For major mergers ($0.25 < q < 1$, the range we consider in our simulations), high spins ($a_1 = a_2 = 0.9$), and randomly-distributed spin orientations, \citet{vanmet10} find that 69.8\% of recoil kicks will be above 500~\kms~and 35.3\% will be above 1000~\kms.  If the BH spin magnitudes are also randomly distributed ($0 \leq a_1, a_2 \leq 1$), the these fractions are 41.6\% and 13.5\%.  Our fiducial-mass merger simulations (described below) have~\vesc~$=  689 - 1248$~\kms, and we also use low-mass simulations with~\vesc~as low as 338~\kms.  The NR results indicate that there is a substantial probability of BHs in these models receiving kicks up to $\sim$~\vesc.  In our high-mass simulations, however,~\vesc~$= 1737 - 2555$~\kms, so the probability of recoil events with~\vk~$\sim$~\vesc~is much lower in these models.  

We also slightly modify the treatment of BH mergers in the code.  In the standard {\footnotesize GADGET} prescription, the BHs merge when they are separated by less than a gravitational softening length ($a_{\rm sep} < R_{\rm soft}$) and have a relative velocity $v_{\rm rel} < 0.5\, c_{\rm s}$, where $c_{\rm s}$ is the local sound speed.  If, as can happen in gas-rich mergers, the central escape velocity is increasing rapidly when the BHs merge, small variations in~\tmrg~can result in very different BH trajectories (see Fig.~\ref{fig:q1tmrg_rmax}).  In order to reliably compare results from different simulations, we therefore force the BHs to merge at a given time that is predetermined as follows.  We define the ``coalescence time"~\tcoal~as the earliest time after which the BH binary is tight enough that it could plausibly merge.  Given the numerical uncertainties near the spatial resolution limit, we generously choose~\tcoal~to be the time when the BH separation falls below $a_{\rm sep} = 10\, R_{\rm soft}$ (and $v_{\rm rel} < 0.5\, c_{\rm s}$, as before).  For reference, $R_{\rm soft}= 80$ pc in our fiducial-mass simulations (see below).  Then, restarting the simulation slightly before~\tcoal, we force the BHs to merge at a predetermined time~\tmrg~$ \ge $~\tcoal. ~\tmrg~can then be varied as a free parameter, allowing us to systematically probe the range of possible BH merger times.  In practice, this is only necessary for nearly equal-mass, gas-rich mergers in which~\vesc~varies significantly throughout the merger.  In other cases the results are insensitive to the choice of merger time, so~\tmrg~is simply chosen to be~\tcoal.  

For the majority of our simulations, we use the same total mass, $1.36\times10^{12}$~\msun, for the primary galaxy, and scale the secondary to yield the desired mass ratio.  The exception is a small subset of simulations with lower and higher total mass, described below.  For all of our ``fiducial mass" simulations, we use a gravitational softening length of $R_{\rm soft} = 80$ pc for baryons and $R_{\rm soft, DM} = 240$ pc for the dark matter (DM).  In each galaxy, 4.1\% of the total mass is in a baryonic disk component.  The primary galaxy has $N_{\rm halo} = 4.8\times10^5$ DM halo particles, and for $f_{\rm gas} = 0.4$ it has $N_{\rm disk} = 3.2\times10^5$ disk particles, with equal numbers of gas and star particles.  In all other models, $N_{\rm halo}$ and $N_{\rm disk}$ are set such that the same particle mass resolution is maintained for each component ($m_{\rm star} = 4.2\times10^5$~\msun, $m_{\rm gas} = 2.8\times10^5$~\msun, \& $m_{\rm halo} = 5.4\times10^6$~\msun).  

Each galaxy is given a single BH particle, which as previously stated, consists of a ``seed" BH along with a gas reservoir from which an accretion rate is calculated smoothly.   We use a small initial seed mass for the BH, $1.43 \times 10^5$~\msun, in accordance with the $M_{\rm BH}$ - $M_{\rm bulge}$ \citep{magorr98} relation, because our initial galaxies are pure disks with no bulge component.  We find that our results are insensitive to the choice of seed mass.  The total mass of this hybrid particle is the ``dynamical" mass of the BH; this is the mass that is used for gravitational force calculations.  This dynamical BH mass is set to $10^{-5}$ of the total galaxy mass initially.  This value is chosen such that the seed mass and dynamical mass have similar values by the time of the recoil kick, to avoid a disparity between the mass used for gravitational forces and that used for accretion physics of the recoiling BH.  To help ensure that the BH remains in the center of the galaxy prior to the BH merger and recoil kick, we set the accretion drag force equal to its value for Eddington-limited accretion prior to the BH merger (though the actual accretion rate is still calculated self-consistently based on the Bondi-Hoyle formula).  Once the BHs merge, the accretion drag force is calculated using the actual Bondi-Hoyle accretion rate for the remainder of the simulation.

Our set of fiducial-mass simulations covers a wide range of galaxy mass ratios and gas fractions, but in order to probe the effects of GW recoil on black hole - host galaxy relationships, as we do in \S~\ref{ssec:msigma}, we require a larger range in total galaxy mass.  We have run a sample of seven high and low mass simulations with a primary galaxy mass ranging from 0.05 - 20 times the fiducial primary mass, which we refer to as $M_{\rm 0}$.  For the lower-mass runs, we maintain the same number of particles as in the fiducial simulations, such that higher mass resolution is achieved.  Accordingly, we reduce the softening lengths in these simulations by $(M / M_{\rm 0})^{1/3}$.  For the higher-mass runs ($M_{\rm 0}\times10$ \& $M_{\rm 0}\times20$), an increase in particle number by this factor would be prohibitively computationally expensive, so we instead increase the particle number by a factor of 5 in each case, compromising a factor of 2 and 4 in mass resolution, respectively.  In the latter case, we maintain a reasonable $M_{\rm BH}/M_{\rm particle}$ ratio by increasing the initial dynamical mass by a factor of 10.  

\begin{figure}
\resizebox{\hsize}{!}{\includegraphics{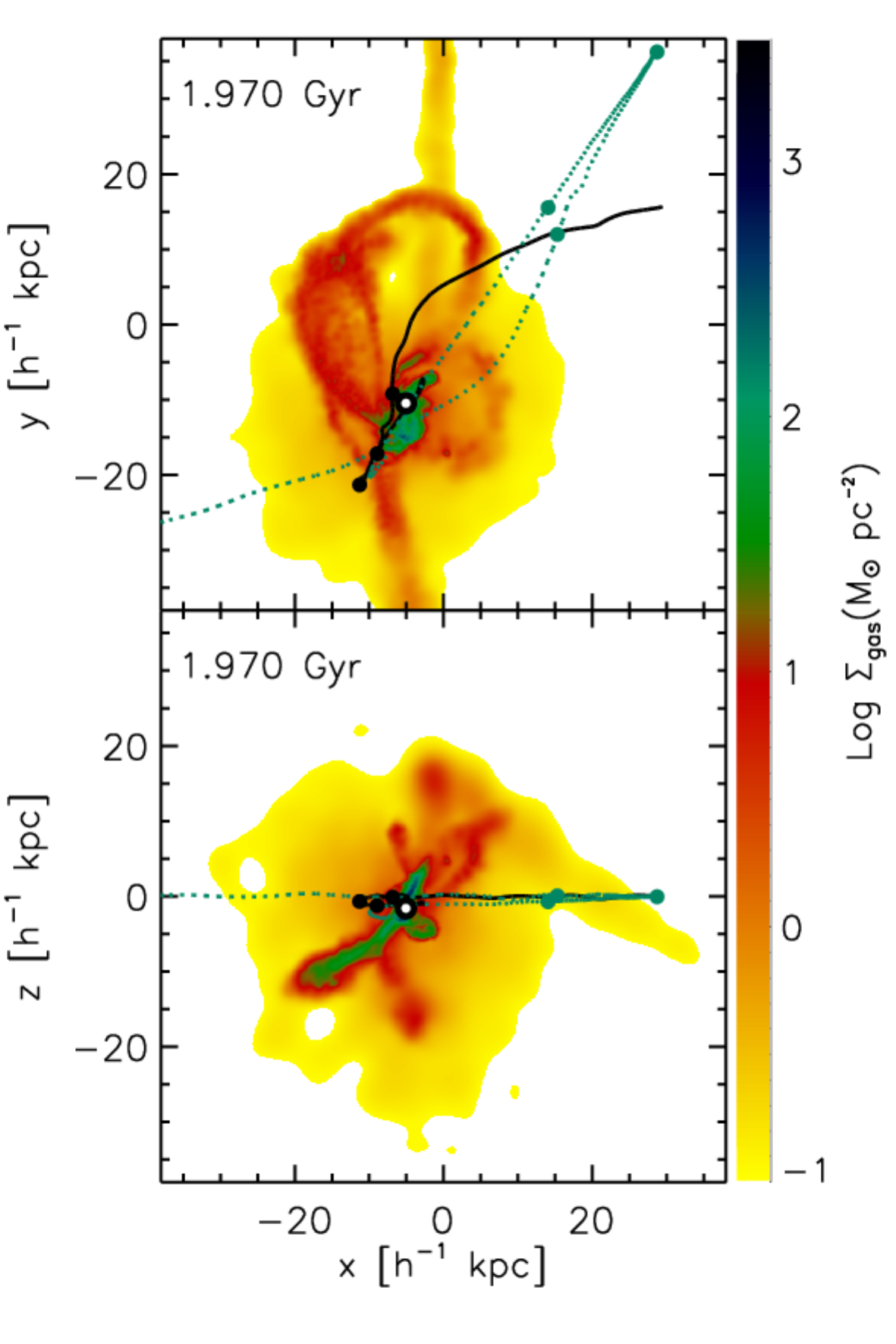}}
%\resizebox{\hsize}{!}{\includegraphics{qhalf_fg30_tc_nokick_197_gas_2proj.png}}
\caption[]{Fiducial example of a galaxy merger with BHs but no recoil kick (model q0.5fg0.3a).  The gas distribution at time of BH merger (1.97 Gyr) is shown in both the $x$-$y$ ({\em top}) and $x$-$z$ ({\em bottom}) projections, with the density scale shown on the right-hand side.  The black solid and green dotted lines show the pre-merger path of the BHs corresponding to the larger and smaller galaxies, respectively. The filled dots denote the BH positions at 500 Myr intervals.  The black-and-white dot indicates the position of the merged BH at the moment of merger.  \label{fig:v0_contour_gas}}
\end{figure}

\begin{figure}
\resizebox{\hsize}{!}{\includegraphics{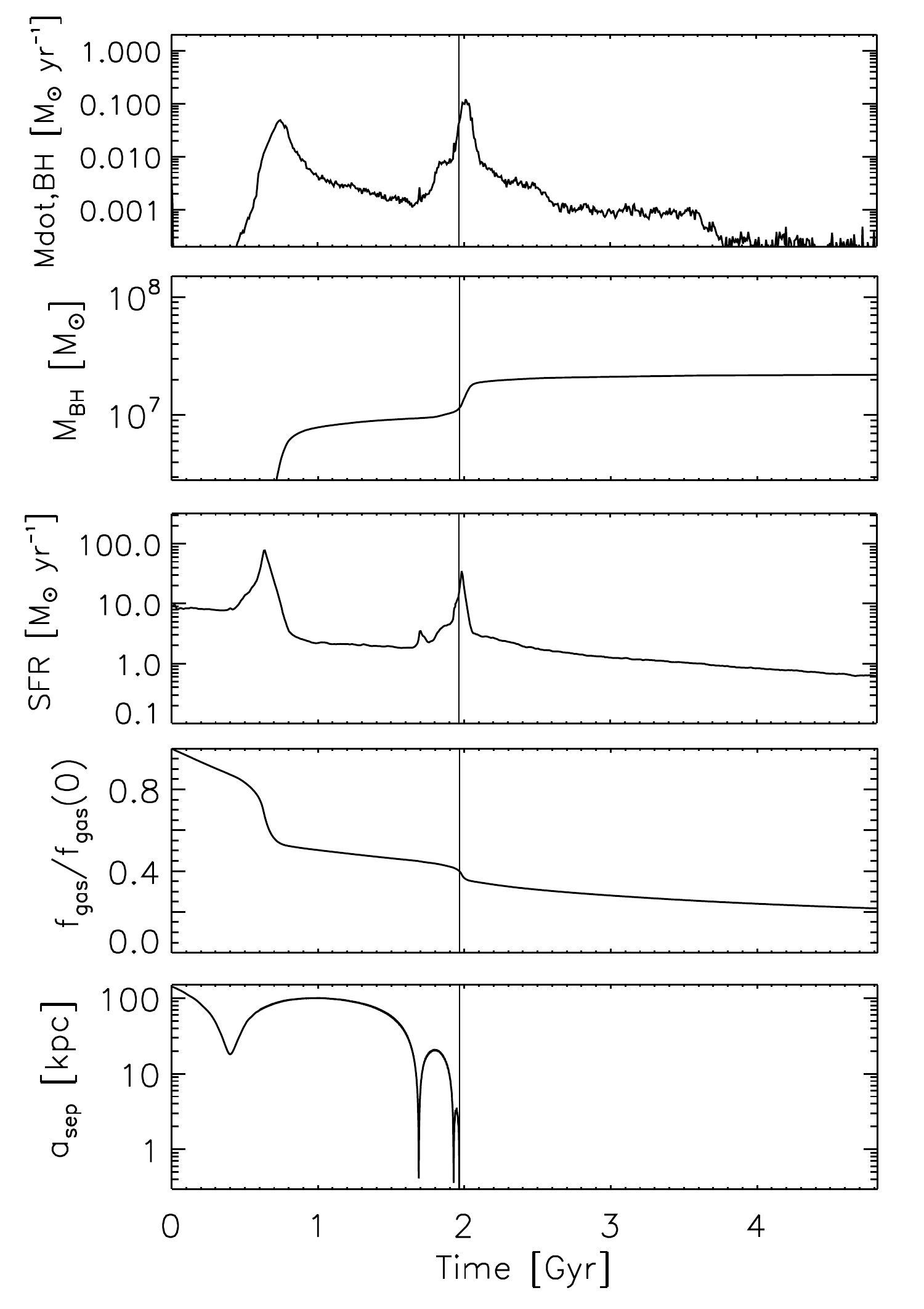}}
%\resizebox{\hsize}{!}{\includegraphics{qhalf_fg30_bh_sep_sfr_fgas.pdf}}
\caption[]{For the same simulation shown in Fig.~\ref{fig:v0_contour_gas} (model q0.5fg0.3a), the evolution of following quantities throughout the simulation is shown, from top to bottom: BH accretion rate ($\dot M_{\rm BH}$), global star formation rate, normalized galaxy gas fraction ($f_{\rm gas}/f_{\rm gas,0}$), BH mass ($M_{\rm BH}$), and BH separation prior to merger. Vertical lines indicate the time of BH merger. \label{fig:v0_sep_sfr_fgas}}
\end{figure}

In order to evaluate the sensitivity of our results to numerical artifacts such as the choice of mass resolution, softening length, and integration accuracy, we have run a number of additional simulations in which we systematically vary these parameters.  In addition to these tests, two of the models included in our results have higher mass resolution (and correspondingly higher spatial resolution) than our fiducial runs: our low-mass merger models have the same number of particles as our fiducial-mass models and thus have 10 and 20$\times$ higher mass resolution.  Higher mass and spatial resolution reduce the noisiness of the potential, and varying the resolution changes the detailed structure of the gas, especially in the highest density regions.  Higher integration accuracy also reduces numerical noise.  Consequently, the exact trajectory, and hence the settling time, of a given recoil event will be affected to some extent by these numerical factors.  However, the variation in recoil trajectories for different choices of resolution and integration accuracy are small compared to the differences for varying kick speed and merger remnant properties.  We also define the settling radius of the BH to be $4 \, R_{\rm soft}$ to avoid sensitivity of our conclusions to the BH motion in the innermost galaxy region where the softened potential may dominate.  Most importantly, for all choices of mass resolution, spatial resolution, and integration accuracy parameters tested, the qualitative behavior of the recoiling BHs and the relative differences between simulations are robust.  Therefore our main results do not depend on these numerical factors. 

Table~\ref{table:models} lists the parameters for the galaxy merger models we have constructed.  We have tested a total of 62 different galaxy merger models in which we vary the galaxy mass ratio ($q$), the gas fraction (\fgas), the orbital configuration.  55 of these models use the ``fiducial" primary galaxy mass, and for easier reference we assign each of these a name given by q[{\em value}]fg[{\em value}][{\em orb}], where ``q" denotes the galaxy mass ratio, ``fg" denotes the initial gas fraction, and each letter {\em orb} is identified with a specific orbital configuration.  The remaining seven models have higher or lower total mass and are referred to by q[{\em value}]fg[{\em value}]M[{\em factor}][{\em orb}], where ``M" denotes the total galaxy mass relative to the equivalent fiducial-mass galaxy.  

In all cases, the orbital configuration is specified by six parameters, and the galaxies are initially set on parabolic orbits.  The choice of impact parameter for the first pericentric passage, $R_{\rm peri}$, and initial galaxy separation, $a_{\rm i}$, determine the initial orbital energy and angular momentum.  In our fiducial-mass models, we set $a_{\rm i} = 143$ kpc and $R_{\rm peri} = 7.1$ kpc, except for orbit ``b", which has $R_{\rm peri} = 14.3$ kpc.  In our high- and low-mass models, $a_{\rm i}$ and $R_{\rm peri}$ are scaled such that $a_{\rm i}$ is the same fraction (0.625) of the virial radius, $R_{\rm 200}$, and  $a_{\rm i}$/$R_{\rm peri} = 20$.  We denote these orbits with letters w - z to differentiate them from the orbits a - k of our fiducial-mass mergers.  Note that the orbital parameters do not remain constant as the merger progresses, owing to energy losses via dynamical friction.  The angles ($\theta_1, \phi_1$) and ($\theta_2, \phi_2$) determine the initial orbital phase and inclination, respectively, of each galaxy.  These angles, as well as $R_{\rm peri}$ for each galaxy model are given in Table~\ref{table:models}.  Our fiducial merger orbit (orbit ``a") is a ``generic" orbit, in that no symmetries exist in the initial orientation angles of the disks.  Note that many of these orbits are identical to those used in \citet{cox06} and \citet{robert06c}.

\begin{figure*}
\resizebox{0.33\hsize}{!}{\includegraphics{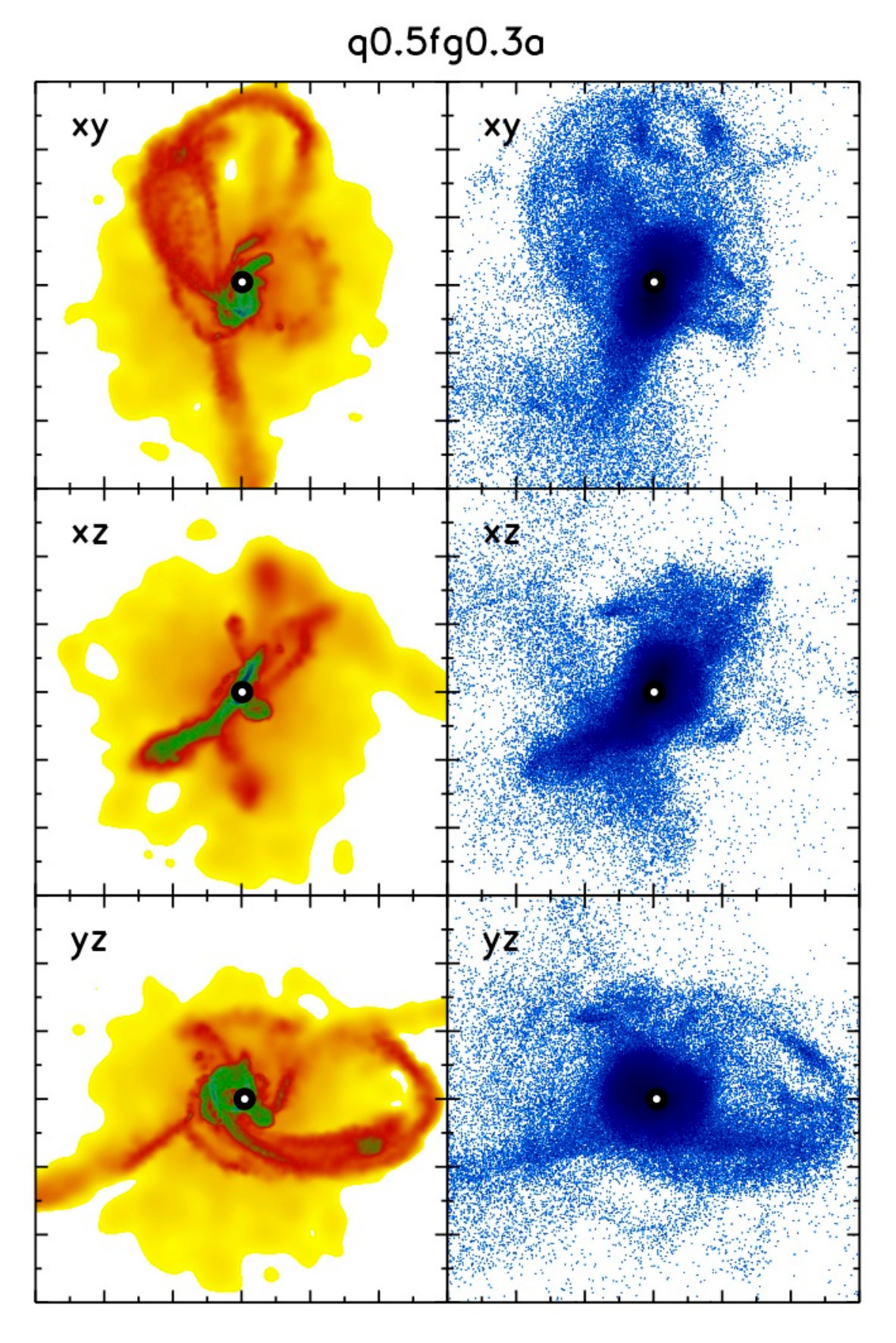}}
\resizebox{0.33\hsize}{!}{\includegraphics{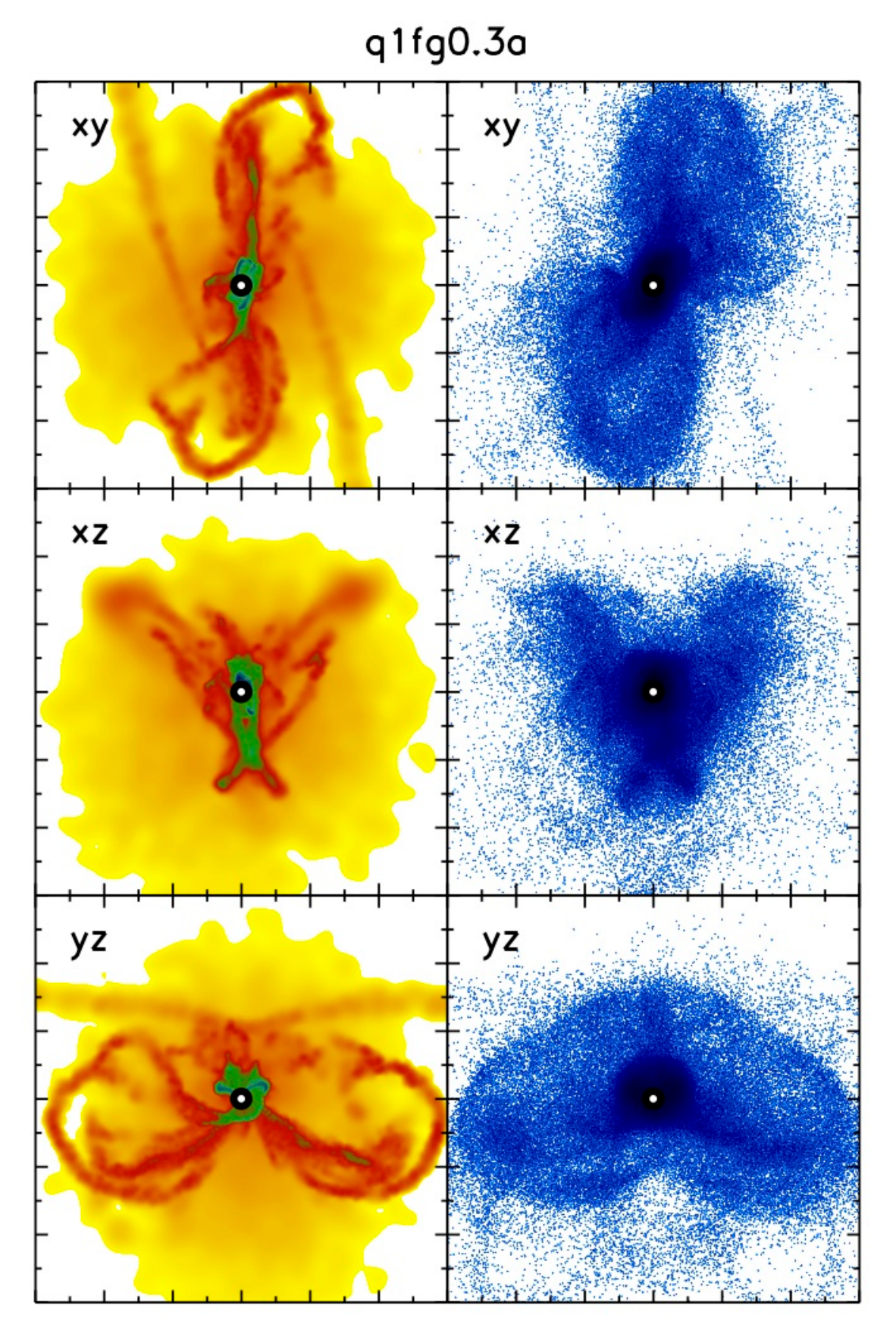}}
\resizebox{0.33\hsize}{!}{\includegraphics{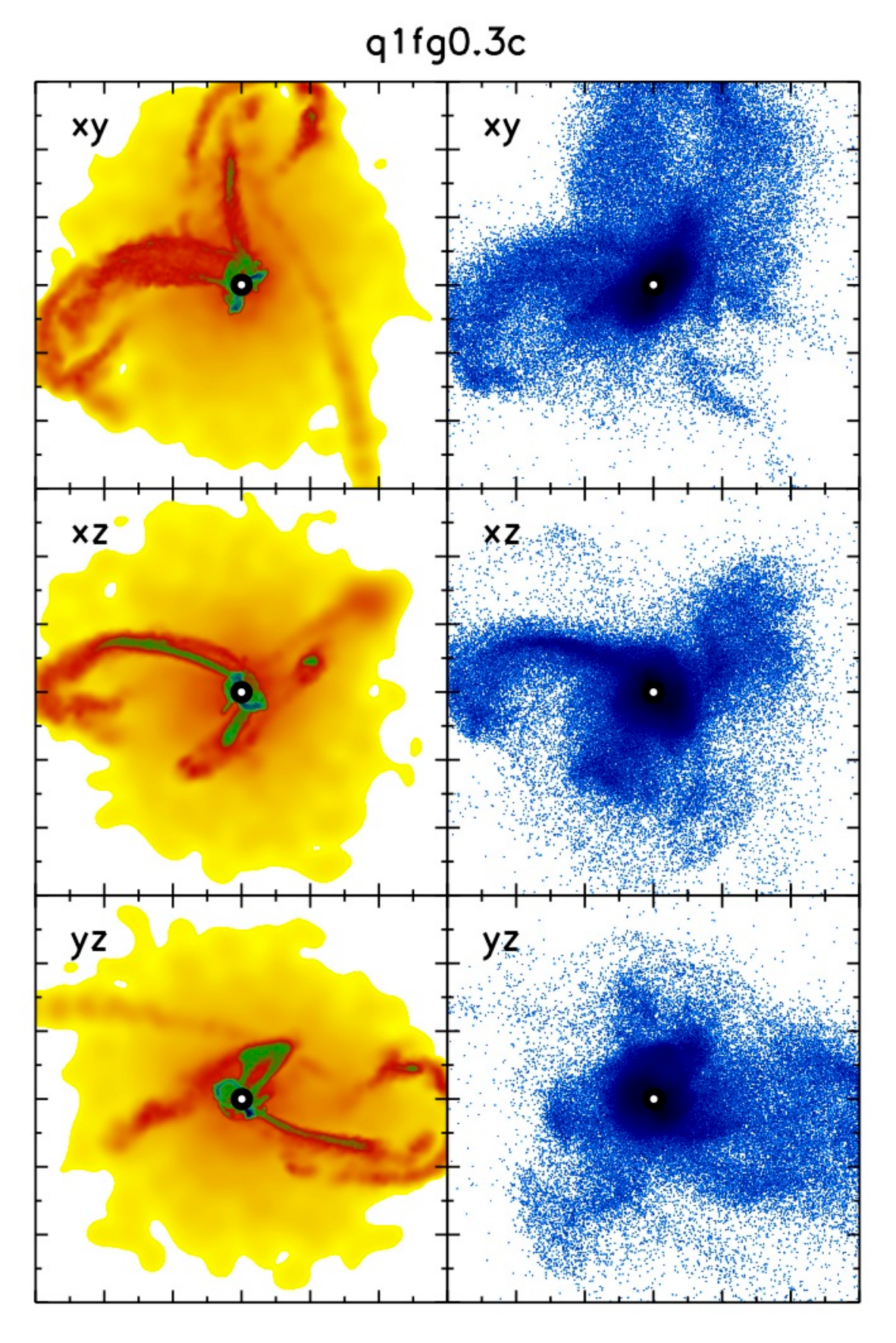}}
\resizebox{0.33\hsize}{!}{\includegraphics{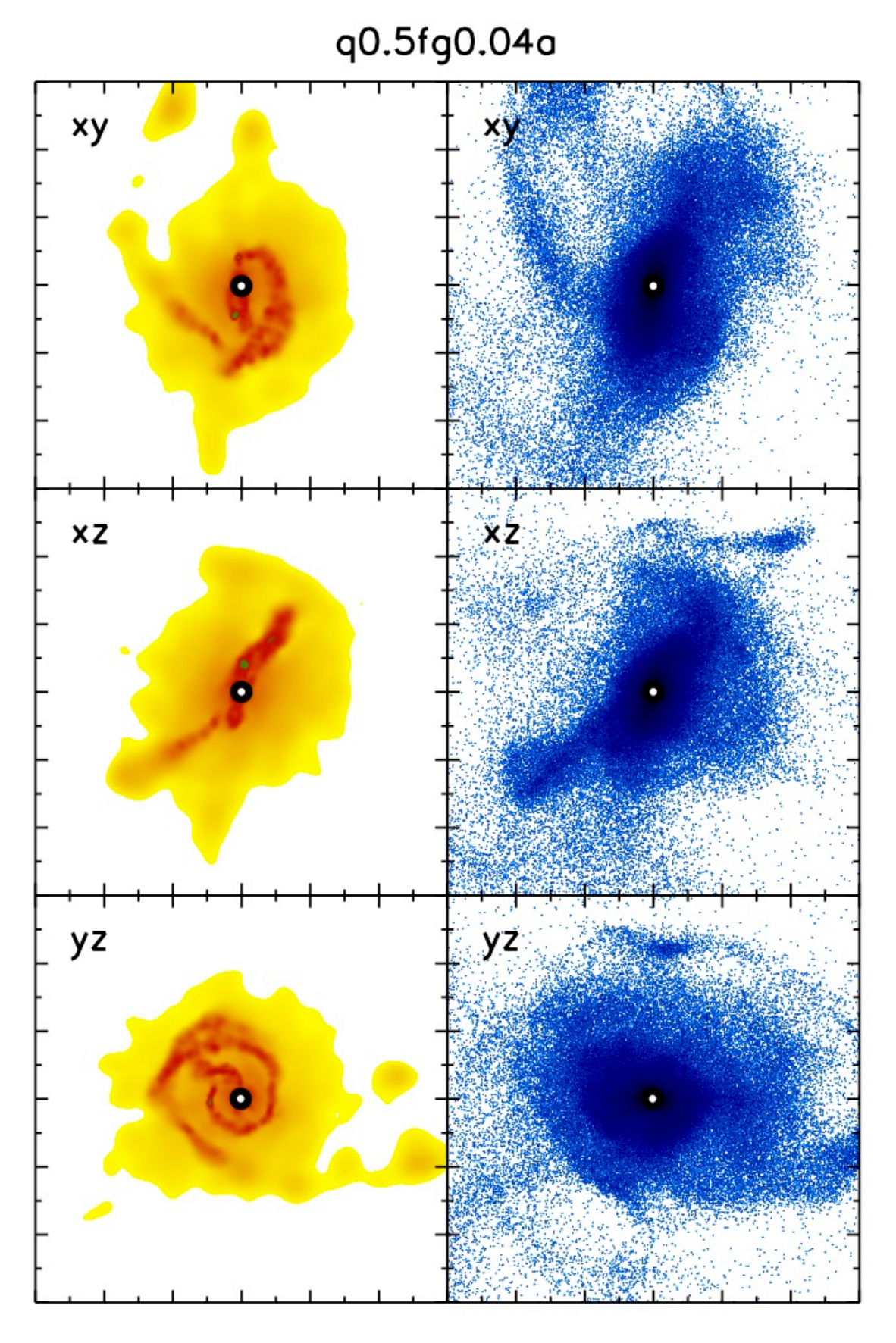}}
\resizebox{0.33\hsize}{!}{\includegraphics{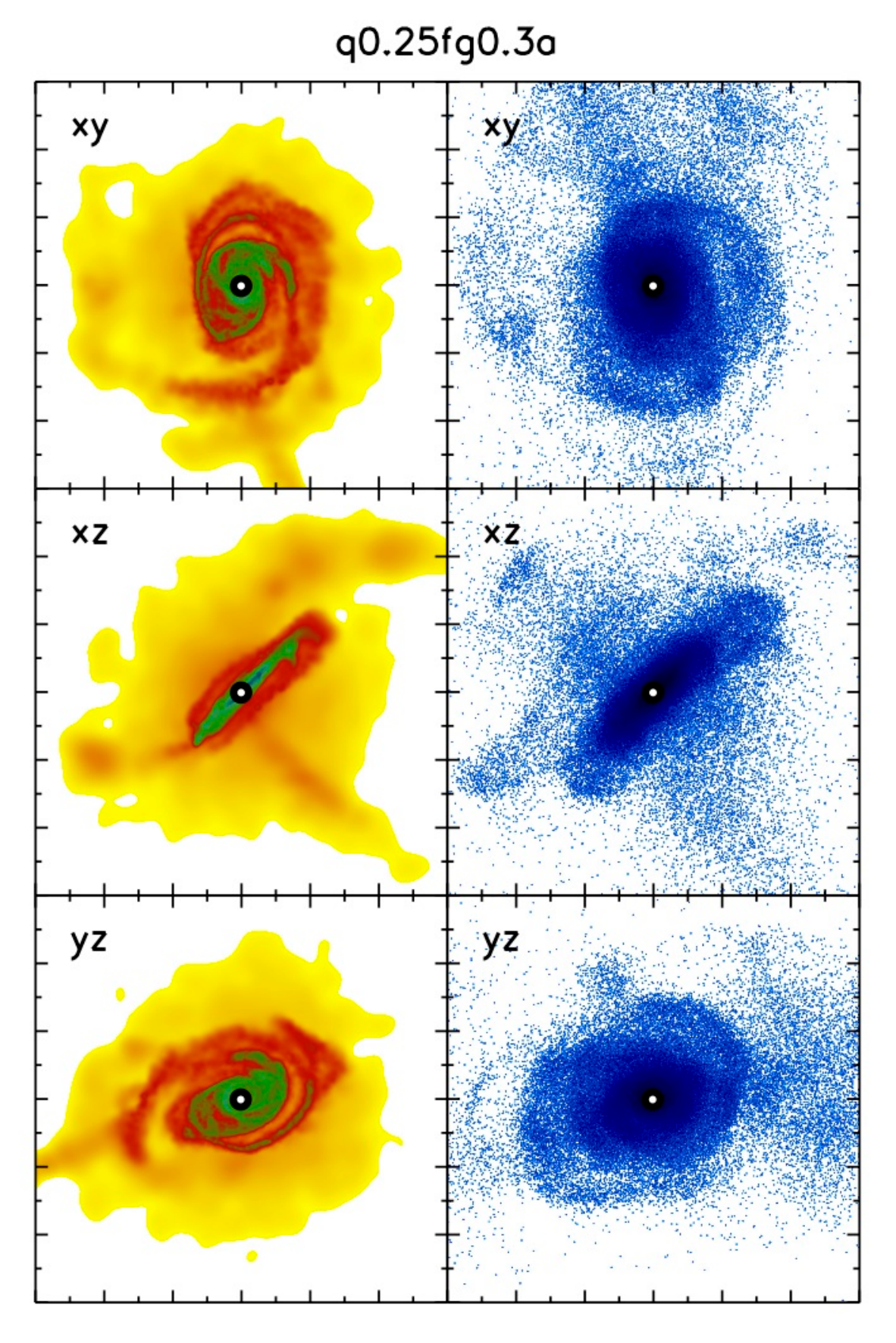}}
\resizebox{0.33\hsize}{!}{\includegraphics{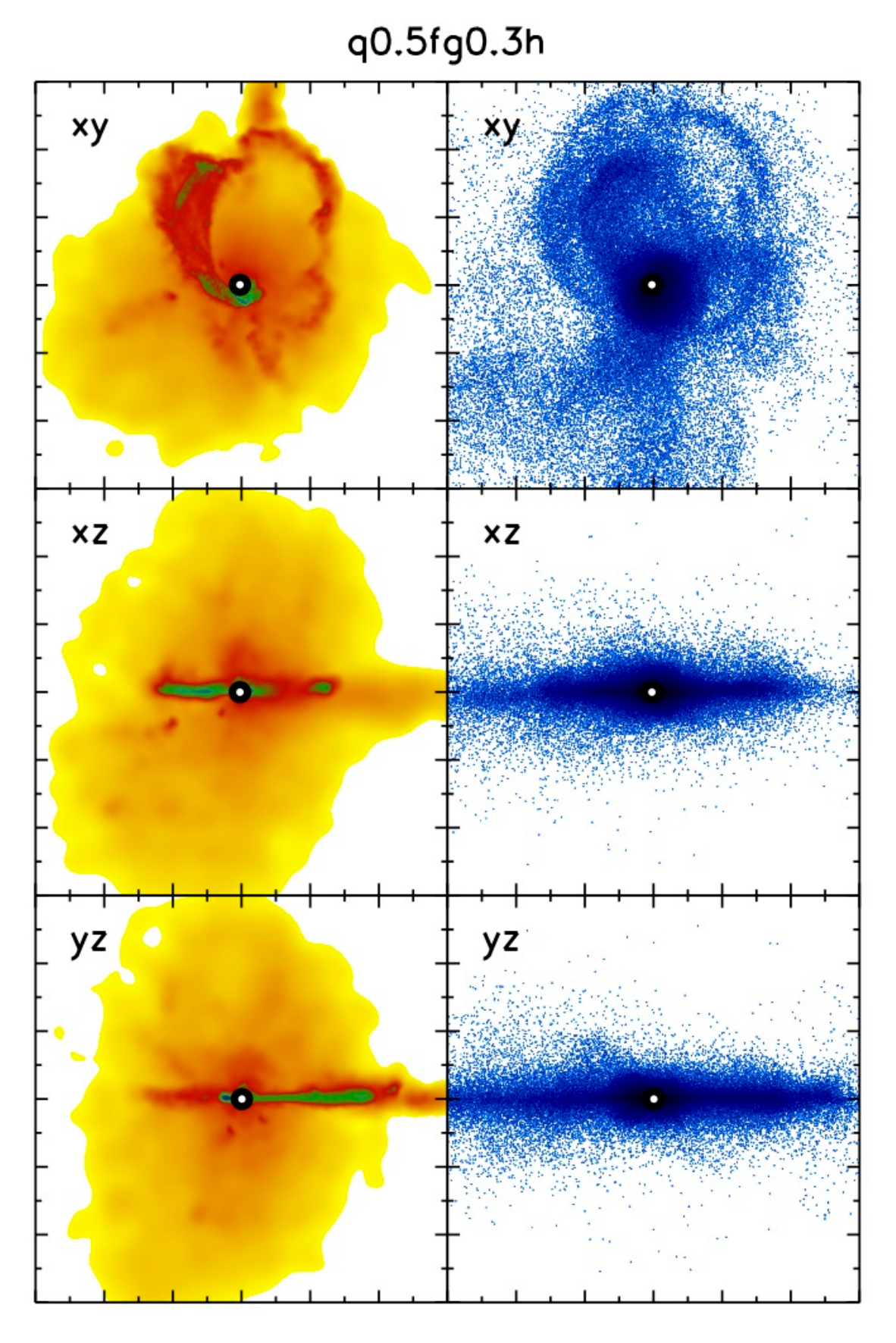}}
%\resizebox{0.33\hsize}{!}{\includegraphics{qhalf_fg30_tc_nokick_197_xlen30_3proj.png}}
%\resizebox{0.33\hsize}{!}{\includegraphics{q1_fg30_tc_nokick_164_xlen30_3proj.png}}
%\resizebox{0.33\hsize}{!}{\includegraphics{q1_fg30_orb3_tc_nokick_170_xlen30_3proj.png}}
%\resizebox{0.33\hsize}{!}{\includegraphics{qhalf_fg04_tc_nokick_219_xlen30_3proj.png}}
%\resizebox{0.33\hsize}{!}{\includegraphics{q025_fg30_tc_nokick_290_xlen30_3proj.png}}
%\resizebox{0.33\hsize}{!}{\includegraphics{qhalf_fg30_orb8_tc_nokick_184_xlen30_3proj.png}}
\caption[]{Projected gas (yellow/red) and stellar (blue) density distributions for six different merger models, shown at the time of BH merger in each case.  The black-and-white dot indicates the position of the BH in each panel.  The spatial scale of all panels is 86 kpc, and the projection ($x$-$y$, $x$-$z$, or $y$-$z$) is labeled on each panel. The models shown are (clockwise from top left, also labeled on each plot) q0.5fg0.3a, q1fg0.3a, q1fg0.3c, q0.5fg0.04a, q0.25fg0.3a, \& q0.5fg0.3h.  \label{fig:3proj}}
\end{figure*}

For each model, we run a merger simulation in which the BHs are not allowed to merge.  From this we can determine~\tcoal, which is used to set~\tmrg.  Then, restarting slightly before~\tmrg, we simulate both a merger with no recoil kick and a merger with~\vk/\vesc~$ = 0.9$.  For the models shown in boldface in Table~\ref{table:models}, we also simulate recoil kicks with~\vk/\vesc~$ = 0.4, 0.5, 0.6, 0.7, 0.8, 1.0, 1.1$, and 1.2.  Each simulation is run for 2.9 Gyr after~\tmrg.  Finally, using the q0.5fg0.3a model, we simulate a small sample of recoil kicks with varying kick orientation ($\theta_{\rm k}, \phi_{\rm k}$), the results of which are discussed in \S~\ref{ssec:kick_orient}.

\section{Mergers Without GW Recoil}
\label{sec:premrg}

\begin{figure*}
\resizebox{0.85\hsize}{!}{\includegraphics{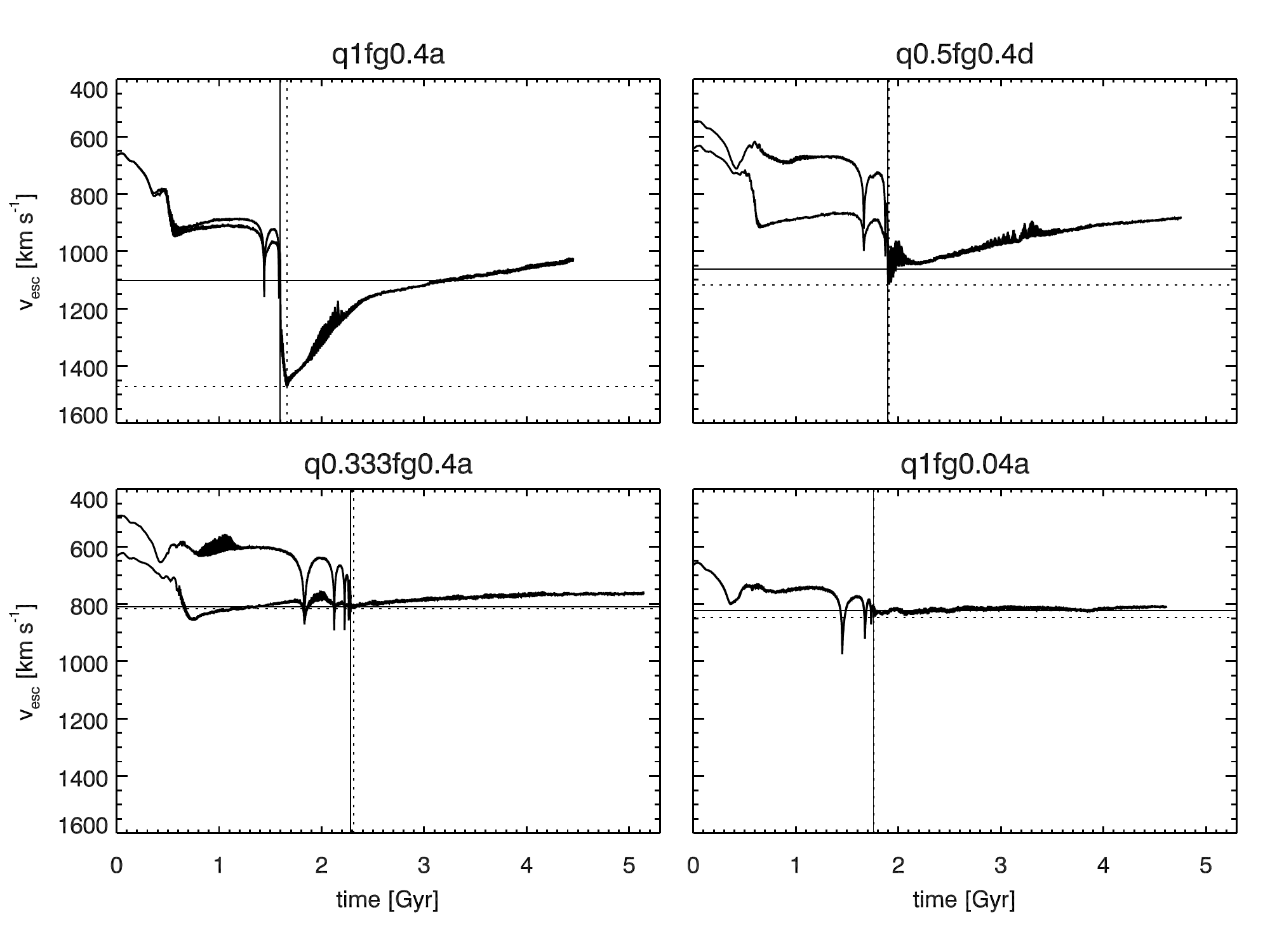}}
%\resizebox{0.85\hsize}{!}{\includegraphics{compare_sample_vesc.png}}
\caption[]{BH~\vesc~vs. time for a sample of four galaxy merger simulations with no recoil kick imparted to the BH.  (\vesc~$\equiv \sqrt{-2\Phi({\bf x}_{\rm BH})}$.)  In each panel, the two curves represent~\vesc~of each BH prior to merger.  After the merger, the escape speed of the remnant BH is shown.  The solid vertical and horizontal lines mark the time of BH merger,~\tmrg, and the BH escape speed at that time,~\vesc(\tmrg).  The dotted vertical and horizontal lines denote the (post-merger) time of the maximum escape speed and its value.  {\em Top left panel:} q1fg0.4a; {\em top right panel:} q0.5fg0.4d; {\em lower left panel:} q0.333fg0.4a; {\em lower right panel:} q1fg0.04a. \label{fig:compare_vesc}}
\end{figure*}

We begin by examining galaxy merger simulations with no GW recoil kick given to the BHs.  Some basic characteristics of a galaxy merger simulation with $q=0.5$ and~\fgas~$ = 0.3$ (our ``fiducial" merger, model q0.5fg0.3a) are illustrated in Figs.~\ref{fig:v0_contour_gas} \& \ref{fig:v0_sep_sfr_fgas}.  Fig.~\ref{fig:v0_contour_gas} shows the gas distribution at the time of BH merger, as well as the paths of the two BHs prior to merger.  The gas distribution is very irregular, with distinct tidal streams.  The inner region of the merger remnant is also lumpy and irregular.  In other words, the initial disk structure of the progenitors has been destroyed by the merger, and the remnant is still highly disturbed at the time of BH coalescence.  Fig.~\ref{fig:v0_sep_sfr_fgas} shows the BH accretion and star formation throughout the simulation as well as the evolution of the BH separation prior to merger.  This example illustrates some characteristics that are generic to our galaxy merger simulations.  Simultaneous bursts of star formation and BH accretion occur after the first close pericentric passage and at final coalescence \citep{mihher94a,mihher96} as the result of gas inflows caused by gravitational torques \citep{barher91,barher96}.  As we shall demonstrate, these gas inflows can greatly influence the dynamics of ejected black holes by deepening the central potential and by providing additional drag force.  Note that because our simulations include star formation throughout, the gas fraction at the time of BH merger is significantly lower than its initial value; in the example shown, 60\% of the initial gas has been consumed by~\tmrg.   
By the end of the simulation, 2.9 Gyr after the merger, almost 80\% of the gas has been depleted by star formation, and the BH accretion rate is very low.   In our other merger models, 47-80\% of the initial gas is depleted by the time of BH merger, and 62-88\% is depleted by the end of the simulation.  The final black hole mass in this example ($2.2\times10^7$~\msun) is also typical of our fiducial-mass mergers.  

Fig.~\ref{fig:3proj} shows gas and stellar density distributions of this fiducial (q0.5fg0.3a) model and five other merger models.  Each plot shows gas and stellar density from three different projections.  These examples are chosen to illustrate the generic nature of the morphological features mentioned above.  In all cases, the merger remnants are visibly disturbed and lumpy.  Tidal tails are ubiquitous, though their size and density varies.  Higher-\fgas~mergers have more compact remnants \citep{robert06b, dekcox06, hopkin08a,hopher10a}.  Lower-$q$ mergers are less strongly disrupted; a disk-like structure can be seen in the $x-z$ orientation of model q0.25fg0.3a.  A special case is shown in the lower-right corner of Fig.~\ref{fig:3proj}.  Model q0.5fg0.3h has a coplanar orbit with both galaxies rotating prograde to the orbital motion.  Due to the highly aligned orbit, the disk structure of the progenitor galaxies is preserved in this merger remnant; tidal features are visible only from the face-on ($x-y$) projection.  Our other ``highly aligned" orbits (i, j, \& k) result in similar remnant morphologies; they are included in our suite of merger models mainly for comparison, as mergers generally will not have such careful alignment.

\begin{figure*}
\resizebox{0.4\hsize}{!}{\includegraphics{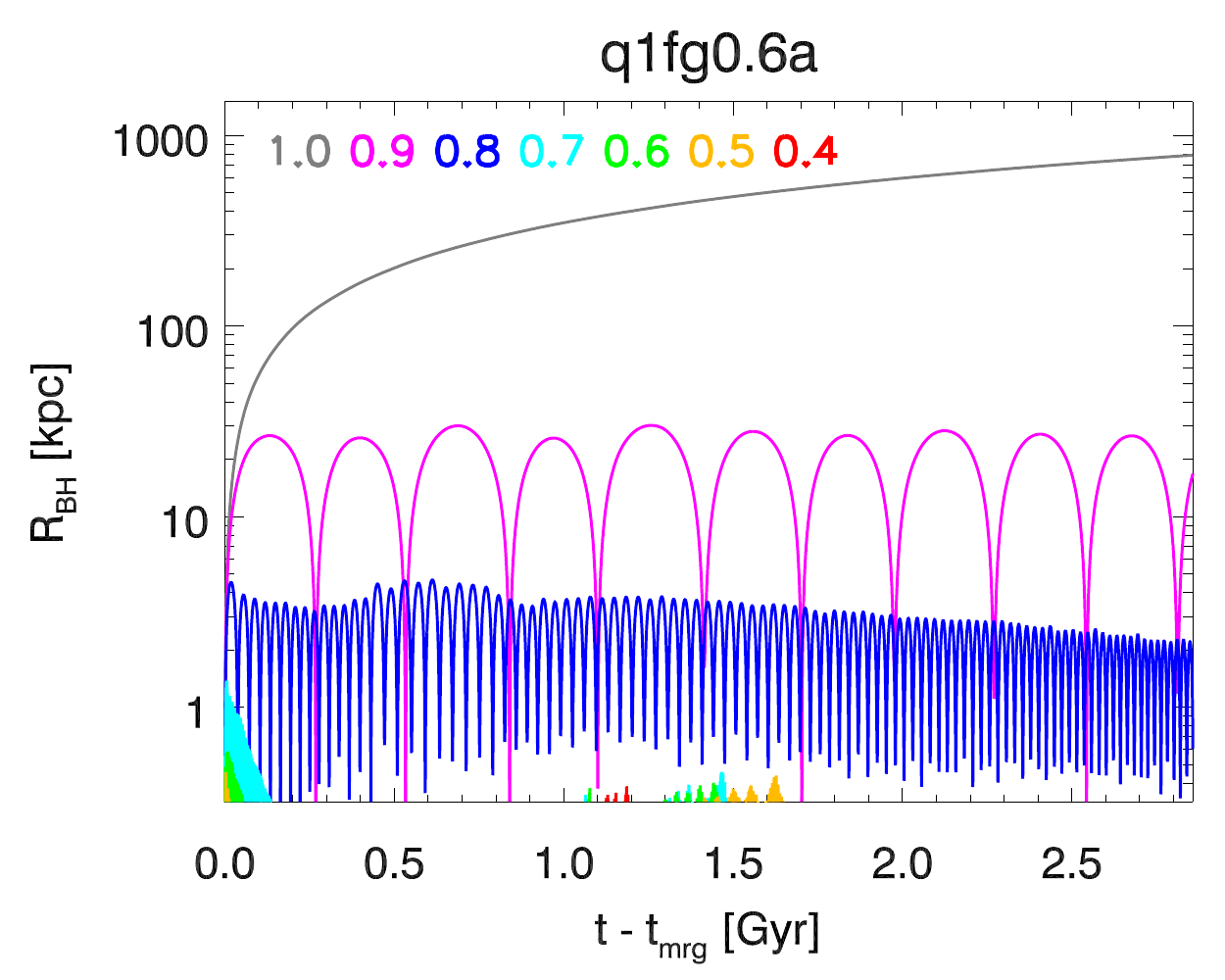}}
\resizebox{0.4\hsize}{!}{\includegraphics{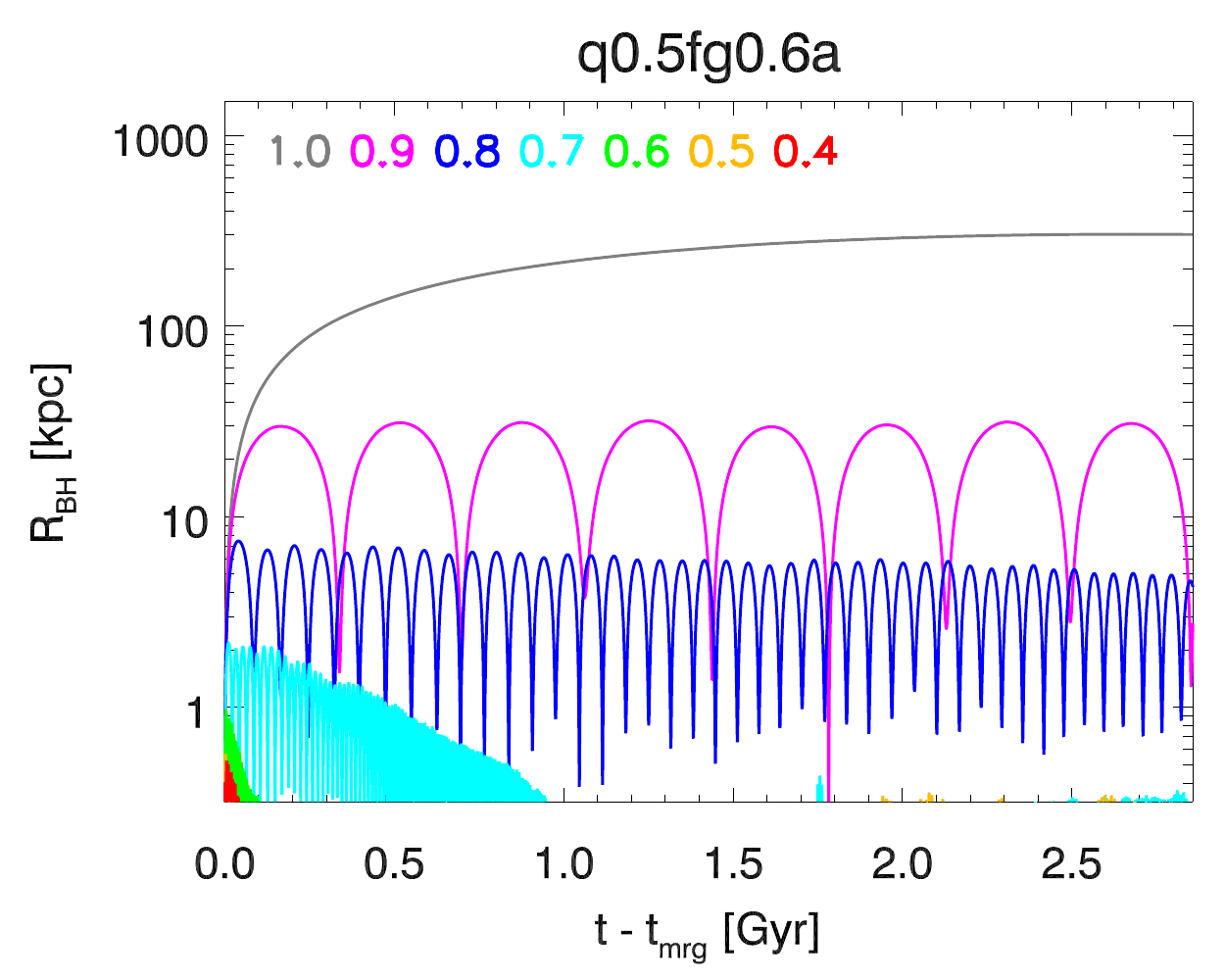}}
\resizebox{0.4\hsize}{!}{\includegraphics{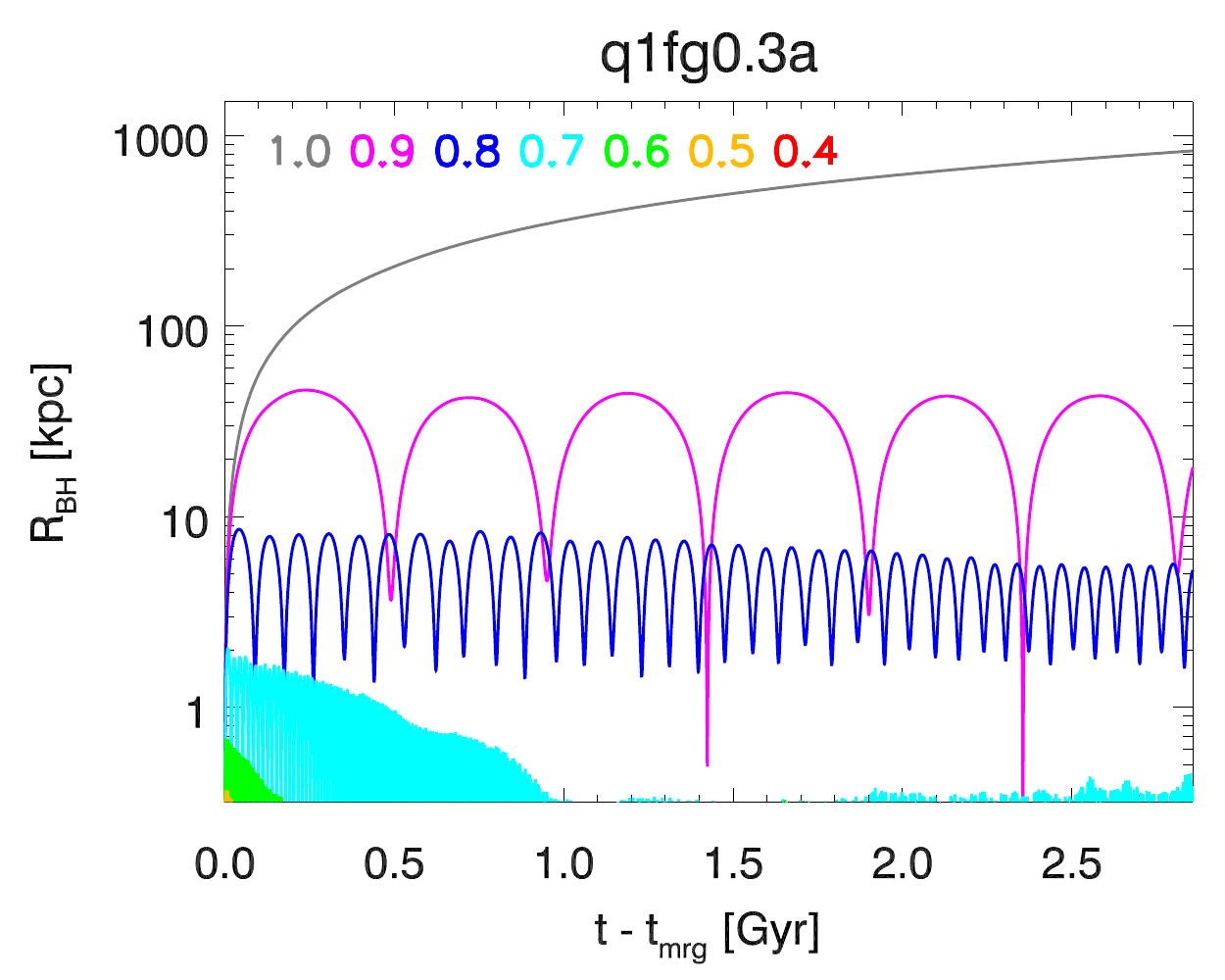}}
\resizebox{0.4\hsize}{!}{\includegraphics{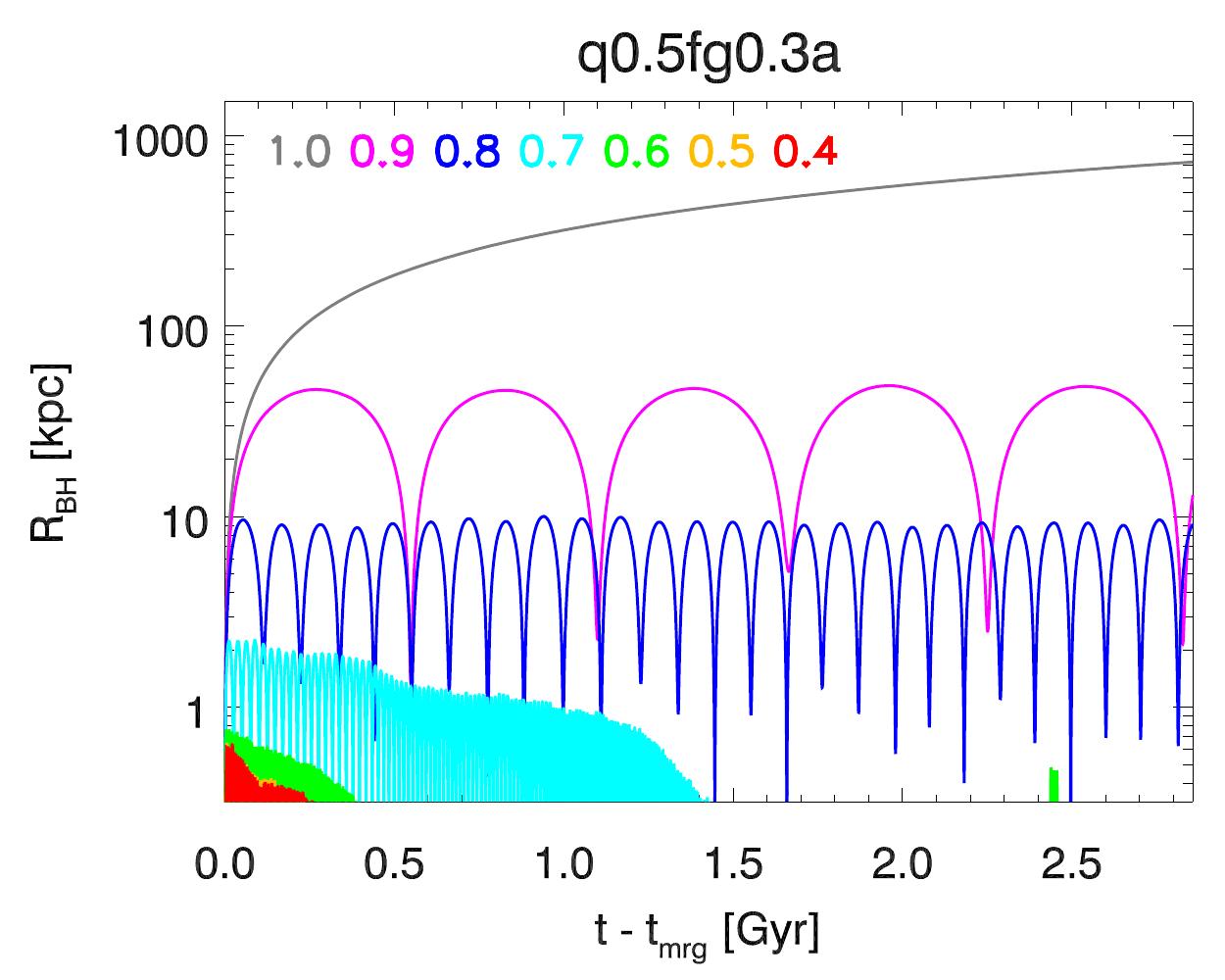}}
\resizebox{0.4\hsize}{!}{\includegraphics{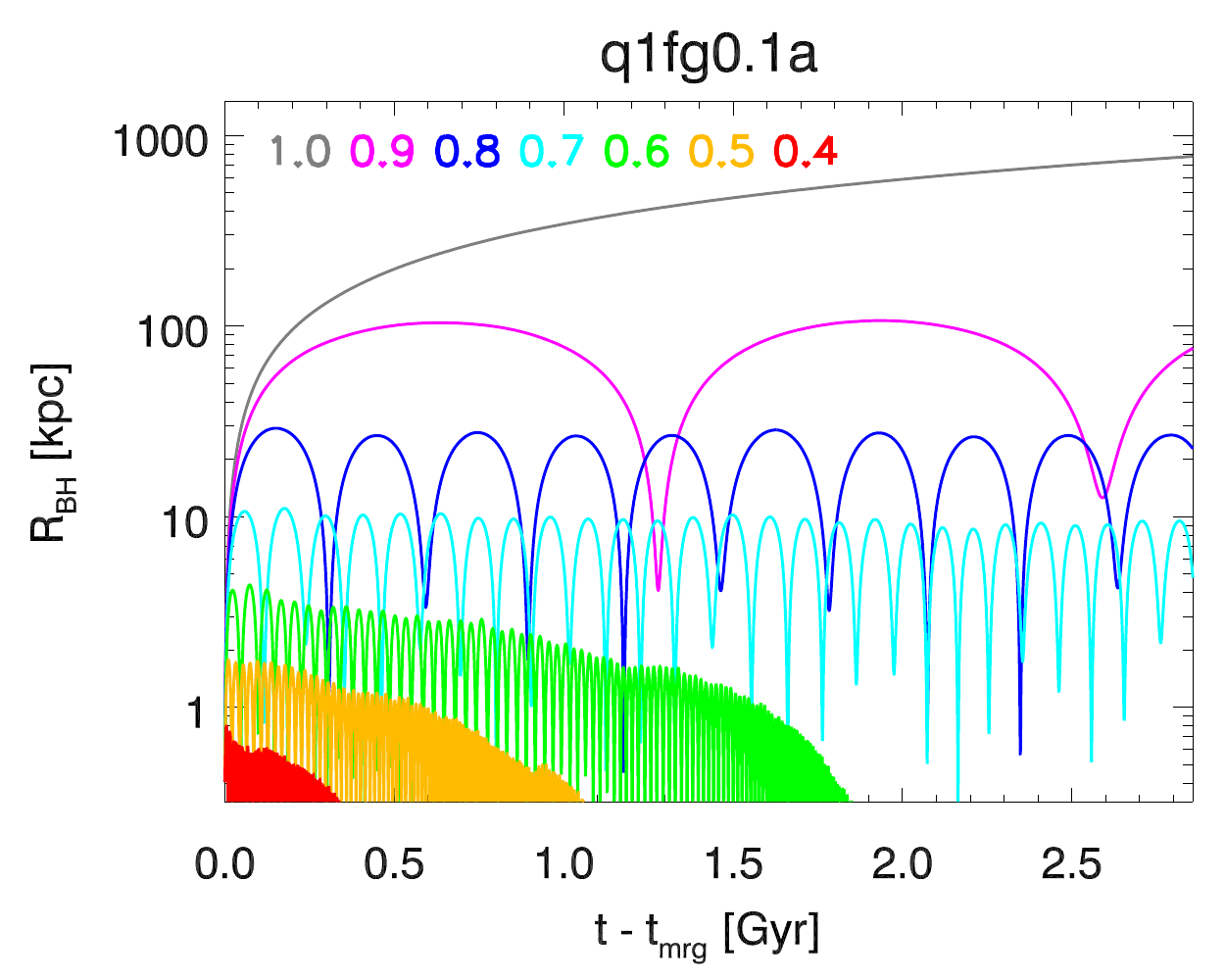}}
\resizebox{0.4\hsize}{!}{\includegraphics{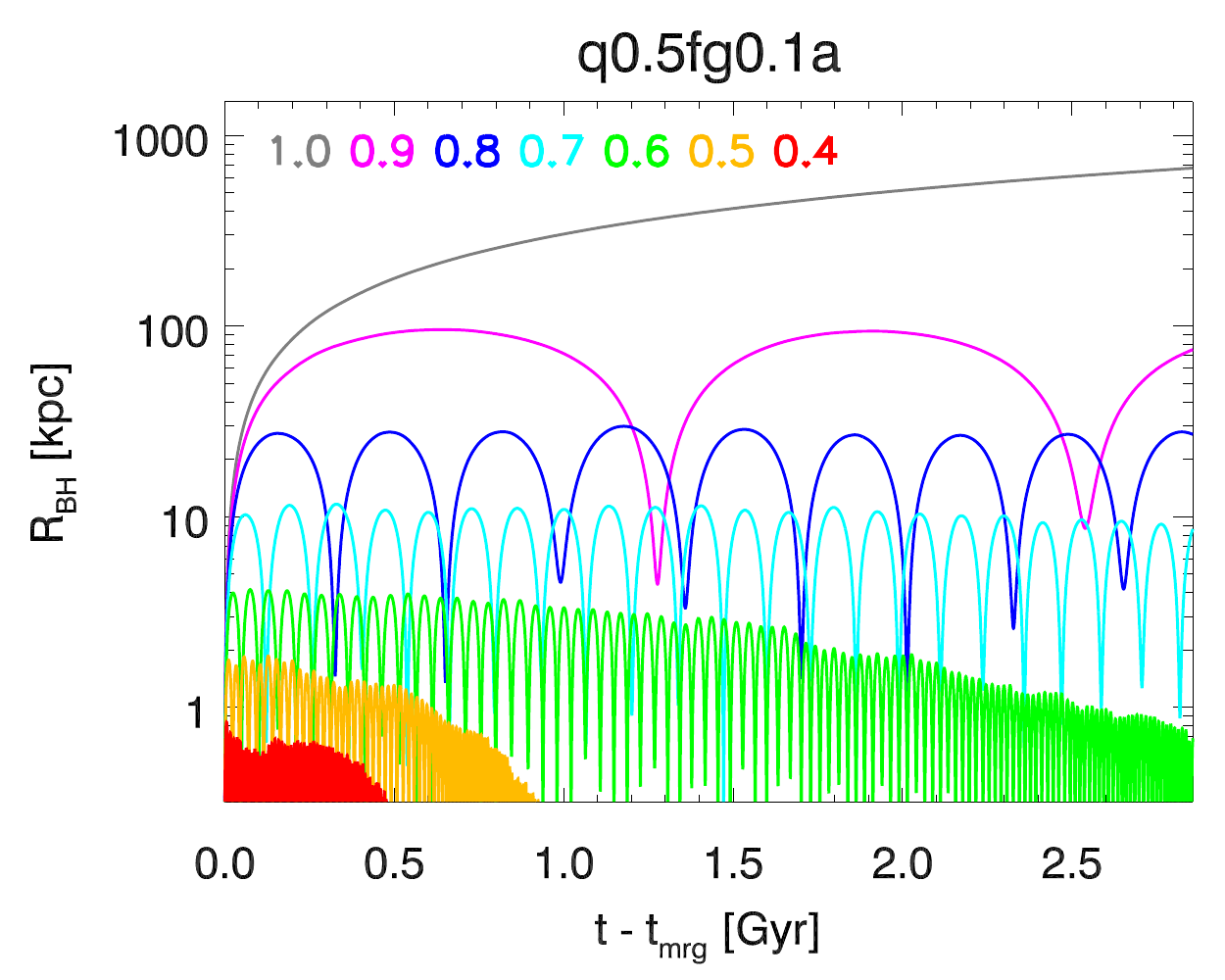}}
\resizebox{0.4\hsize}{!}{\includegraphics{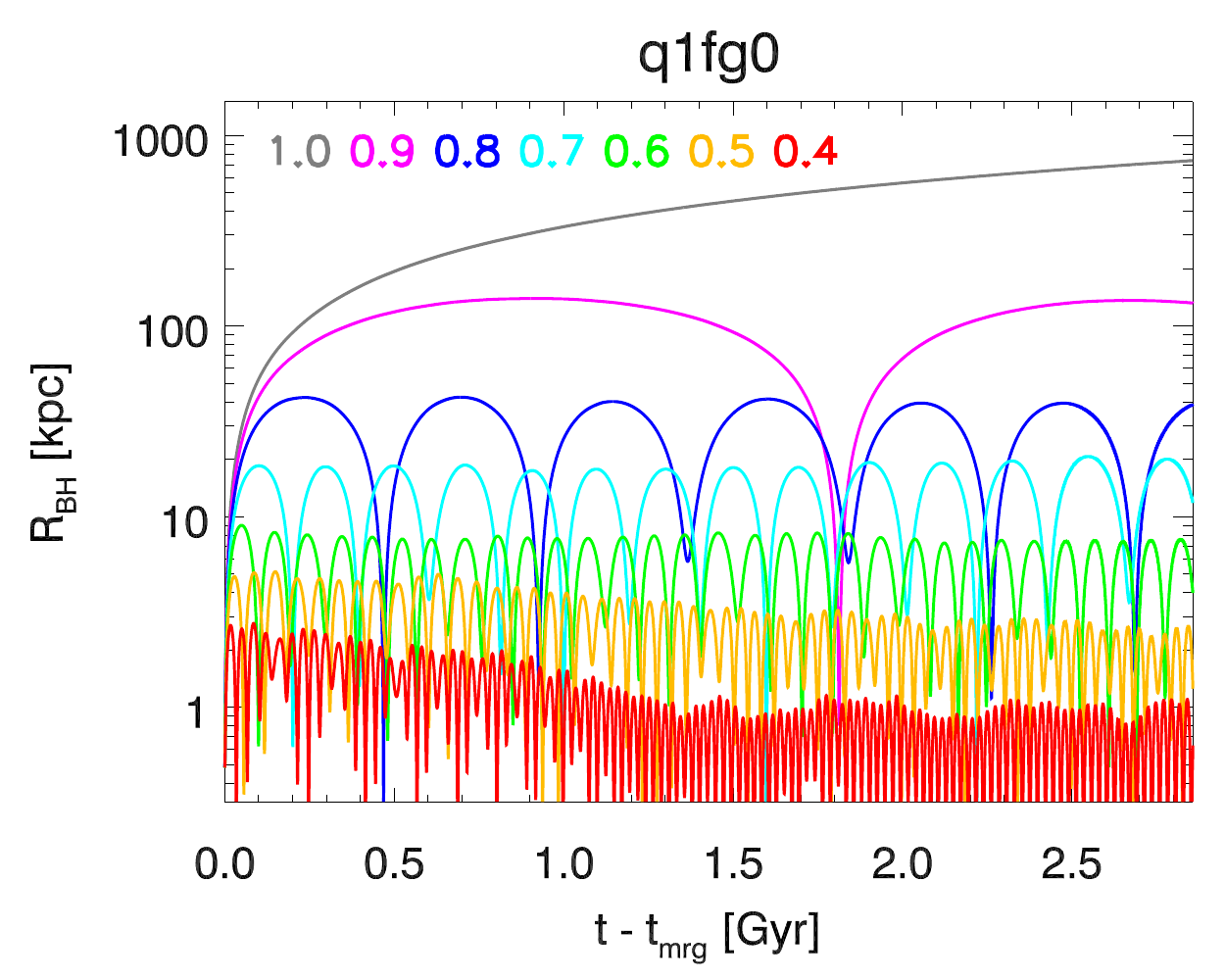}}
\resizebox{0.4\hsize}{!}{\includegraphics{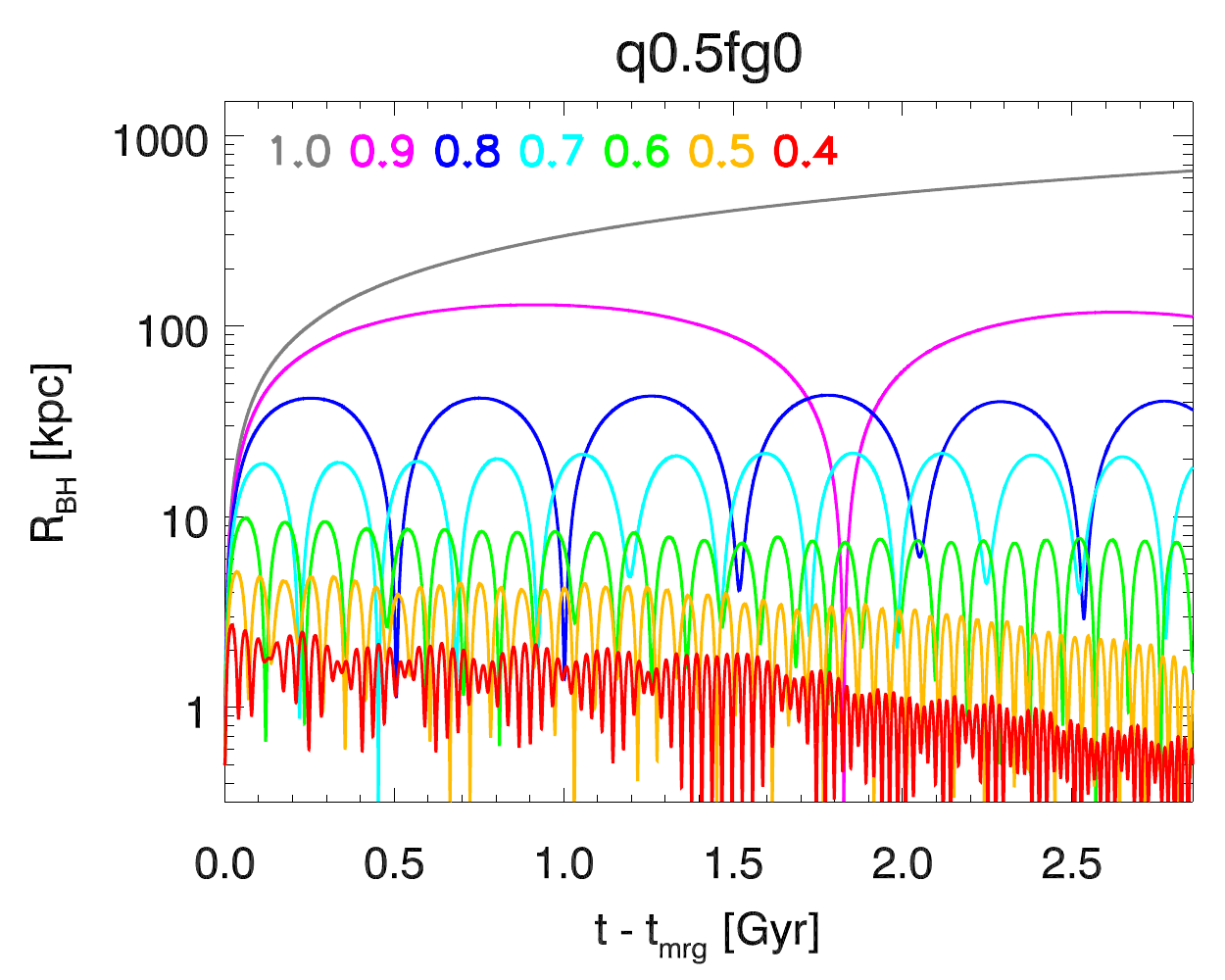}}
%\resizebox{0.4\hsize}{!}{\includegraphics{q1_fg60_all_rhist_center_bh1_rlim4.pdf}}
%\resizebox{0.4\hsize}{!}{\includegraphics{qhalf_fg60_all_rhist_center_bh1_rlim4.pdf}}
%\resizebox{0.4\hsize}{!}{\includegraphics{q1_fg30_all_rhist_center_bh1_rlim4.pdf}}
%\resizebox{0.4\hsize}{!}{\includegraphics{qhalf_fg30_all_rhist_center_bh1_rlim4.pdf}}
%\resizebox{0.4\hsize}{!}{\includegraphics{q1_fg10_all_rhist_center_bh1_rlim4.pdf}}
%\resizebox{0.4\hsize}{!}{\includegraphics{qhalf_fg10_all_rhist_center_bh1_rlim4.pdf}}
%\resizebox{0.4\hsize}{!}{\includegraphics{q1_fg0_all_rhist_center_bh1_rlim4.pdf}}
%\resizebox{0.4\hsize}{!}{\includegraphics{qhalf_fg0_all_rhist_center_bh1_rlim4.pdf}}
\caption[]{In each plot, BH recoil oscillation amplitudes are shown for varying kick velocities within a single model.  The $x$-axis is the time after the BH merger, $t - t_{\rm mrg}$.  The color-coded numbers on each plot indicate~\vk/\vesc~for each curve.  Galaxy models shown are, from left to right and top to bottom: q1fg0.6a, q0.5fg0.6a, q1fg0.3a, q0.5fg0.3a, q1fg0.1a, q0.5fg0.1a, q1fg0a, q0.5fg0a.\label{fig:alltraj}}
\end{figure*}

As mentioned in \S~\ref{sec:simulations}, in nearly equal-mass, gas rich mergers, the potential well in which the BH sits may deepen rapidly during final coalescence.  Once the merger is complete, the remnant central potential will become shallower as the central stellar region begins to relax.  The phase of rapid~\vesc~fluctuation coincides with the time of BH coalescence and may therefore affect recoiling BH dynamics.  Fig.~\ref{fig:compare_vesc} shows the evolution of the BH escape speed (\vesc) for four merger simulations without recoil kicks.  The solid lines denote~\tmrg~and~\vesc(\tmrg), while the dotted lines denote the maximum (post-merger) value of~\vesc~and the time at which it occurs.  The difference between~\vesc(\tmrg) and $v_{\rm esc, max}$ is small in all except the top left panel, which shows the q1fg0.4a merger.  It is clear that the escape speed, and hence the trajectory, of a kicked BH in this model will depend on whether the kick occurs at~\tcoal~or at some slightly later time.  However, the sharp increase of~\vesc~occurs {\em only} in gas-rich, nearly-equal-mass mergers.  The other three panels in Fig.~\ref{fig:compare_vesc} show~\vesc~vs. $t$ for a sample of simulations with lower $q$ and~\fgas; in these models,~\vesc~is much less volatile during and after the BH merger.  Moreover, the example shown in Fig.~\ref{fig:compare_vesc} is not the most gas-rich of our merger models but is chosen because of the large difference between~\vesc(\tcoal) and $v_{\rm esc, max}$, 34\%.  Larger differences between these quantities yield larger uncertainty in our results for BH dynamics when the recoil kick is assigned at~\tmrg~$=$~\tcoal.  The final four columns in Table~\ref{table:models} give~\tmrg,~\vesc(\tmrg), $v_{\rm esc,max}$, and the fractional difference between~\vesc(\tmrg) and $v_{\rm esc,max}$ for each merger model.  All mergers with~\fgas~$ < 0.3$ have $\la 10\%$ difference between~\vesc(\tmrg) and $v_{\rm esc,max}$, and mergers with  $q < 0.5$ have $< 1\%$ difference.  Therefore, while our results for nearly equal-mass, gas-rich mergers are subject to the assumption that the BH merger occurs on a short timescale, the results for our other models are insensitive to the merger time.  We explore further the case of gas-rich, $q \sim 1$ mergers in \S~\ref{ssec:tmrg}.  

\section{Recoiling Black Hole Dynamics}
\label{sec:bh_dynamics}

\subsection{Characterization of Recoil Trajectories}
\label{ssec:bh_traj}

A universal feature of our recoil trajectories is that they have low-angular-momentum orbits, which occurs because the centrally-concentrated baryonic component of the galaxy dominates the BH trajectories even when they extend far into the halo.  Consequently, we refer throughout the paper to the trajectories as ``oscillations" of the BH about the galactic center.  We see this clearly in Fig.~\ref{fig:alltraj}, which shows the trajectories of BHs kicked with~\vk/\vesc~$ = 0.4 - 1.0$ for eight different merger models.  Also readily apparent in these plots is the variation in BH oscillation amplitude and duration between models.  For a fixed value of~\vk/\vesc, BHs travel further from the galactic center in mergers with lower $q$ and~\fgas.  In the q1fg0.6a model, BHs kicked with~\vk~$ = 0.7$~\vesc~($= 878$~\kms) travel $< 2$ kpc from the galactic center, while in the collisionless q1fg0a model, the same~\vk/\vesc~allows the BH to travel ten times further.  Note that scaling the kick speeds to~\vesc~is important for determining the trends between galaxy models;~\vesc(\tmrg) for all of our models ranges from $338 - 2555$~\kms~($689-1258$~\kms~for the fiducial-mass models).  The trend toward smaller recoil oscillation amplitude for higher~\fgas~occurs because higher gas fractions result in more compact remnants that have higher central densities, and these remnants also have larger available supplies of gas.  Both factors contribute to steeper central potentials and increased gas drag and dynamical friction.  We can also see from Fig.~\ref{fig:alltraj} that recoil oscillations are slightly larger for lower $q$, though the trend is more evident in higher-\fgas~remnants.  Higher mass ratio mergers create stronger perturbations that drive gas more efficiently to the galaxies' central regions, further contributing to the steep potentials and high densities.

\begin{figure}
\center{\resizebox{\hsize}{!}{\includegraphics{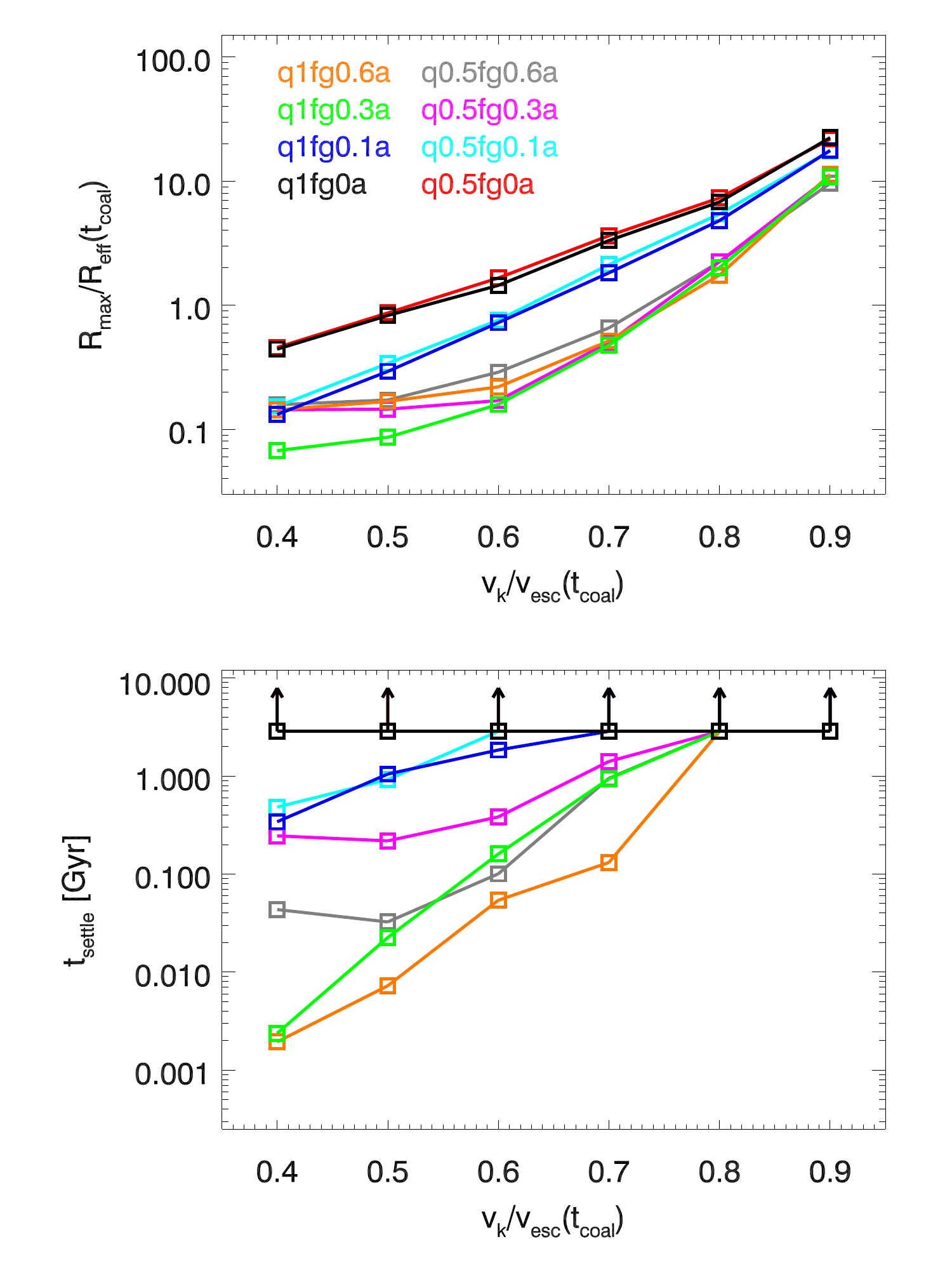}}}
%\center{\resizebox{\hsize}{!}{\includegraphics{all_rmax_tsettle_nohiacc.pdf}}}
\caption[]{$R_{\rm max}/R_{\rm eff}$ ({\em top window}) and $t_{\rm settle}$ ({\em bottom window}) are plotted versus~\vk/\vesc~for the eight models whose trajectories are shown in Fig.~\ref{fig:alltraj}.   $t_{\rm settle}$ is defined as the time after the BH merger at which the apocentric distance of the BH orbits drops below $4\, R_{\rm soft}$.  Arrows denote lower limits on $t_{\rm settle}$ in cases when the BH has not settled by the end of the simulation (defined as~\tmrg~$+\, 2.9$ Gyr). \label{fig:vk_rmax}}
\end{figure}

We can see these trends more clearly in Fig.~\ref{fig:vk_rmax}.  The top plot shows the maximum galactocentric distance of each orbit in Fig.~\ref{fig:alltraj}, normalized to the half-mass effective radius of the galaxy ($R_{\rm max}$/$R_{\rm eff}$).  The bottom plot shows the settling time, $t_{\rm settle}$, defined as the time when the apocentric distance of the BH oscillations, relative to the stellar center of mass, falls below $4\, R_{\rm soft}$.  This definition is chosen to avoid following BH trajectories below a scale where their oscillations may be dominated by numerical noise or gravitational softening.  We find that our conclusions are not sensitive to the exact definition of $t_{\rm settle}$.  Note that for some high-velocity recoils (for all recoils in the collisionless simulations), we have only lower limits on $t_{\rm settle}$, because in these cases the BH did not settle by the end of the simulation.  

In the top panel of Fig.~\ref{fig:vk_rmax}, the curves for a given~\fgas~lie nearly on top of each other.  This supports our assertion that the initial gas fraction is more important than the mass ratio in determining $R_{\rm max}$/$R_{\rm eff}$ for a given~\vk/\vesc.  By normalizing these quantities. we essentially isolate the effects of gas drag or potential shape from variation in galaxy size or central potential depth.  One can easily see from Fig.~\ref{fig:vk_rmax} that, especially for lower~\vk/\vesc, a gas fraction $\ge 30\%$ can reduce the amplitude of BH oscillations by up to an order of magnitude with respect to a purely collisionless system.

Note that $R_{\rm max}$/$R_{\rm eff}$ is at least as large for~\fgas~$= 0.6$ as it is for~\fgas~$= 0.3$, opposite the trend we expect if gas drag dominates the BH trajectories.  However, galaxies with higher~\fgas~generally have higher star formation rates, so systems with initial gas fractions of $0.3-0.6$ will have similar amounts of gas by the time of the BH merger.  Furthermore, as previously stated, higher gas fractions yield more concentrated merger remnants (smaller $R_{\rm eff}$).  Combined, these effects result in similar $R_{\rm max}$/$R_{\rm e}$ for recoils in mergers with initial~\fgas~$= 0.3-0.6$.  

As discussed in \S~\ref{sec:simulations}, the recoil trajectories may be sensitive to numerical parameters such as resolution or integration accuracy.  In particular, because $R_{\rm eff}$ is of the order of a few kpc, recoiling BHs with $R_{\rm max}/R_{\rm eff} \la 0.1$ spend substantial time at or below the spatial resolution limit.  $R_{\rm max}$ and $t_{\rm settle}$ are therefore approximate in these cases.  The BH settling times are also susceptible to possible inaccuracies in dynamical friction forces due to our finite mass resolution, which is likely to result in underprediction of the true amount of dynamical friction.  However, we can predict that the effect of this uncertainty on our results is likely to be small; in our lower-mass galaxy mergers, which have $10\times$ and $20\times$ better mass resolution, BHs kicked near the escape speed still experience very little dynamical friction and have very long wandering times.  Even if dynamical friction is still underestimated in these higher-resolution runs, it is clear that in general, recoils near the escape speed will result in wandering times of at least a few Gyr, and that in gaseous mergers, recoiling BHs with~\vk~$\la 0.5$~\vesc~remain within the central kpc of the galaxy.  Finally, because we use the same mass resolution in all of our simulations (excepting the low/high mass runs), the relative trends that we find between different galaxy models are robust. 

\begin{figure}
\resizebox{\hsize}{!}{\includegraphics{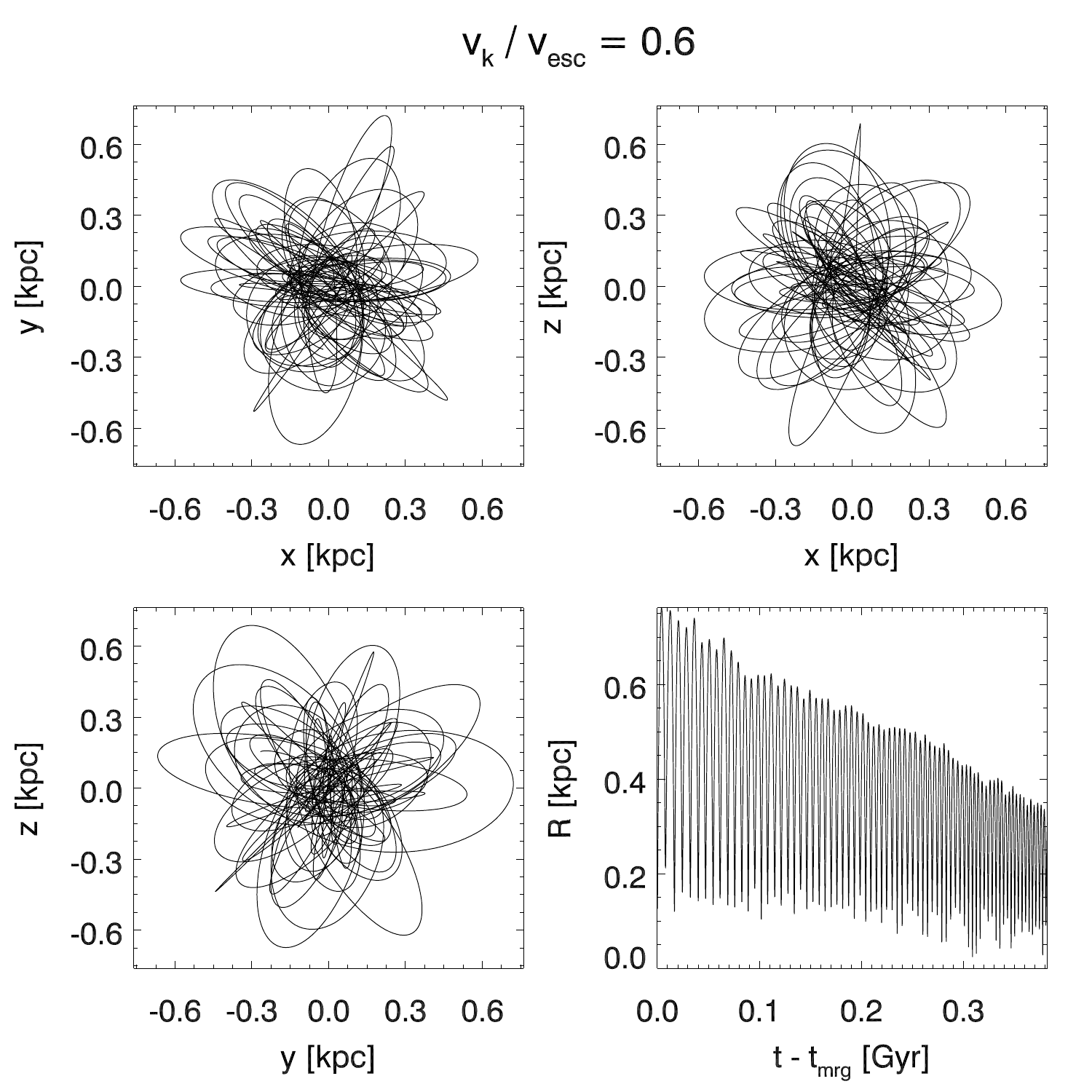}}
\resizebox{\hsize}{!}{\includegraphics{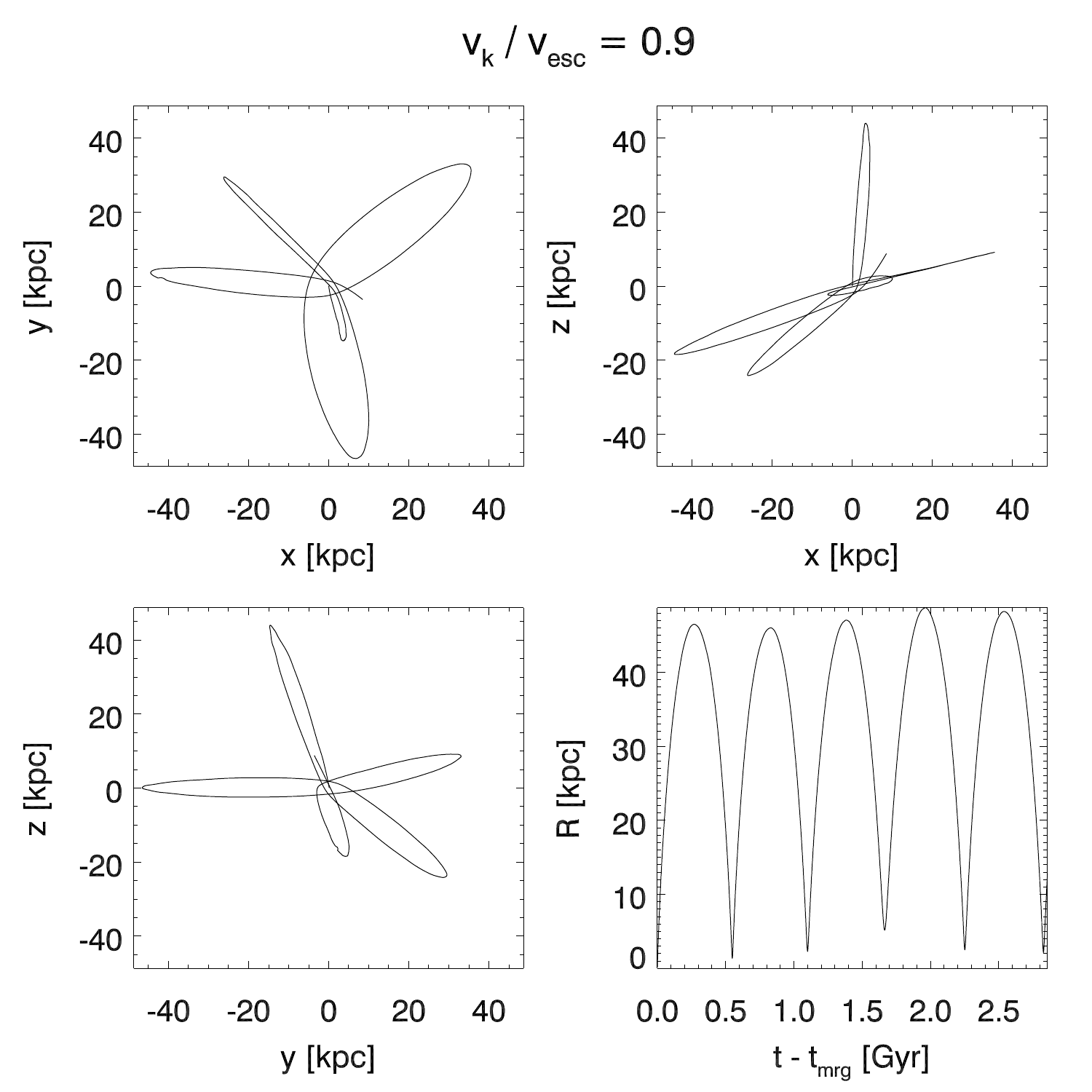}}
%\resizebox{\hsize}{!}{\includegraphics{qhalf_fg30_tc_v60_bh_center_traj_cut.pdf}}
%\resizebox{\hsize}{!}{\includegraphics{qhalf_fg30_tc_v90_bh_center_traj.pdf}}
\caption[]{Top four windows: trajectory for the~\vk/\vesc~$=0.6$ recoil simulation with the q0.5fg0.3a model. Bottom four windows: trajectory for the~\vk/\vesc~$=0.9$ recoil simulation with the q0.5fg0.3a model.  In each case, the four windows show the trajectories in the $x-y$, $x-z$, and $y-z$ projections, as well as the BH's galactocentric distance versus time.  Note the greatly different spatial scales of the two simulations.  \label{fig:3dtraj}}
\end{figure}

We expect some variation in $R_{\rm max}$ for different merger orbits as well.  In the set of eleven different orbital configurations we simulated with $q=0.5$ \&~\fgas~$=0.3$, the central escape speeds at~\tmrg~vary from $805 - 1046$~\kms.  Since the total mass in all of these simulations is constant, the variation in~\vesc~reflects differences in the central concentration, and hence the steepness of the potential well.  We simulate recoil trajectories in each of these with~\vk/\vesc~$= 0.9$, and find that $R_{\rm max}$/$R_{\rm eff} = 4 - 20$.  Based on this example, we see that the merger orbit alone can influence the remnant central potential well enough to change the amplitude of recoil trajectories by a factor of $\sim 5$.  

Fig.~\ref{fig:3dtraj} shows two examples of recoil trajectories in three different projections as well as $R(t)$.  The top plot shows a recoil event with~\vk/\vesc~$=0.6$, which in this model (q0.5fg0.3a) corresponds to~\vk~$\approx 600$~\kms. Note that the orbit is highly non-planar; it looks similar in all three projections.  The orbit is also centrophilic, which as we have noted is common to all of our recoil simulations.  The average orbital period is short, $\sim 7\times10^5$ yr, and $t_{\rm settle} \sim 400$ Myr, such that the BH settles to the center well before the end of the simulation.   

Recoil events with velocities near~\vesc~produce long-lived ($ > $ Gyr), large-amplitude BH oscillations.  Many of these oscillations still have large amplitudes at the end of the simulation, 2.9 Gyr after the BH merger, so we have only lower limits on $t_{\rm settle}$.  The bottom plot in Fig.~\ref{fig:3dtraj} illustrates this type of recoil event, again for the q0.5fg0.3a model.  The trajectory shown has a kick speed of~\vk/\vesc~$=0.9$, so~\vk~$\approx 900$~\kms.  The amplitude of these oscillations is $\sim 50$ kpc, almost two orders of magnitude greater than in the lower-\vk~example.  Although the BH only completes five orbits before the end of the simulation, we can see that again there is no evidence for a preferred orbital plane; the orbit is three-dimensional.  Because the oscillation amplitude is so large and shows no sign of damping out by the end of the simulation, it is likely that such an orbit will not return to the galaxy center within a Hubble time, and certainly not within the mean time between major galaxy mergers.  Such BHs essentially can be considered ``lost" to the galaxy.  If in other cases the BHs do return to the galactic center after several Gyr and another galaxy merger has taken place in the meantime, the returning and new BHs could form another binary.  

In recoil events with $v > $~\vesc, the rapidly recoiling BH can only take a small amount of gas with it; only gas that is orbiting the BH with $v_{\rm orb} \ga $~\vk~stays bound to the BH when it is ejected.  The accretion timescale and luminosity of these ejected disks is discussed in \S~\ref{ssec:ejdisk}.  Once this supply is exhausted, the naked BH is unlikely to become active again, as this would require a serendipitous passage through a dense gas cloud, as well as a sufficiently low relative velocity that substantial accretion could occur.  (Recall that Bondi-Hoyle accretion scales as $v_{\rm BH}^{-3}$.)  These escaping BHs are therefore likely to wander indefinitely, undetected, through intergalactic space.

\subsection{BH Merger Time}
\label{ssec:tmrg} 

\begin{figure}
\resizebox{1.1\hsize}{!}{\includegraphics{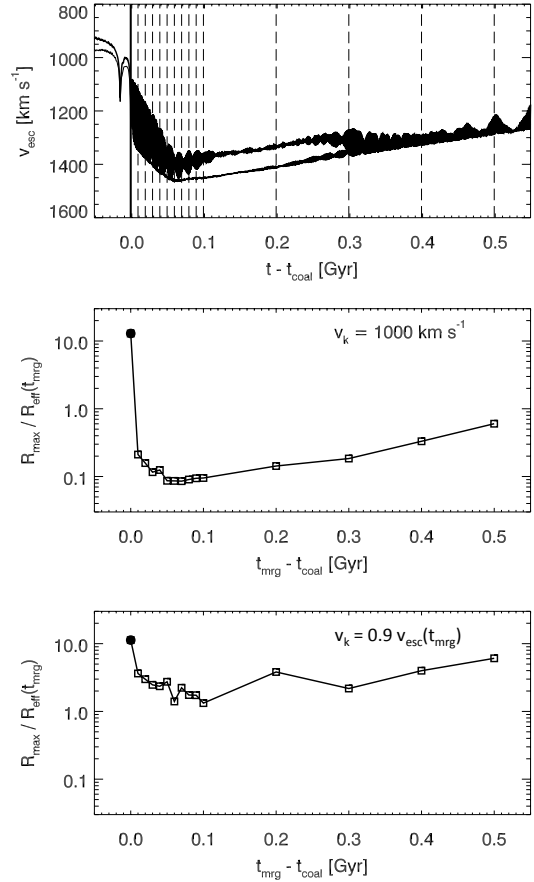}}
%\resizebox{\hsize}{!}{\includegraphics{bh_vesc_zoom.pdf}}
%\resizebox{\hsize}{!}{\includegraphics{q1_tmrg_1000_rmax.pdf}}
%\resizebox{\hsize}{!}{\includegraphics{q1_tmrg_rmax.pdf}}
\caption[]{{\em Top window:}~\vesc~is plotted as a function of $t -$~\tcoal~for the q1fg0.4a merger simulation.  The portion shown here is a zoom-in of the top-left window in Fig.~\ref{fig:compare_vesc}.  The solid vertical line denotes~\tcoal, and the dashed vertical lines denote the subsequent merger times used to test variation in~\tmrg.  {\em Middle window:} $R_{\rm max}/R_{\rm eff}$ is plotted versus BH merger time for the set of delayed-merger simulations done using model q1fg0.4a. Each recoiling BH in these simulations is assigned a recoil kick of $1000$~\kms.  {\em Bottom window:}  Same as middle window, but for a set of simulations in which the kick speed in~\kms~varies, but is held to a constant value of~\vk/\vesc; in every case~\vk/\vesc~$= 0.9$.  The $x$-axis ranges from~\tmrg~$=$~\tcoal~to~\tmrg~$=$~\tcoal~$+ \, 500$ Myr. \label{fig:q1tmrg_rmax}}
\end{figure}

As illustrated in Fig.~\ref{fig:compare_vesc},~\vesc~can increase significantly during coalescence in equal-mass, gas-rich mergers.  To better understand how this may affect BH trajectories, we have run a series of simulations with the BH merger (and recoil kick) occurring at sequentially later times using model q1fg0.4a.  The~\vesc~evolution of model q1fg0.4a is replotted in the top window of Fig.~\ref{fig:q1tmrg_rmax}, with the fiducial merger time,~\tmrg~$=$~\tcoal, marked by the solid black line and the subsequent merger times tested marked by black dashed lines.  We have tested simulations with each of these merger times twice, once with~\vk~$= 1000$~\kms~and once with~\vk/\vesc(\tmrg)~$= 0.9$.  By conducting the second experiment we are normalizing to the evolving depth of the central potential well, which helps us to separate the gravitational and drag effects on the trajectories.  

Fig.~\ref{fig:q1tmrg_rmax} shows $R_{\rm max}$ for the BH orbits in each of these varying-\tmrg~simulations, with the~\vk~$= 1000$~\kms~simulations in the middle panel and the \vk/\vesc(\tmrg)~$= 0.9$ simulations in the bottom panel.  As~\vesc~increases from 1100~\kms~at~\tcoal~to almost 1500~\kms~at its maximum, the amplitudes of BH trajectories from constant-\vk~recoil events decrease by more than two orders of magnitude.   However, the bottom panel of Fig.~\ref{fig:q1tmrg_rmax} shows that when~\vk~is set to a constant {\em fraction} of the escape speed, the BHs kicked from deeper potential wells (larger~\vesc) still have smaller-amplitude trajectories, but the variation is much less severe than when~\vk~$=$ constant.  The variation that persists even for fixed~\vk/\vesc~can be attributed to the enhanced gas drag and dynamical friction, as well as the steeper central potential well, that result from the rapid formation of the central dense cusp.  At very late merger times, $100 - 500$ Myr after~\tcoal, the opposite trend is seen.  The amplitude of constant-\vk~trajectories increases by almost an order of magnitude from the minimum at~\tcoal$+\, 70$ Myr to the value at~\tcoal$+\, 500$ Myr.  This is partly because the central potential becomes shallower as the merger remnant begins to relax.  The same trend occurs at late times in the constant-\vk/\vesc~simulations, because the potential well also becomes less steep, and gas drag becomes less efficient as more of the gas is consumed in star formation.  We note that merger times up to~\tcoal~$+\, 500$ Myr are included here only for illustrative purposes; in reality, such delayed mergers are probably unrealistic in gas-rich systems.  

If the BHs are able to merge rapidly, before the central potential reaches its maximum depth, another interesting effect can occur.  During the period of rapid central potential evolution, the timescale on which~\vesc~increases is typically much shorter than the timescale for a recoiling BH orbit to decay via gas drag or dynamical friction.  Consequently, on its first several pericentric passages through the central region, the BH encounters an increasingly deep potential well, and its velocity is actually boosted by a small amount, such that the ratio $v_{\rm BH}$/\vesc~is roughly constant on subsequent pericentric passages.  Because only $v_{\rm BH}$ in creases, not $v_{\rm BH}$/\vesc, the BH's galactocentric distance is not boosted beyond its initial maximum, $R_{\rm max}$, and once the rapid potential evolution ceases, drag forces become dominant and the velocity begins to decay.  This effect is by design short-lived, and it is strongest when the BH orbital timescale is short, such that the BH orbit is dominated by the dynamics of the inner galaxy region.  These velocity boosts are therefore mostly an interesting illustration of the sensitivity of recoil dynamics to the BH merger time, although we will see some implications of this effect for the recoiling AGN lifetimes calculated in \S~\ref{ssec:offsetagn}.

The primary implication of our BH merger time analysis is that in nearly equal-mass, gas-rich mergers such as the example shown in Fig.~\ref{fig:q1tmrg_rmax}, a delay of even a few $\times 10^7$ yr in the BH merger may significantly reduce the amplitude and duration of recoiling BH oscillations.  Based on results from numerous hydrodynamic simulations of binary BH inspiral \citep[e.g.,][]{escala05,dotti07}, we have good reason to believe that BHs in gas-rich systems merge efficiently in general, but the merger time of a given BH binary is impossible to predict with this precision.  By assuming~\tmrg~$=$~\tcoal~in our simulations, we exclude the possibility of delayed mergers, but we also take steps to ensure that this has a minimal effect on our results.  First, only a small number of our simulations have large variation in~\vesc.  In these cases, we see by comparing Figs.~\ref{fig:vk_rmax} \&~\ref{fig:q1tmrg_rmax} that the variation in oscillation amplitudes for a given~\vk/\vesc~due to evolving central density (less than a factor of 10) is generally much less than the variation for different values of~\vk/\vesc~(up to three orders of magnitude).  All of the analysis in this study is presented in terms of~\vk/\vesc.  Therefore, the dramatic suppression of recoil trajectories with fixed~\vk, while an important result in itself, does not affect our other conclusions.

\begin{figure}
\resizebox{\hsize}{!}{\includegraphics{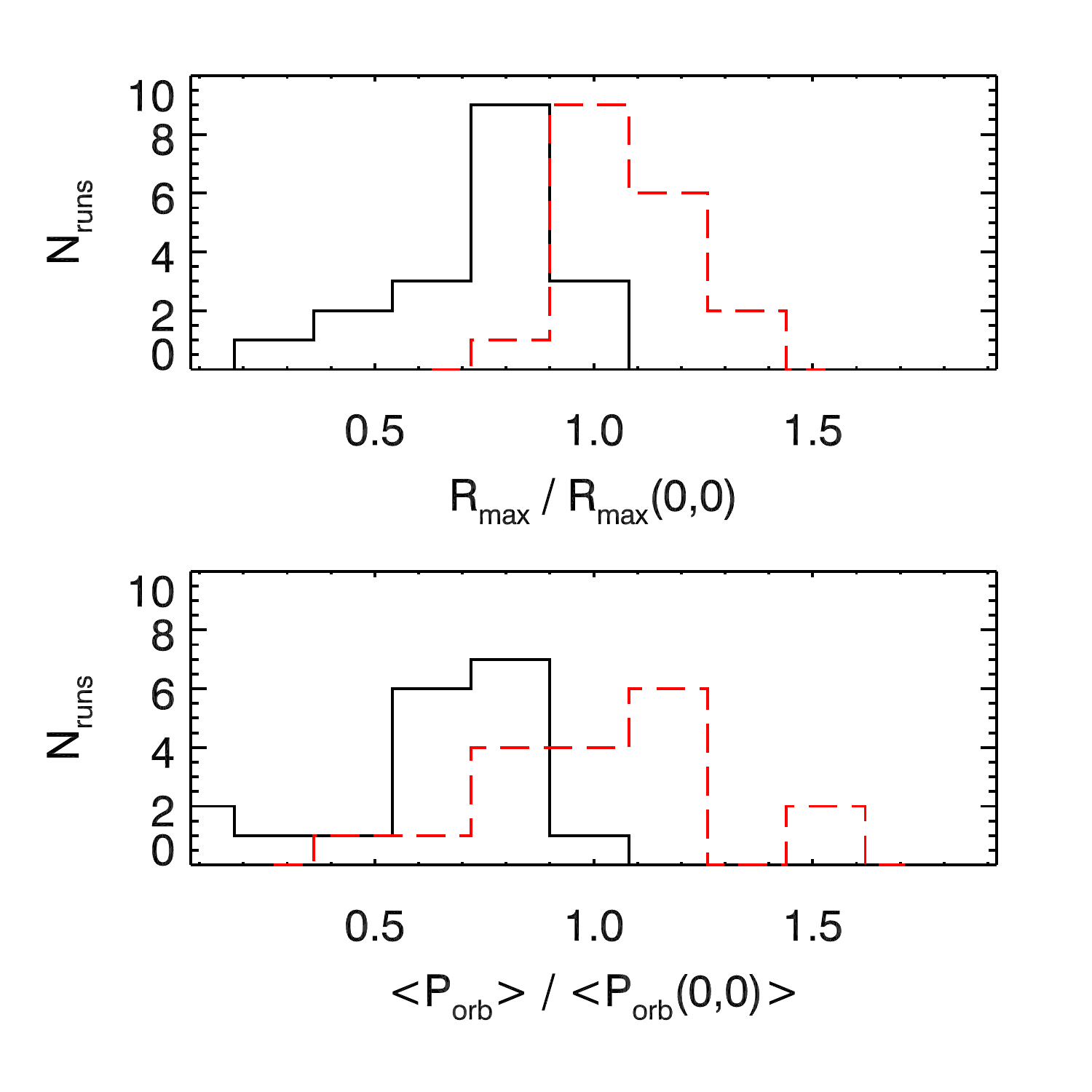}}
%\resizebox{\hsize}{!}{\includegraphics{kick_direction_hist.pdf}}
\caption[]{Top window: distribution of $R_{\rm max}$ for our set of ($\theta,\phi$) $\neq$ (0,0) simulations, normalized to $R_{\rm max}$(0, 0) for each corresponding~\vk/\vesc~value.  The set of simulations includes runs with~\vk/\vesc~$=$ 0.6, 0.8, \& 0.9.  The black solid line is the distribution of simulations with the q0.5fg0.3a model, and the red dashed line corresponds to the q1fg0.3a model.  Bottom window: the distributions of the average orbital period are shown for the same simulations as in the top window, normalized to $\langle P_{\rm orb}(0, 0) \rangle$.  \label{fig:kickdir}} 
\end{figure}

\subsection{Recoil Kick Direction}
\label{ssec:kick_orient} 

Most of our recoiling BHs have~\vk~$ \ga 500$~\kms, in which case their kicks will be oriented out of the orbital plane of the binary \citep{campan07a, campan07b, louzlo09}.  However, because the BH binary inspiral occurs on sub-resolution scales in our simulations, the binary orbital angular momentum vector cannot be predicted.  Furthermore, the galactic disk structure of the merger progenitors is disrupted by major mergers (see Fig.~\ref{fig:3proj}), except when the galaxies have nearly coplanar orbits.  In generic mergers, there is no obvious reference direction in which to orient the recoil kick with respect to the progenitors' initial structure.  Clearly, on certain recoil trajectories (i.e., along tidal tails or central clumps) a BH will encounter denser gaseous regions than on others.  However, it is unclear {\em a priori} whether these density variations are sufficient to cause significant variation in BH recoil trajectories.  In most of our simulations, we simply orient the recoil kick along the $z$-axis of the simulation, $(\theta_{\rm k}, \phi_{\rm k}) = (0,0)$, in the coordinate system with respect to which the initial galaxy orbits are assigned.  Here, we examine a subset of simulations with varying kick orientation to determine the sensitivity of our results to this choice.

We test a sample of different kick orientations with three different values of~\vk/\vesc~(0.6, 0.8, \& 0.9) in our q1fg0.3a and q0.5fg0.3a merger models.  As a gauge of how the BH trajectories differ among this sample, we calculate the maximum galactocentric distance ($R_{\rm max}$) and average orbital period ($\langle P_{\rm orb} \rangle$) of the BH in each simulation.  Fig~\ref{fig:kickdir} shows the distributions of these quantities, normalized to $R_{\rm max}$ and $\langle P_{\rm orb} \rangle$ of the ($\theta_{\rm k}, \phi_{\rm k}$) $=$ (0, 0) simulations for each corresponding~\vk/\vesc~value.  The q0.5fg0.3a distributions peak slightly below 1.0 and the q1fg0.3a distributions peak slightly above 1.0, but the combined histogram is distributed fairly evenly around unity.  In other words, the ($\theta_{\rm k}, \phi_{\rm k}$) $=$ (0, 0) kick orientation is not a ``special" direction.  The overall variation in $R_{\rm max}$/$R_{\rm max}$(0, 0) for each model is less than a factor of 4. In contrast, Fig.~\ref{fig:alltraj} shows that $R_{\rm max}$ is almost two orders of magnitude higher for~\vk/\vesc~$=0.9$ than for $0.6$.  Clearly, the amplitude and duration of BH oscillations depends much more strongly on the kick {\em magnitude} than its direction, and we need not concern ourselves too much with our choice of kick orientation.

\section{Black Hole Accretion and Feedback}
\label{sec:accr_fb}

\subsection{Bondi Accretion and AGN Lifetimes}
\label{ssec:bondi_only}

\begin{figure*}
\resizebox{\hsize}{!}{\includegraphics{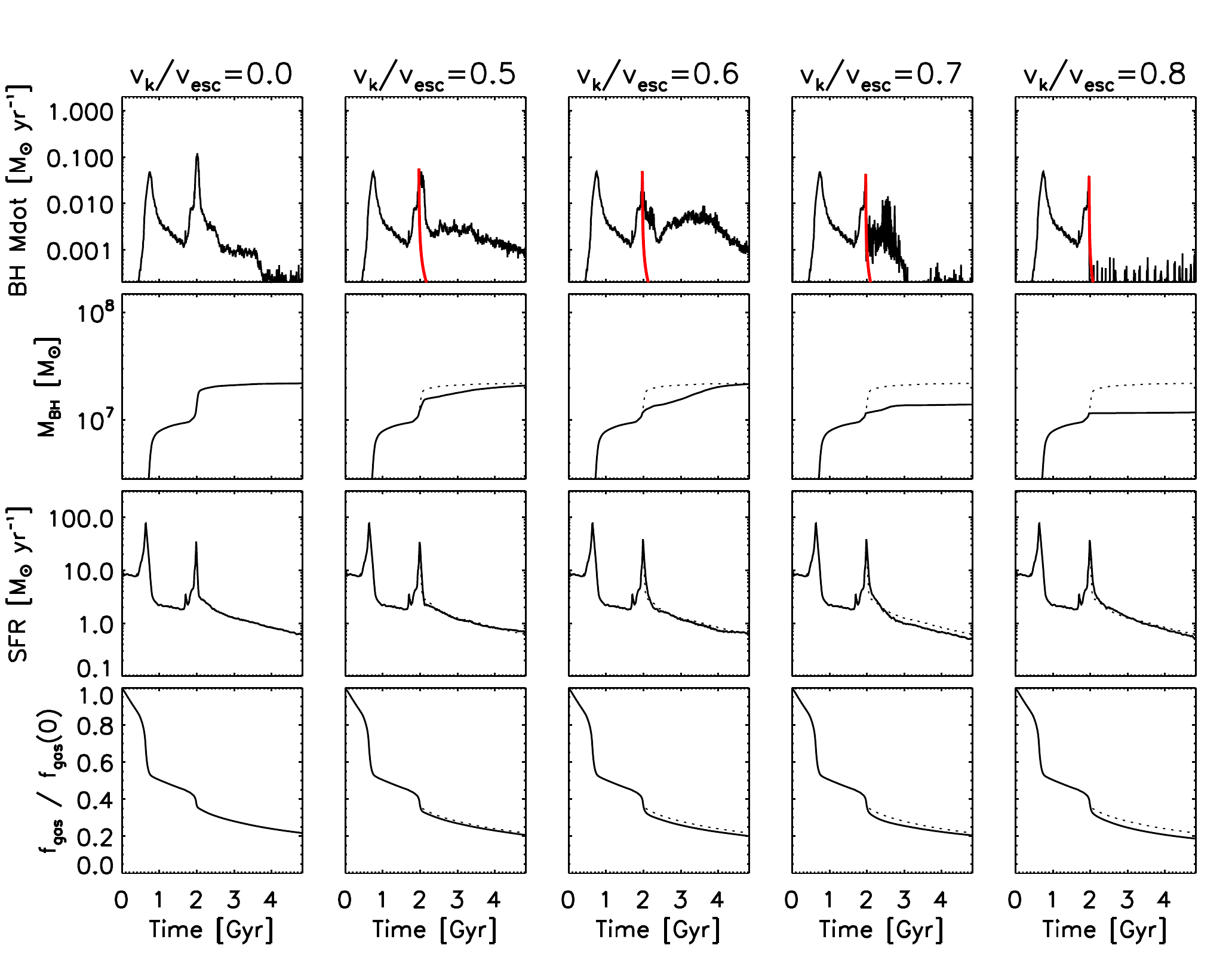}}
%\resizebox{\hsize}{!}{\includegraphics{combine_sfr_fgas_bhs.pdf}}
\caption[]{The same quantities for model q0.5fg0.3a are plotted as in Fig.~\ref{fig:v0_sep_sfr_fgas}, but here they are shown for simulations with~\vk/\vesc~$ = 0.5-0.8$ in addition to the~\vk~$=0$ case.  The red lines in the $\dot M_{\rm BH}$ plots (top row) correspond to the accretion rate calculated from our ejected-disk model as described in \S~\ref{ssec:ejdisk}. In the~\vk~$\neq 0$ plots, the~\vk~$=0$ curves are drawn again with dashed lines, except for the $\dot M_{\rm BH}$ plots where the dashed lines are omitted for clarity.  \label{fig:sfr_fgas_vks}}
\end{figure*}

\begin{figure*}
\resizebox{0.9\hsize}{!}{\includegraphics{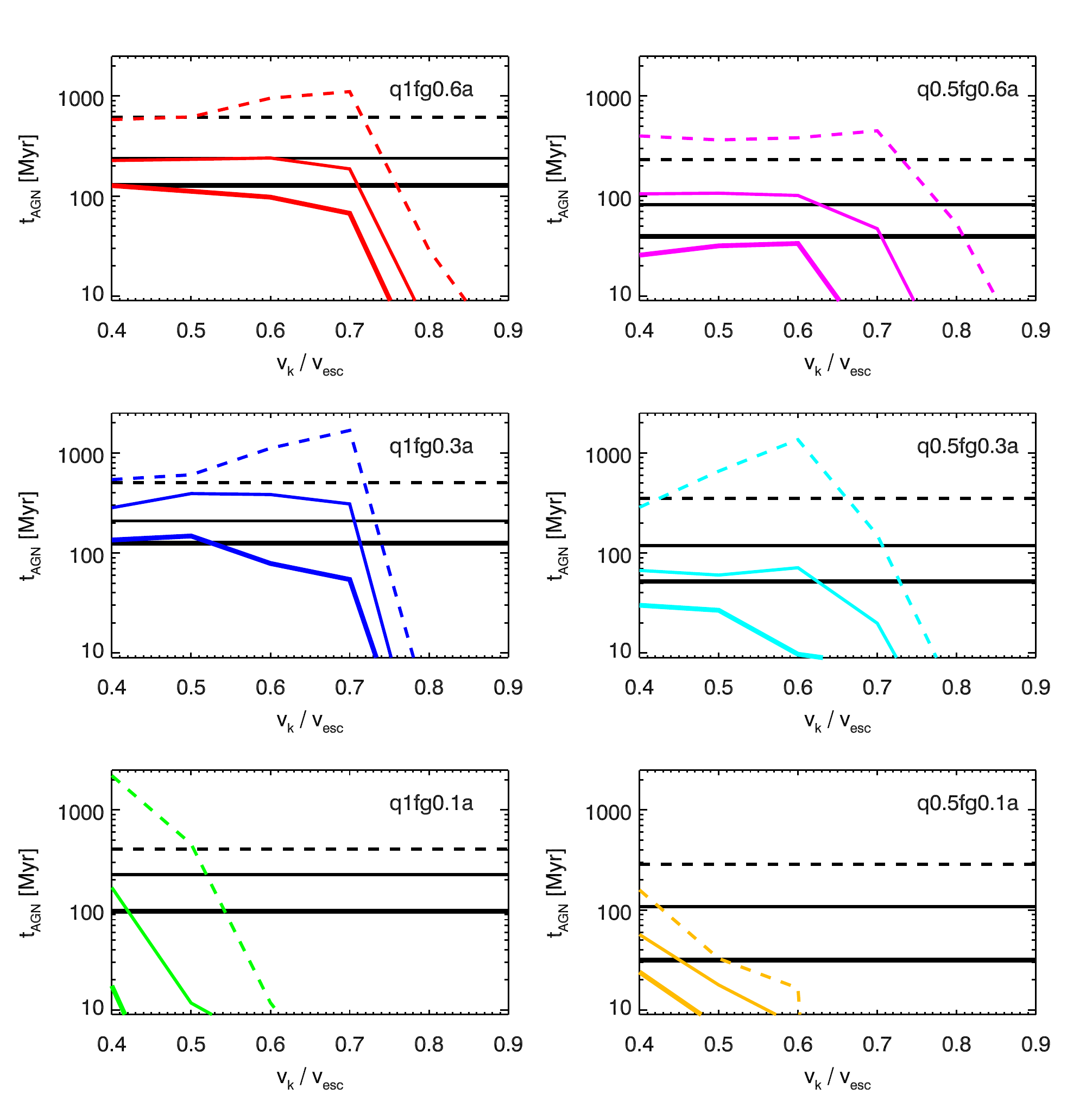}}
%\resizebox{0.9\hsize}{!}{\includegraphics{nk_compare_tagn.pdf}}
\caption[]{AGN lifetimes (\tagn) of recoiling BHs with~\vk/\vesc~$= 0.4-0.9$, compared to~\tagn~for~\vk~$=0$ in six different galaxy models.  In each plot, the black lines indicate the stationary AGN lifetimes and the colored lines indicate the recoiling BH lifetimes at each value of~\vk/\vesc.  The models shown are (from top left): q1fg0.6a (red), q1fg0.3a (blue), q1fg0.1a (green), q0.5fg0.6a (magenta), q0.5fg0.3a (cyan), and q0.5fg0.1a (gold), as indicated in the plot labels.  The different line types denote~\tagn~according to three different definitions of an ``active" BH.  For the thin and thick solid lines,~\tagn~is defined as the time for which the BH luminosity $L_{\rm bol} > 3\%$ and $10\%\, L_{\rm Edd}$, respectively.  The dashed lines show~\tagn~for which $L_{\rm bol} > L_{\rm min}$, where $L_{\rm min} = 3.3\times10^9$ L$_{\odot}$, or $0.1 L_{\rm Edd}$ for a $10^6$~\msun~BH.  In all cases $L_{\rm bol}$ is calculated from the Bondi-Hoyle accretion rate as described in the text.  No cuts have been applied to~\tagn~based on the BH settling time, $t_{\rm settle}$, so in some cases~\tagn~includes accretion that occurs after the BH settles back to the galactic center. \label{fig:compare_nokick}}
\end{figure*}

Fig.~\ref{fig:sfr_fgas_vks} compares the BH accretion and star formation histories for simulations of the q0.5fg0.3a merger model with~\vk~$= 0$ and~\vk/\vesc~$= 0.5 - 0.8$.  The top windows in Fig.~\ref{fig:sfr_fgas_vks} show that even kicks with~\vk/\vesc~$ = 0.5 - 0.6$ produce qualitatively different AGN lightcurves (as inferred from $\dot M_{\rm BH}$) than in the zero-kick case, although the final BH masses are similar.  In these cases, the peaks in accretion rate at coalescence ($t \approx 2$ Gyr) are slightly lower due to the sudden removal of the BHs from the highest-density region.  However, the BHs remain within 1 kpc of the galactic center, and accretion continues through the end of the simulation at higher rates than in the zero-kick case.  Thus, although the total mass accreted is about the same for~\vk/\vesc~$=0-0.6$, the active lifetime is longer when a recoil kick occurs.  

In the~\vk/\vesc~$ = 0.7$ simulation, the BH accretion cuts off more sharply at~\tmrg~and is subsequently highly variable, due to the larger-amplitude, longer-lived BH oscillations.  Unlike the BHs with lower-velocity recoils, this BH attains a noticeably lower final mass than its zero-kick counterpart.

For recoil kicks $\ga 0.8$~\vesc, the BH accretion cuts off sharply at~\tmrg, and the BH subsequently accretes very little gas as it travels on large, long-period orbits that extend well into the halo.  These high-velocity recoil events therefore reduce the AGN lifetime.  Additionally, the BH mass is essentially frozen at its value at~\tmrg, which is about half that of the stationary BH mass.  If the BH does return to the galaxy center after several Gyr, most of the galaxy's gas will have been consumed in star formation, so the BH will remain undermassive.  In the galaxy merger models we have used, the BH mass deficits for high-velocity recoils versus no recoil range from a factor of $\sim 1$ (almost no deficit) to $\sim 5$.  These BH mass deficits and their implications are discussed in \S~\ref{ssec:msigma}.

To quantify the amount of enhancement or reduction in AGN activity owing to GW recoil, we calculate AGN lifetimes (\tagn) for both stationary and recoiling BHs in six different galaxy models for which we have varied~\vk.  Fig.~\ref{fig:compare_nokick} shows the results, with colored lines indicating~\tagn~for recoiling BHs as a function of~\vk~and black horizontal lines indicating~\tagn~for a stationary BH in the corresponding merger model.  In order to differentiate between bright AGN and low-luminosity AGN, we calculate~\tagn~assuming three different definitions of an ``active" BH.  The thick solid lines in the figure are calculated assuming the BH is an AGN when $L_{\rm bol} > 10\% \, L_{\rm Edd}$, and the thin solid lines assume $L_{\rm bol} > 3\%\, L_{\rm Edd}$.  Because these quantities depend on the BH mass, we use an absolute luminosity value for our third AGN definition: $L_{\rm bol} > L_{\rm min}$ (denoted by the dashed lines in Fig.~\ref{fig:compare_nokick}).  We choose $L_{\rm min} = 3.3\times10^9$ L$_{\odot}$, which is $0.1 \, L_{\rm Edd}$ for a $10^6$~\msun~BH. 

In these models, the low-luminosity AGN lifetime is generally enhanced by moderate-velocity recoil events, and the bright AGN lifetime is generally reduced.  Although the details depend on the model parameters and kick speed, in all models,~\tagn~falls off steeply at high~\vk/\vesc, when the BH is ejected far enough from the center on large enough orbits that Bondi accretion becomes ineffective.  This corresponds to the accretion behavior seen in the last column of Fig.~\ref{fig:sfr_fgas_vks} (\vk/\vesc~$=0.8$).  

For smaller~\vk, such that the BH is confined to the central few kpc, the dependence of~\tagn~on the nature of the merger remnant is more complex.  We examine first the relatively gas-poor models (those with initial~\fgas~$=0.1$).  In the q1fg0.1a model,~\tagn~is enhanced only for~\vk/\vesc~$\le 0.5$ and falls precipitously at higher~\vk.  In the q0.5fg0.1a model, recoil events always reduce the AGN lifetime.  Merger remnants with less gas have smaller central gas reservoirs to fuel the BH during pericentric passages, and their shallower central potentials result in larger BH orbits, which have fewer pericentric passages.  Consequently, in gas-poor remnants,~\tagn~will generally be lower for recoiling BHs than for stationary BHs.  

In gas-rich merger remnants,~\tagn~is more readily enhanced by GW recoil.  Stationary BHs in gas-rich mergers typically experience a bright AGN phase due to the rapid inflow of cold gas to the central region, which is terminated when the AGN feedback energy heats the surrounding gas enough to quench the BH fueling.  Fig.~\ref{fig:compare_nokick} shows that although GW recoil shortens this bright AGN phase, the low-luminosity~\tagn~is enhanced for recoil kicks as high as~\vk/\vesc~$= 0.7$.  In these cases, the recoiling BHs are on tightly-bound orbits and can accrete gas during their numerous passages through the central dense region.  Because they remain in motion, however, their feedback energy is spread over a much greater volume than for stationary BHs.  As a result, the gas encountered by the recoiling BHs never heats up enough to completely cut off the BH fuel supply, and the BH has a longer lifetime as a low-luminosity AGN.  In contrast to the traditional ``feast or famine" model of merger-triggered AGN fueling, these short-period recoiling BHs are more inclined to ``nibble".  

We can see evidence of this displaced AGN feedback through its effect on star formation.  The bottom row of Fig.~\ref{fig:sfr_fgas_vks} shows the depletion of gas throughout each simulation, normalized to the initial gas fraction, and the dotted lines show the~\vk~$ = 0$ curve in each of the other plots.  It is evident that slightly less gas remains at the end of each higher-\vk~simulation.  In the merger with a stationary BH, this gas is accreted or expelled from the galactic center via feedback, but when the central BH is removed, the gas continues be consumed by star formation.  Although in this example the star formation rate enhancement is small and difficult to discern from examining the SFR evolution alone, the effect can be much larger in higher-$q$,~\fgas~mergers.  We explore in~\S~\ref{ssec:centralevol}~the implications of enhanced SFRs for our equal-mass, gas-rich mergers.

Although we have found that~\tagn~may be enhanced by GW recoil for low-luminosity AGN, we emphasize that the bright AGN phase ($L > 10\% L_{\rm Edd}$) is always shorter for recoiling BHs than for stationary BHs, or at best roughly equal.  This is a direct result of the peak accretion episode being disrupted at the time of the recoil kick.  The important consequence of this is that the total accreted BH {\em mass} is never enhanced by GW recoil in our simulations.  For moderate recoil speeds, the BH mass deficit relative to a stationary BH is generally slight, but we find that the effect of recoil is {\em always} to decrease the final BH mass.  The implications of this finding are discussed in \S~\ref{ssec:msigma}.
 
Finally, we note a possible observational consequence of longer AGN lifetimes.  Fig.~\ref{fig:sfr_fgas_vks} clearly shows that when~\vk~$=0$, the two main accretion episodes are associated with the periods of rapid star formation.  This means that the AGN luminosity may be difficult to distinguish from the starburst luminosity, and the AGN may also be dust-obscured for at least part of its active phase.  If GW recoil allows the AGN to remain active long after the starburst is complete, such an AGN might be more readily observable, because it would no longer be competing with the starburst luminosity and may also be less dust-obscured.

\subsection{Ejected Gas Disk}
\label{ssec:ejdisk}

We have seen that moderate-velocity recoils in gas-rich remnants may allow efficient accretion from ambient gas.  However, a BH moving rapidly or in a low-gas-density region is unlikely to sweep up much gas from its surroundings, and we would expect the probability of observing this type of active BH to be low.  Indeed, in our simulations, the Bondi accretion rate of the BH rapidly declines following a large recoil kick (see Fig.~\ref{fig:sfr_fgas_vks}).  The possibility not accounted for in our simulations is that, if the BH is surrounded by an accretion disk at the time of recoil, some amount of this gas may remain bound to the recoiling BH.  We therefore implement an analytic model to calculate the BH accretion rate from this ejected gas disk, based on the BH mass and accretion rate at the moment of recoil.  

The BH carries with it the part of the accretion disk where $v_{\rm orb} \ga $~\vk; i.e., it will carry a disk with radius approximately $R_{\rm ej} \sim G M_{\rm BH} / $\vk$^2$.  This is a reasonable approximation if the BH is kicked directly out of the orbital plane of the disk, as is the case when~\vk~$ \ga 500$~\kms~and the BH orbital plane is aligned with that of the disk.  If the accretion rate from this ejected disk is high enough, such a system could potentially be seen as an offset AGN via either resolved spatial offsets or offset spectral lines.  BL08 predicted accretion timescales of $\la 1 - 10$ Myr for the type of recoiling BHs we consider here ($M_{\rm BH} \sim 10^6 - 10^8$~\msun;~\vk~$ \sim 400 - 1000 $~\kms), assuming a constant-$\dot M$ model.  Here, we improve on these estimates by calculating a time-dependent accretion rate for the ejected disk.

We assume the innermost region of the gas disk is well-modeled by a viscous, Keplerian, thin disk (specifically, an $\alpha$-disk, \citet{shasun73}).  The evolution of the disk surface density, $\Sigma$, can be described by a diffusion equation:
\begin{equation}
{\partial \over \partial t} \Sigma(R,t) = {3 \over R} {\partial \over \partial R} \left [ R^{1/2} {\partial \over \partial R} \left ( \nu \Sigma R^{1/2} \right ) \right ],
\label{eqn:diff}
\end{equation}
where $\nu$ is the kinematic viscosity \citep[see, e.g.,][]{lynpri74,pringl81}.  The steady-state solution is recovered when the left-hand side of the equation vanishes, but this is not the case for our ejected disk, which is no longer fed at its outer radius.  Exact analytic solutions to Eq.~(\ref{eqn:diff}) can be found when $\nu$ scales only with radius \citep{lynpri74}, but in the $\alpha$-disk model, $\nu$ is a function of both $R$ and $\Sigma$.  \citet{pringl91} has derived self-similar solutions to this equation for viscosity scaling as $\nu \propto \Sigma^m R^n $.  These solutions assume a disk with no inner radius, so for our case we must neglect the finite inner disk radius at the innermost stable circular orbit of the BH ($R_{\rm ISCO}$), as well as the possibility of a circumbinary gap in the disk that persists after the BH merger.  This should not affect our results, however, as the ejected disk will be much larger than the ISCO ($R_{\rm ej}/R_{\rm ISCO} > 10^3$), and the gap is expected to refill on a timescale of years \citep{milphi05,tanmen10}.  For an $\alpha$-disk in which viscosity is assumed to scale with gas pressure and the opacity equals the electron-scattering value ($\kappa_{\rm es} \approx 0.4$ cm$^2$ g$^{-1}$), the viscosity is given by
\begin{eqnarray}
\nu = C \,R \, \Sigma^{2/3} \,  ,\\ 
C \equiv \left ( { k_B \, \alpha \over \mu \, m_{\rm p} } \right )^{4/3} \left ( { \kappa_{\rm es} \over 48 \, \sigma_{\rm SB} \, G \, M_{\rm BH} } \right )^{1/3}.  \nonumber 
\end{eqnarray}
In this equation $k_{\rm B}$ is Boltzmann's constant, $\alpha$ is the thin-disk viscosity parameter, $\mu$ is the mean molecular weight, $m_{\rm p}$ is the proton mass, and $\sigma_{\rm SB}$ is the Stefan-Boltzmann constant. This corresponds to the self-similar solution of \citet{pringl91} with $m=2/3, n=1$ \citep[see also][]{canniz90}:
\begin{eqnarray}
\label{eqn:disksoln} 
{\Sigma(R,t) \over \Sigma_{\rm 0} } = \left ( { t \over t_{\rm 0} } \right )^{-15/16} f\left[ \left( {R \over R_{\rm 0}} \right) \left( {t \over t_{\rm 0}} \right)^{-3/8} \right],\\ 
f[u] \equiv (28)^{-3/2}\, u^{-3/5} \,(1 - u^{7/5})^{3/2}. \nonumber
\end{eqnarray}
The arbitrary constants $\Sigma_{\rm 0}$, $t_{\rm 0}$, and $R_{\rm 0}$ satisfy
\begin{equation}
t_{\rm 0}^{-1} = { C\, \Sigma_{\rm 0}^{2/3} \over R_{\rm 0} }.
\label{eqn:consts}
\end{equation}
Integrating Eq.~(\ref{eqn:disksoln}) yields a time-dependent expression for disk mass:
\begin{equation}
M_{\rm d}(t) = (28)^{3/2} \, {4\pi \over 7} \, R_{\rm 0}^2\, \Sigma_{\rm 0} \, \left( {t \over t_{\rm 0}} \right)^{-3/16}.
\end{equation}
From this we obtain the time-dependent accretion rate:
\begin{equation}
\dot M_{\rm d}(t) =  - {3\over16}\, (28)^{3/2} \,{4\pi \over 7} \, {R_{\rm 0}^2\, \Sigma_{\rm 0} \over t_{\rm 0}} \,\left( {t \over t_{\rm 0}} \right)^{-19/16}.
\end{equation}
We determine $\Sigma_{\rm 0}$, $t_{\rm 0}$, and $R_{\rm 0}$ using Eq.~(\ref{eqn:consts}) along with the conditions that $M_{\rm d}(t') = M_{\rm ej}$ and $\dot M_{\rm d}(t') = \dot M_{\rm mrg}$, where $t' = t_{\rm 0} + t_{\rm sim} - t_{\rm mrg}$, $t_{\rm sim}$ is the time in the simulation, $M_{\rm ej}$ is the ejected disk mass, and $\dot M_{\rm mrg}$ is the Bondi accretion rate calculated in the simulation at the time of merger.  The mass evolution of the disk is therefore given by:
\begin{eqnarray}
M_{\rm d}(t') = M_{\rm ej} \, \left( {t' \over t_{\rm 0}} \right)^{-3/16}, \\
\dot M_{\rm d}(t') = \dot M_{\rm mrg} \, \left( {t' \over t_{\rm 0}} \right)^{-19/16}, \\
t_{\rm 0} = {3 \over 16} \, { M_{\rm ej} \over \dot M_{\rm mrg} }.
\label{eqn:mdot}
\end{eqnarray}

We can calculate the initial radius and mass of the ejected disk using the multi-component, self-gravitating disk model of BL08.  The calculation takes as input the kick velocity, BH accretion rate, and BH mass at the time of the merger.  Note that because we use the accretion rate calculated in the code as input, the normalization of the ejected-disk accretion rate is subject to the same assumptions as is the Bondi-Hoyle accretion rate (see \S~\ref{sec:simulations}).  We refer the reader to BL08 for the details of the self-gravitating disk model, but we note that for the BH masses and accretion rates considered here, $R_{\rm ej}$ lies in the region of the disk that is self-gravitating but still strongly dominated by the BH potential (i.e., Zone II of the BL08 model).  Quantitatively, $R_{\rm ej}$ for our fiducial-mass simulations is between $10^{3.9 - 5.7}$ Schwarzschild radii, or $\sim 0.02 - 0.6$ pc.  The corresponding mass of the ejected disk, $M_{\rm ej}$, is $10^{3.5 - 7.5}$~\msun~($10^{4.6 - 6.3}$~\msun~for our fiducial-mass models).  

We use these quantities as initial conditions for the ejected disk model outlined above.  The timescale for the accretion rate (and mass) decay, $t_{\rm 0}$, ranges between $10^{5.6 - 7.6}$ yr.   $t_{\rm 0}$ depends on  $M_{\rm ej}$ and $\dot M_{\rm mrg}$, which in turn depend on~\vk~and the galaxy model parameters.  Specifically,  $t_{\rm 0}$ decreases with increasing~\vk,~\fgas, and $q$.  After a time $t_{\rm 0}$, the accretion rate has dropped by a factor of $2^{-19/16} \approx 0.44$ (note that the time in Eqns.~\ref{eqn:mdot} is defined such that $t' = t_{\rm 0}$ at~\tmrg).  By the end of our simulations, about $55 - 80\%$ of the disk mass has been consumed.  This corresponds to an  increase in the BH mass of $\Delta M = 0.3 - 6\%$ ($0.3 - 3\%$ for~\vk/\vesc~$=0.9$ simulations only).  As expected, these values of $\Delta M$ are somewhat lower than the result of BL08, $\Delta M \approx 10\%\, M_{\rm BH}$.  BL08 considered kick speeds as low as 100~\kms; their result also assumed that the entire disk mass was accreted and that the accretion was Eddington-limited at the time of BH merger.

\begin{figure*}
\resizebox{0.49\hsize}{!}{\includegraphics{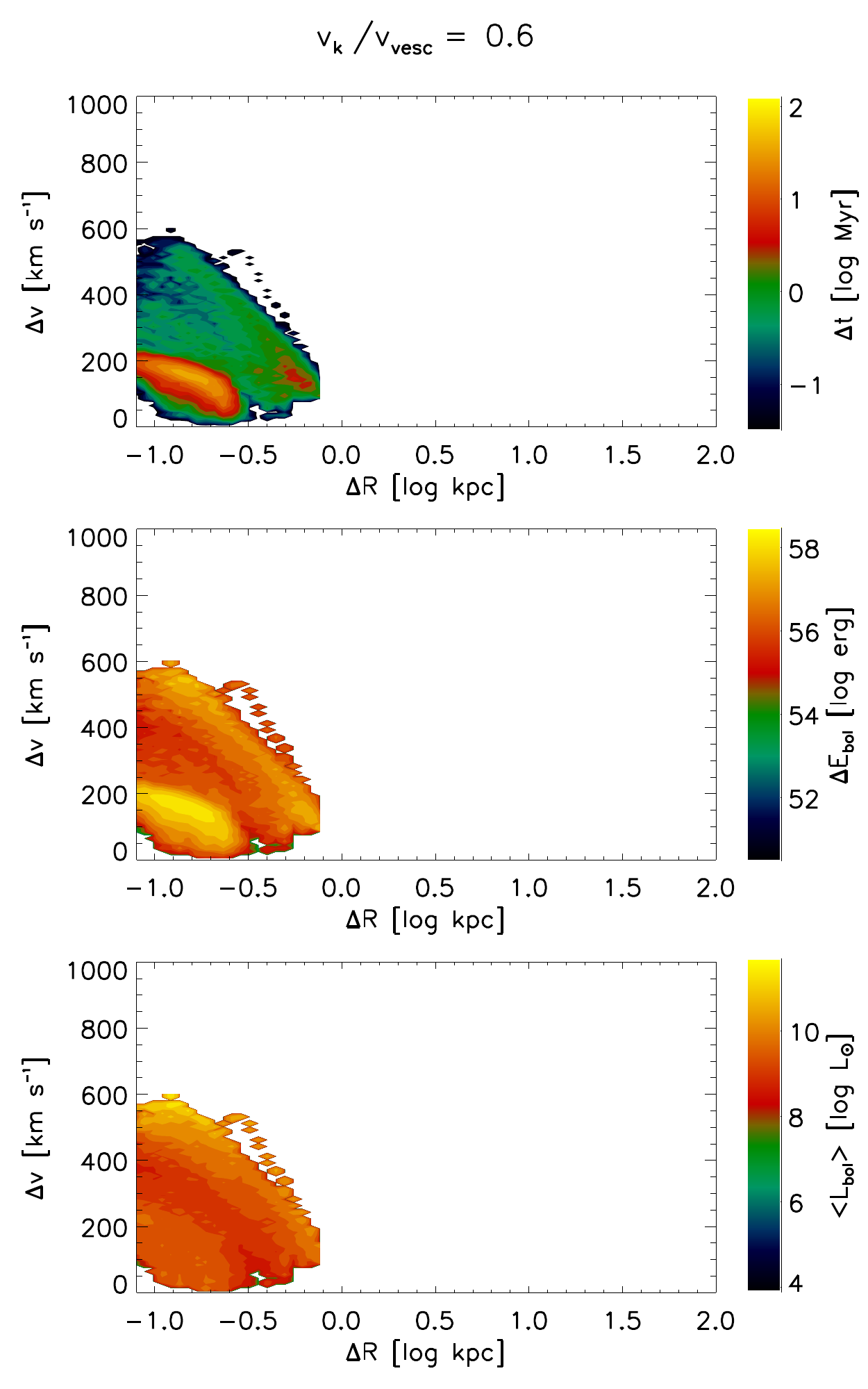}}
\resizebox{0.49\hsize}{!}{\includegraphics{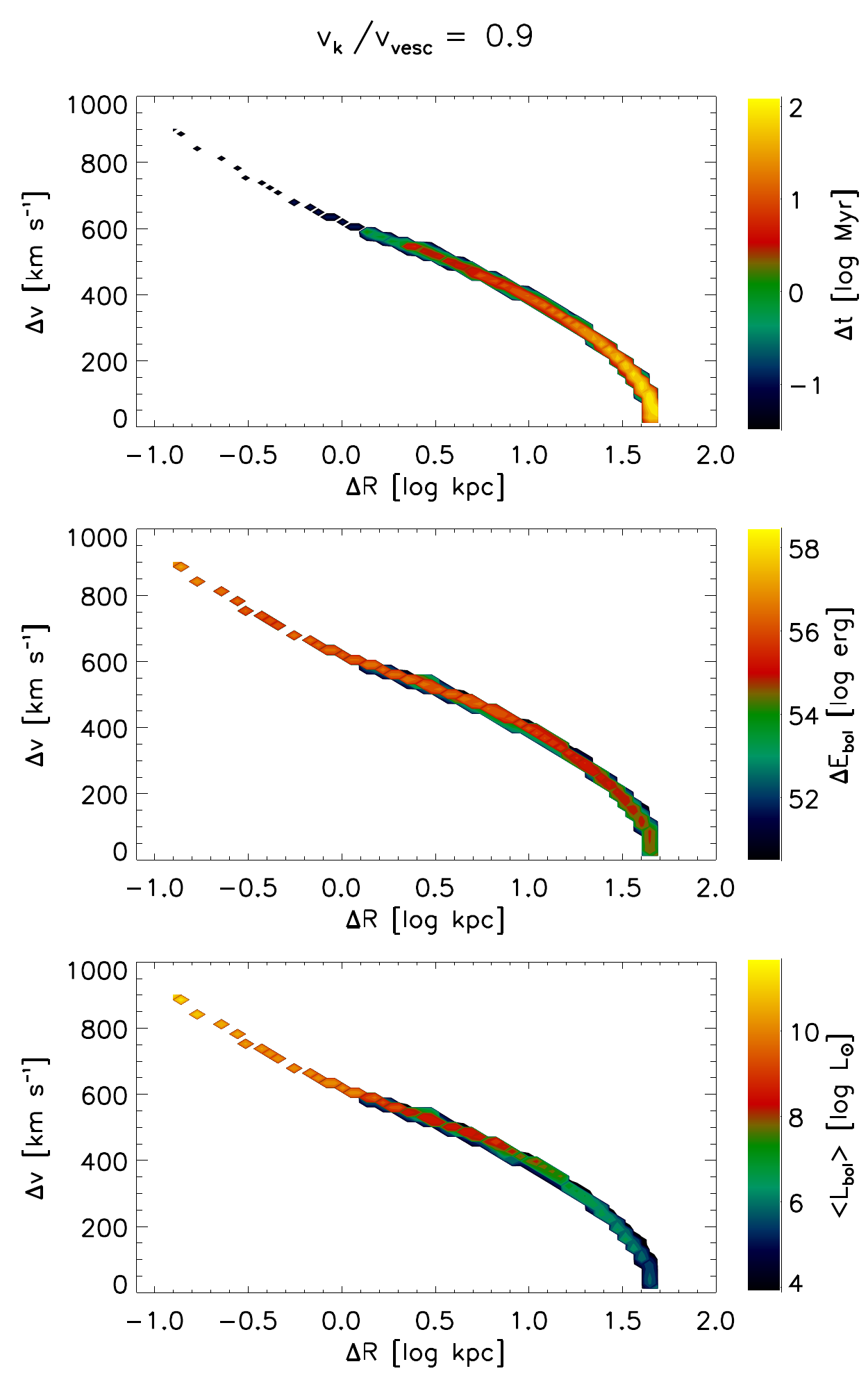}}
%\resizebox{0.49\hsize}{!}{\includegraphics{qhalf_fg30_v60_traj_dist_lg.pdf}}
%\resizebox{0.49\hsize}{!}{\includegraphics{qhalf_fg30_v90_traj_dist_lg.pdf}}
\caption[]{Contour plot of recoiling BH quantities $\Delta t$ (top panels), $\Delta E_{\rm bol}$ (middle panels), and $\langle L_{\rm bol} \rangle$ (bottom panels) in $\Delta R$, $\Delta v$ space.  $\langle L_{\rm bol} \rangle$ is the average bolometric BH luminosity per bin.  The left panels are calculated from the q0.5fg0.3a merger simulation with~\vk/\vesc~$= 0.6$, and the right panels are for the same merger model but with~\vk/\vesc~$= 0.9$.  \label{fig:trajdist}}
\end{figure*}

Fig.~\ref{fig:sfr_fgas_vks} demonstrates that except in high-\vk~simulations where the average orbital timescale is long (\vk~$\ga 0.8$~\vesc), the ejected-disk accretion is typically insignificant compared to the Bondi accretion of the oscillating BH.  Of course, the BH may still carry an accretion disk during this phase, but the accretion will be more accurately characterized by the highly-variable Bondi rate calculated from the local density and sound speed than by an isolated-disk model.  Therefore, our analytic ejected-disk model becomes important only for high-velocity recoils in which the Bondi accretion rate drops sharply after the kick.  Below, we discuss the AGN lifetimes resulting from the combined accretion models, as well as the feasibility of observing offset AGN.

\subsection{Offset AGN}
\label{ssec:offsetagn}

In Fig.~\ref{fig:trajdist}, we show 2 examples of the $\Delta R$, $\Delta v$ phase space occupied by recoiling BHs.  The left and right panels show simulations of the q0.5fg0.3a model with~\vk/\vesc~$= 0.6$ \& 0.9, respectively.  $\Delta R$ is the binned spatial offset of the BH from the stellar center of mass, and $\Delta v$ is the binned kinematic offset from the stellar center-of-mass velocity. The contour shading shows, for each $\Delta R$, $\Delta v$ bin, the amount of time spent at that location ($\Delta t$, top windows), the total AGN energy output ($\Delta E_{\rm bol}$, middle windows), and the average bolometric luminosity $\langle L_{\rm bol} \rangle$ (bottom windows).   $L_{\rm bol}$ is determined at each timestep from the relation $L_{\rm bol} = \epsilon_{\rm rad} \, \dot M_{\rm BH}\, c^2$, where the radiative efficiency $\epsilon_{\rm rad}$ is 0.1 unless $\dot M_{\rm BH} < 0.01 \dot M_{\rm Edd}$, in which case the system is considered radiatively inefficient, and $\epsilon_{\rm rad} = 0.1\, \dot M_{\rm BH} / ( 0.01 \dot M_{\rm Edd} )$ \citep{narmcc08}.

\begin{figure*}
\resizebox{\hsize}{!}{\includegraphics{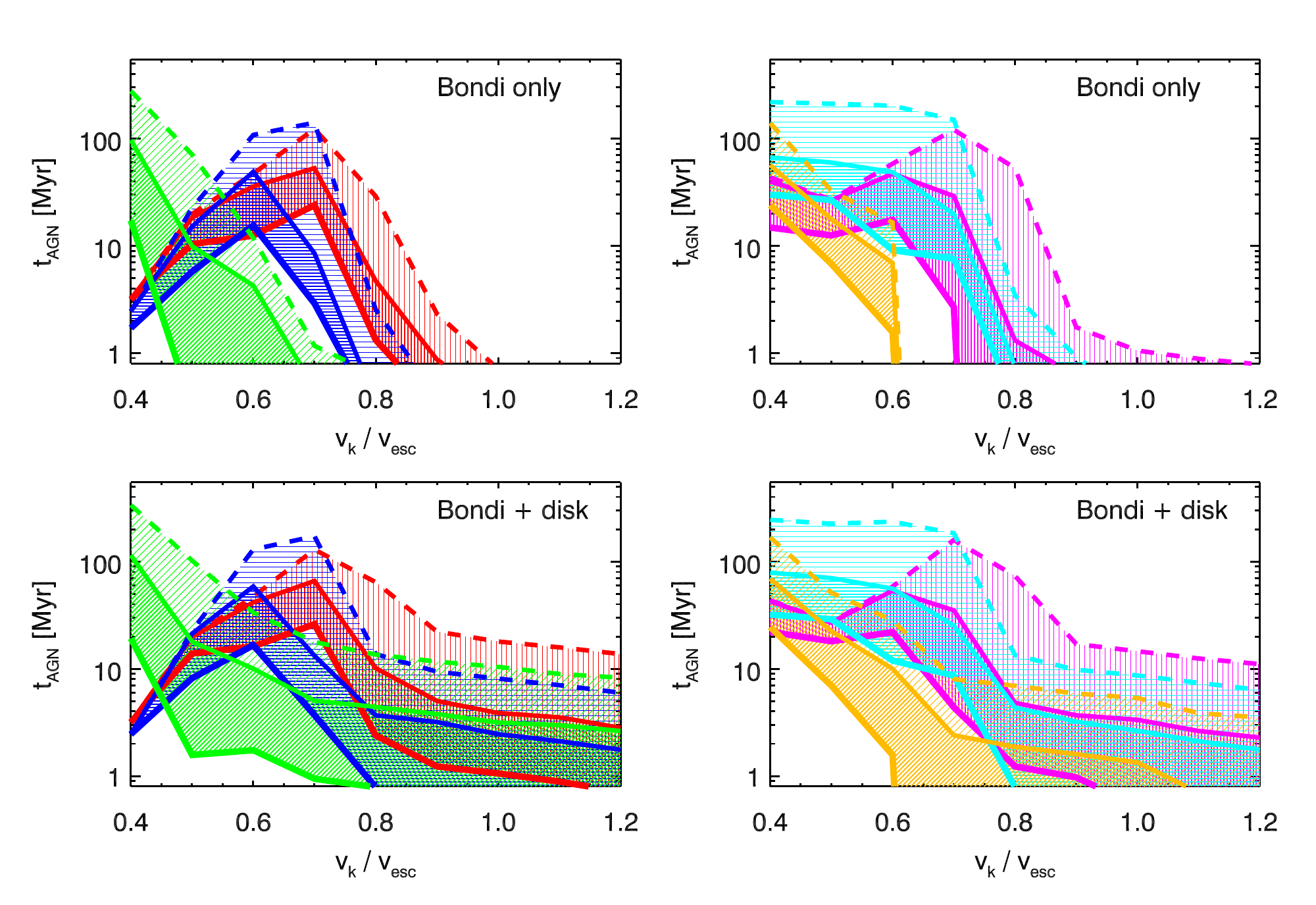}}
%\resizebox{\hsize}{!}{\includegraphics{compare_ejdisk.pdf}}
\caption[]{Active lifetimes of recoiling BHs as a function of~\vk.  Results are shown for the same six galaxy models shown in Fig.~\ref{fig:compare_nokick}, separated by mass ratio for clarity ($q=1$ on the left, $q=0.5$ on the right).  On the left: q1fg0.6a (red), q1fg0.3a (blue), q1fg0.1a (green); on the right: q0.5fg0.6a (magenta), q0.5fg0.3a (cyan), and q0.5fg0.1a (gold).  For each model, simulations were run for~\vk/\vesc~$ = 0.4 - 1.2$.  In all plots, the solid and dashed lines denote the three different definitions of~\tagn~discussed in the text, and the shaded regions denote the range of values between these bounds.  The upper and lower solid lines denote~\tagn~for $L_{\rm bol} > 3\%$ and $10\% \, L_{\rm Edd}$, respectively.  The dashed lines show~\tagn~for $L_{\rm bol} > L_{\rm min}$, where $L_{\rm min} = 3.3\times10^9$ L$_{\odot}$.  Unlike Fig.~\ref{fig:compare_nokick}, here we have applied cuts to~\tagn~such that only the active time~\tagn~$< t_{\rm settle}$ is shown.  In the top two windows, $L_{\rm bol}$ is calculated from the Bondi accretion rate only, while the bottom two windows show~\tagn~with the ejected-disk accretion model included as well.  \label{fig:tagn_compare_ejdisk}}
\end{figure*}

 The~\vk/\vesc~$= 0.6$ recoiling BH, which settles back to the galaxy center in $\sim 400$ Myr, fills a corner of the phase space.  The $\Delta t$ contours peak at low velocities, as these represent apocentric passages where the BH spends the most time.  $\langle L_{\rm bol} \rangle$ is above $10^8 L_{\rm \odot}$ throughout the simulation, with peak values $> 10^{11} L_{\rm \odot}$.  Unlike the $\Delta t$ contours, $\langle L_{\rm bol} \rangle$ peaks at high velocities, because the BH has the highest accretion rates immediately after the kick and during pericentric passages.  The total bolometric energy output per $\Delta R$, $\Delta v$ bin, which is the product of these two distributions, accordingly peaks at both high $\Delta R$, $\Delta v$ and low $\Delta R$, $\Delta v$.  Instead of emitting all of its AGN feedback energy from a single central point in the galaxy, the BH distributes this energy throughout the central kpc, as discussed in \S~\ref{ssec:bondi_only}.  Consequently, this BH has a longer AGN lifetime than the~\vk~$= 0$ BH in the same merger model (see Fig.~\ref{fig:compare_nokick}).
  
In contrast to this example, the~\vk/\vesc~$= 0.9$ recoiling BH occupies a narrow track in the phase space, as it completes only a few orbits and does not experience much drag.  The $\Delta t$ and $\langle L_{\rm bol} \rangle$ contours in this example illustrate an inherent difficulty in observing rapidly-recoiling BHs: the brightest luminosities occur just after the kick, when the BH velocities are highest, but this is also where the BH spends the least amount of time.  Still, the BH has luminosities $\ga 10^8 L_{\rm \odot}$ for $\Delta R$ up to 10 kpc.  Further analysis is required to determine the observability of the recoiling BHs in both of these examples, as well as those in our other simulations.

 Fig.~\ref{fig:tagn_compare_ejdisk} shows the AGN lifetimes,~\tagn, in six merger models for which we have varied~\vk.  AGN lifetimes in the top two windows include only the Bondi accretion rate, while~\tagn~in the bottom two windows includes the accretion from our ejected-disk model as well.   Unlike the lifetimes in Fig.~\ref{fig:compare_nokick}, here we also apply a cut to the AGN lifetime such that only activity that occurs while the BH is recoiling is shown:~\tagn~$< t_{\rm settle}$.  We calculate~\tagn~using each of the three AGN definitions described previously.  The solid lines in the figure correspond to AGN defined as $L_{\rm bol} > (3\%, 10\%) L_{\rm Edd}$, and the dashed lines correspond to $L_{\rm bol} > L_{\rm min}$.  The shaded regions of the plot show the range of values between these definitions.  As before, we choose $L_{\rm min} = 3.3\times10^9$ L$_{\odot} = 1.3 \times 10^{43}$ erg s$^{-1}$. 

A notable feature in Fig.~\ref{fig:tagn_compare_ejdisk} is that the peak AGN lifetimes for recoiling BHs are quite long, $> 100$ Myr, even for kick speeds near 1000~\kms in some cases.  Note also that in the lower-\fgas~mergers,~\tagn~peaks at the lowest~\vk/\vesc, but in the gas-rich mergers the peak value is~\vk/\vesc~$= 0.7 - 0.8$.  This reflects the fact that we have defined~\tagn~$< t_{\rm settle}$ in these plots.  In the gas-rich mergers, gas drag quickly damps out the low-\vk~trajectories, and~\tagn~is limited by $t_{\rm settle}$.  For higher~\vk/\vesc, gas drag is less efficient and~\tagn~is instead limited by the gas supply available to the BH.

\begin{figure*}
\resizebox{\hsize}{!}{\includegraphics{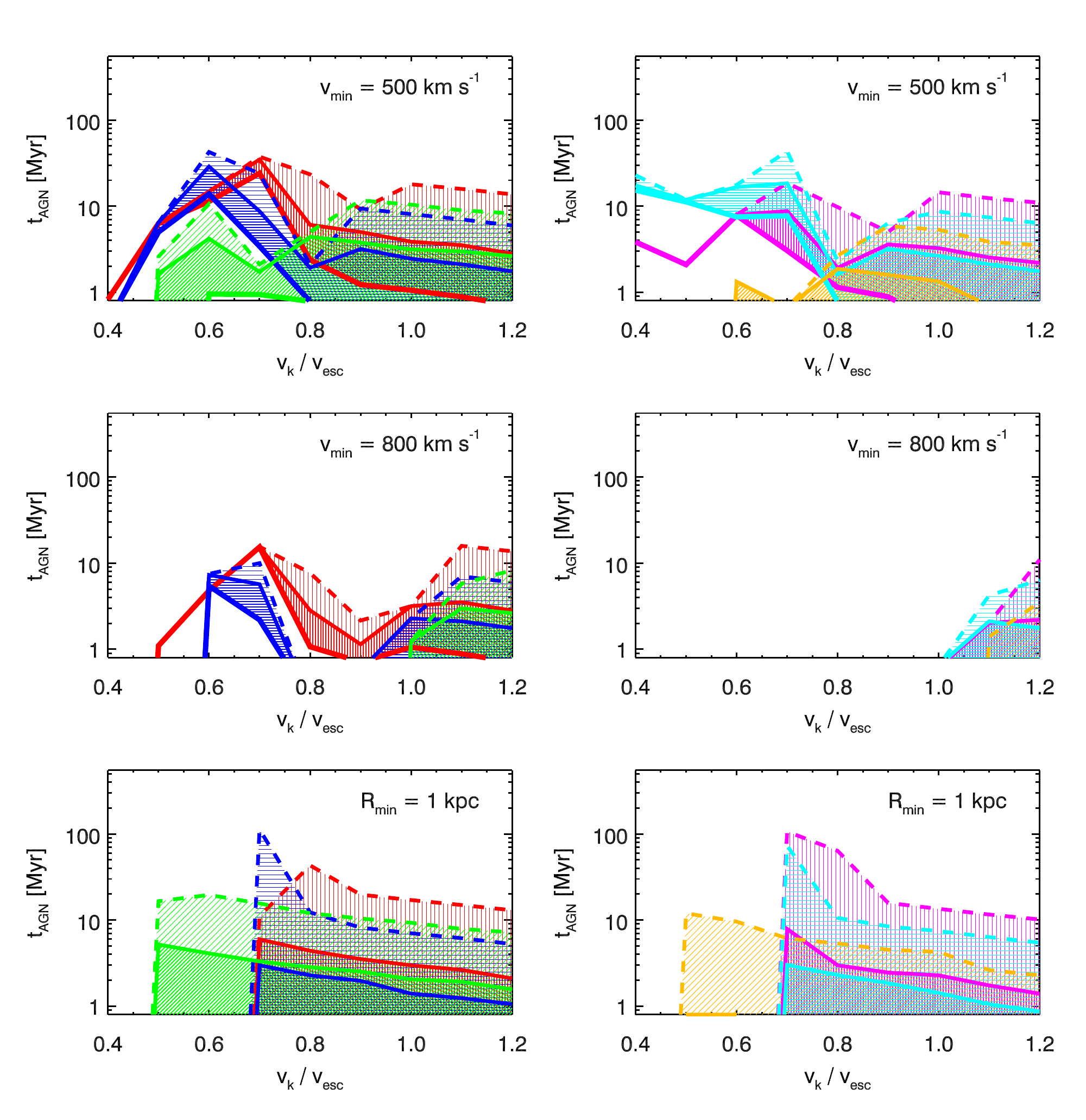}}
%\resizebox{\hsize}{!}{\includegraphics{offset_tagn.pdf}}
\caption[]{ The BH active lifetime is plotted in the same manner as in Fig.~\ref{fig:tagn_compare_ejdisk} and for the same galaxy models, but after minimum-velocity or minimum-separation cuts have been imposed.  In all cases, both the Bondi accretion and the ejected-disk model are included in calculating $L_{\rm bol}$.  The top two windows show the timescale for which the BHs are active and also have a velocity greater than 500~\kms~relative to the host galaxy's stellar center of mass.  The middle two windows show~\tagn~for a minimum velocity of 800~\kms.  The bottom two windows show~\tagn~for when the BH is spatially offset from the stellar center of mass by $> 1$ kpc.\label{fig:tagn_vmin}}
\end{figure*}

In all cases, the AGN lifetimes that include only Bondi accretion (Fig.~\ref{fig:tagn_compare_ejdisk}, top windows) fall sharply to $< 1$ Myr for large~\vk/\vesc.  This demonstrates that Bondi accretion is inefficient when the BH is kicked far from the dense central region.  However, when we also allow for BH accretion from its ejected disk (Fig.~\ref{fig:tagn_compare_ejdisk}, bottom windows), the ejected-disk accretion enhances the active lifetime for high~\vk/\vesc~recoils by more than an order of magnitude.  In the examples shown,~\tagn~for low-luminosity AGN is 3 - 14 Myr even for kicks as high as 1.2~\vesc.  For even higher kick speeds,~\tagn~will slowly decay as the trend in the plot indicates, owing to the decreasing ejected disk mass.  The relatively slow decay of~\tagn~with increasing~\vk~occurs partly because, as opposed to a constant-$\dot M$ accretion model, here~\tagn~depends on the fraction of disk accreted before the BH luminosity falls below the ``AGN" limit.  In addition, the disk mass in the self-gravitating regime has a fairly weak ($r^{1/2}$) dependence on radius, so $M_{\rm ej} \propto 1 / $\vk.  Recall, however, that the probability of recoil events with kick velocities $\gg 1000$~\kms~is quite small \citep{schbuo07, campan07a, baker08, lousto10a, vanmet10}; this more than the BH's fuel supply limits the chances of observing a recoil event with~\vk~$\gg$~\vesc.  

These findings illustrate the importance of BH accretion from the ejected disk for the lifetimes of rapidly-recoiling AGN.  Also crucial, however, is the time-dependent nature of this accretion.  If we were to assume a simple constant-$\dot M$ model for the ejected-disk accretion, we would greatly overpredict the lifetime of the bright AGN phase, and in some cases we would underpredict the lifetime of the low-luminosity phase.  We can calculate the discrepancies between~\tagn~for a constant-$\dot M$ model and the lifetimes shown in Fig.~\ref{fig:tagn_compare_ejdisk}, restricting the comparison to~\vk/\vesc~$\ge 0.9$ to capture only the simulations in which ejected-disk accretion is the dominant mode.  $\dot M_{\rm BH}$(\tmrg) $\approx 10\% - 50\% \dot M_{\rm Edd}$ for all the models shown in Fig.~\ref{fig:tagn_compare_ejdisk} except q0.5fg0.1a.  If we assumed the accretion continues at this rate for a time $M_{\rm ej} / \dot M_{\rm BH}$(\tmrg) in these models, we would overpredict~\tagn~for bright ($L>10\%L_{\rm Edd}$) AGN by a factor of $\sim 4 - 40$.  Conversely,~\tagn~for $L>L_{\rm min}$ would be {\em underpredicted} by a factor up to $\sim 7$.  In the q0.5fg0.1a model, $\dot M_{\rm BH}$(\tmrg) $\approx 5\% \dot M_{\rm Edd}$.  In this case,~\tagn~for $L>3\%L_{\rm Edd}$ would be overpredicted by a factor of $\sim 10 - 40$, depending on kick speed, and~\tagn~for $L>L_{\rm min}$ would {\em also} be overpredicted by a factor of $\sim 2-4$.  Therefore, it is clear that inclusion of more realistic, time-dependent model for ejected-disk accretion, such as the one presented here, is important for calculating the lifetimes and luminosities of rapidly recoiling AGN.  Similarly, for lower-velocity recoils, the BH + ejected-disk system clearly cannot be treated as isolated, because accretion from ambient gas is the dominant mode of AGN fueling.

To determine whether these recoiling AGN could be distinguished from their stationary counterparts, we need more information than simply the active lifetimes.  Recoiling AGN could be identified as such if during their active phase they are either kinematically or spatially offset from their host galaxies; we will consider both possibilities.  

The gas carried along with the recoiling BH is expected to include at least part of the BH's broad-line (BL) region, because BL clouds reside deep within the potential well of the BH.  The narrow-line (NL) region lies much further from the BH, so in general this gas will not be bound to the recoiling BH and will remain at the redshift of the host galaxy.  A kinematically offset AGN could therefore be seen in a spectrum as a broad-line (BL) feature offset from a narrow-line (NL) feature, if the recoiling AGN is able to illuminate NL clouds as it leaves the center of the galaxy.  Otherwise, such a system might have a spectrum with no NLs and BLs that are offset from the redshift of the host galaxy's stellar light.  Because the physics of the NL region are not included in our analysis, we consider the velocity offset between the recoiling BH and the the stellar center of mass.  We calculate the AGN lifetime as before, based on the BH luminosity at each timestep, but in this case we are only interested in the lifetime of {\em offset} AGN; that is, we calculate the time for which the BH is active {\em and} exceeds a minimum velocity offset $\Delta v_{\rm min}$.  We note that the limit $\Delta v_{\rm min}$ is imposed on the physical velocity difference between the BH and the host galaxy.  The line-of-sight velocity seen by a typical observer will be lower, so the velocity-limited AGN lifetimes shown here are upper limits.  

Fig.~\ref{fig:tagn_vmin} shows the results of our analysis for $\Delta v_{\rm min} = 500$~\kms~(top windows) and 800~\kms~(middle windows).  For  $\Delta v_{\rm min} = 500$~\kms, the offset AGN lifetime for low-\fgas, low-\vk~mergers is much smaller than the total AGN lifetime in Fig.~\ref{fig:tagn_compare_ejdisk}. ~\vesc~is relatively low in these models, so kick speeds are also lower for fixed~\vk/\vesc.  In the q0.5fg0.1a merger, for example,~\vesc$= 777$~\kms, so the BH never exceeds the 500~\kms limit for~\vk/\vesc~$\leq 0.6$.  The offset AGN lifetimes for low~\vk/\vesc~in gas-rich models are a factor of 5-10 lower than the total AGN lifetimes, but the peak lifetimes are still $\sim 10-50$ Myr.  At high~\vk/\vesc~($\ge 0.8$), there is little difference between~\tagn~with or without the velocity cut.  In these cases, the ejected disk dominates the BH accretion, and the BH speed is above 500~\kms~for nearly all of the AGN phase.  

When $\Delta v_{\rm min}$ is increased to 800~\kms, the AGN lifetimes drop sharply for most merger models and kick speeds (see the middle windows in Fig.~\ref{fig:tagn_vmin}).  Only in the equal-mass, gas-rich models shown here (q1fg0.6a \& q1fg0.3a) do the BHs appear as bright offset AGN, and for only a narrow range of kick speeds.  These two models have~\vesc~$> 1100$, such that BHs kicked with $\sim 800$~\kms~may have tightly-bound orbits.  In these cases the BHs make many rapid pericentric passages through the dense central region, where they encounter an ample gas supply from which to accrete.  Interestingly, in the q1fg0.6a and q1fg0.3a models, kicks $< 800$~\kms ($\leq 0.6$~\vesc) have nonzero lifetimes in this plot.  These are examples of the phenomenon discussed in \S~\ref{ssec:tmrg}, in which the BHs are kicked before~\vesc~reaches its maximum, so the BHs receive mild velocity boosts on subsequent passages through the increasingly deep central potential well.  

In the other (lower-$q$,~\fgas) models, the central region is less dense, such that bound BHs have lower kick speeds and lower accretion rates during pericentric passages.  In these models, only ejected-disk accretion is efficient at producing an offset AGN with $\Delta v > 800$~\kms, which occurs only when the BHs are ejected entirely from the host galaxy.  These ejected AGN have lifetimes $< 10$ Myr and low luminosities (\tagn~$< 1$ Myr for $L > 10\% \, L_{\rm Edd}$).   

To be observed as a spatially-offset recoiling AGN, a recoiling BH must travel far enough from the galactic center to be resolved as distinct point source before exhausting its fuel supply.  As with the velocity offsets, we define the spatial offsets relative to the stellar center of mass.  The maximum galactocentric distance achieved by a recoiling AGN in any of our fiducial-mass simulations is $\sim 15$ kpc by our lowest-luminosity AGN definition and only $\sim 1$ kpc by the strictest definition.  In our high-mass simulations, which have larger ejected disks, offsets $\ga 50 - 100$ kpc may be achieved while $L > L_{\rm min}$, but the maximum offset while $L > 3 - 10\% \, L_{\rm Edd}$ is still only a few kpc.  Offsets of $\sim 1$ kpc correspond to an angular separation of $\sim 0.55"$ at $z=0.1$; these could be resolved with {\em HST/JWST} as long as they are not observed edge-on.  Offsets $\ga \, 3 - 4$ kpc could be resolved at this redshift with {\em SDSS} and {\em Chandra}.  To determine the lifetimes of the spatially-offset AGN in our simulations, we impose a minimum-distance cut of $\Delta R_{\rm min}  = 1$ kpc from the stellar center-of-mass, and we identify an offset AGN as one that simultaneously has $\Delta R > \Delta R_{\rm min}$ and meets our AGN-luminosity criteria.  The bottom windows of Fig.~\ref{fig:tagn_vmin} show the results of this analysis.  We find that above~\vk/\vesc~$= 0.5 - 0.7$ (physically, the kick speed above which $R_{\rm max}$ exceeds 1 kpc), the offset AGN lifetimes can be long, but only for low-luminosity AGN.  In these models,~\tagn~ranges from $2 - 110$ Myr for $L > L_{\rm min}$, and $0.3 - 8$ Myr for $L > 3\% \,L_{\rm Edd}$.  In none of our simulations does the offset exceed $R_{\rm min}$ for more than 1 Myr while its luminosity is $> 10\% \, L_{\rm Edd}$.  The reason that spatially-offset AGN are only seen at low luminosities is as follows.  For recoiling BHs with short-period oscillations, Bondi accretion is dominant, and most of the bright AGN activity occurs during pericentric passages with small $\Delta R$.  For BHs on long-period orbits or escaping BHs, the ejected disk accretion dominates.  In this case the brightest AGN phase occurs as the BH leaves of the central dense region, and once the BH passes $\Delta R_{\rm min}$, its luminosity has already begun a monotonic decay.  

\begin{figure*}
\resizebox{0.49\hsize}{!}{\includegraphics{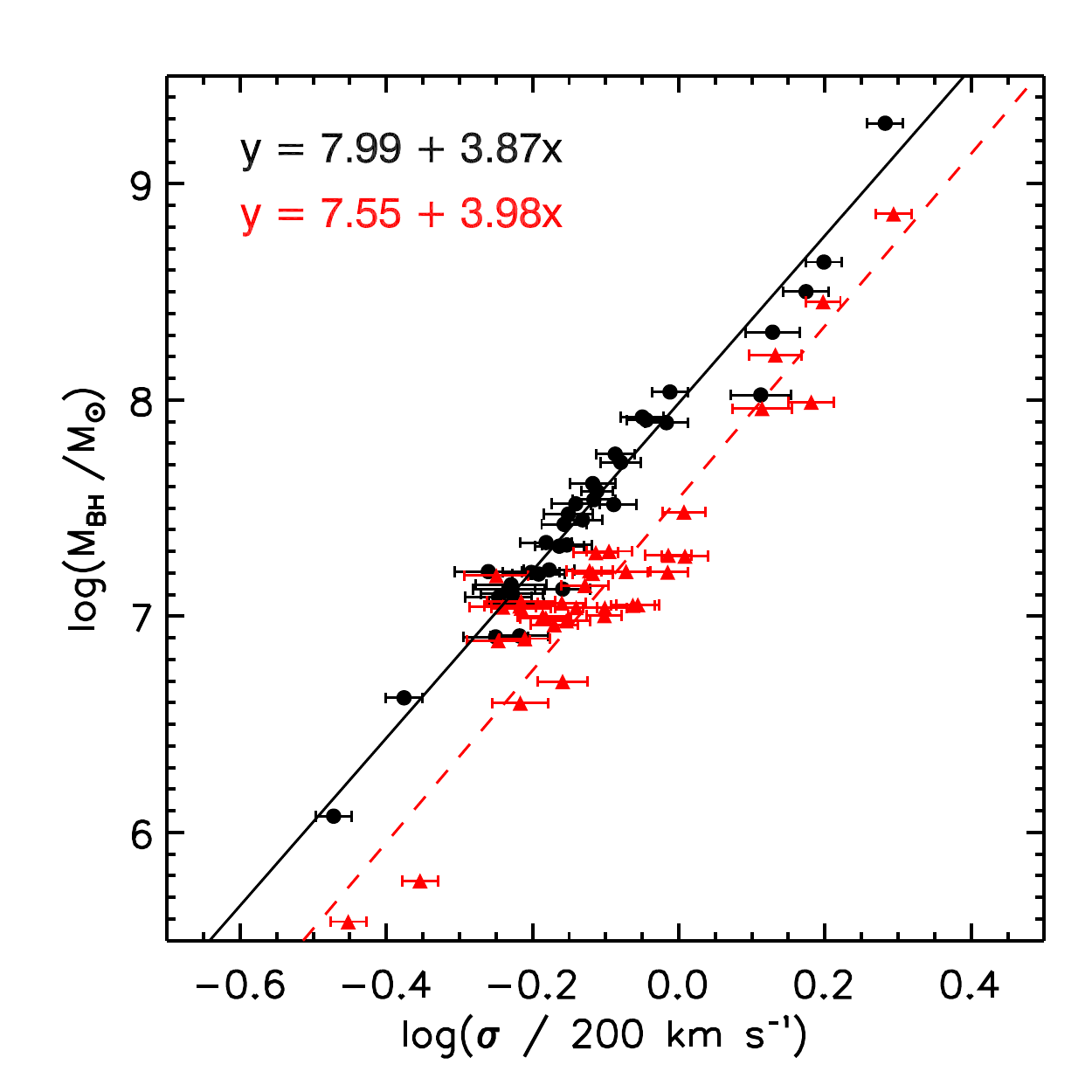}}
\resizebox{0.49\hsize}{!}{\includegraphics{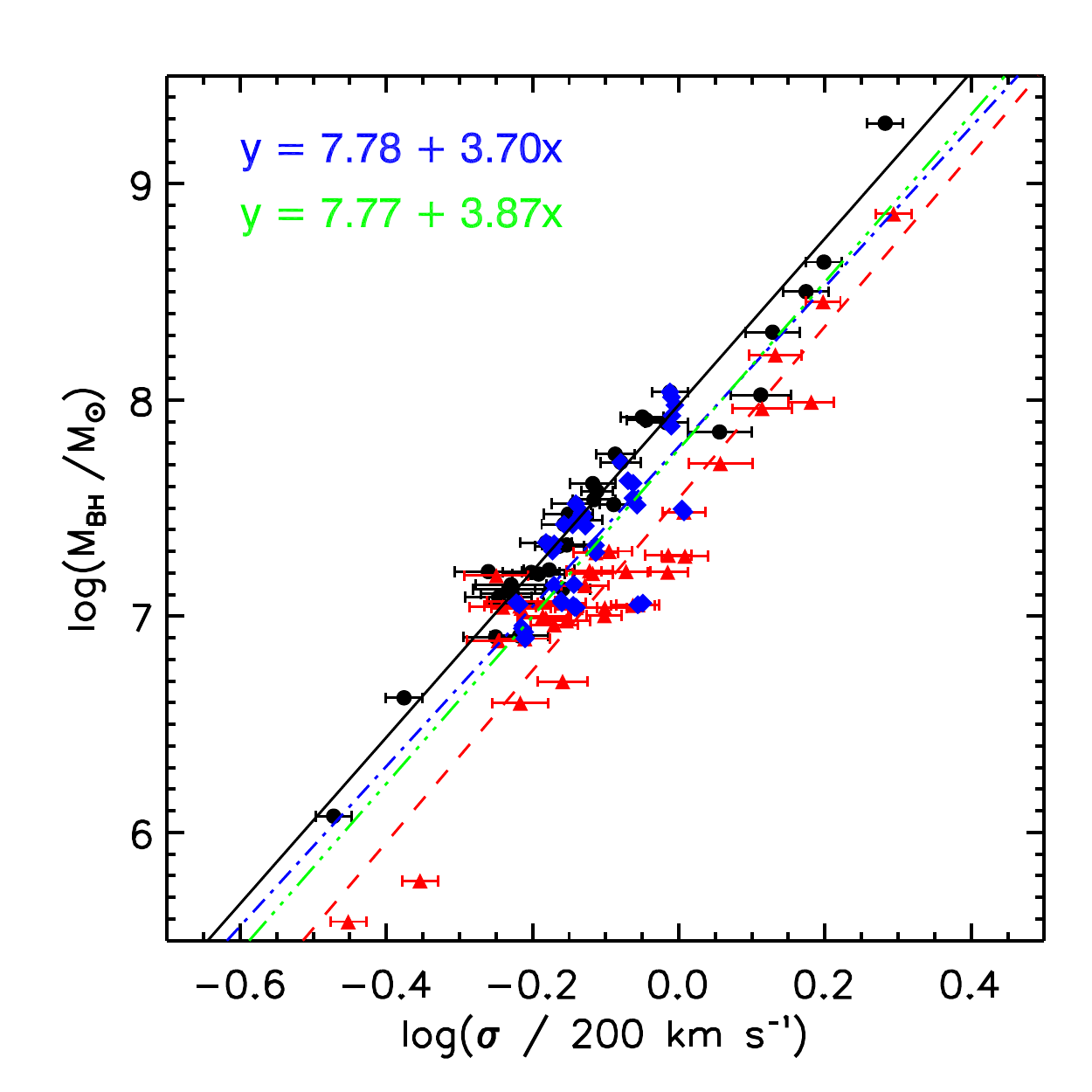}}
%\resizebox{0.49\hsize}{!}{\includegraphics{msigma_pubstyle_nkv90.pdf}}
%\resizebox{0.49\hsize}{!}{\includegraphics{msigma_pubstyle_all.pdf}}
\caption[]{Left panel: $M_{\rm BH}-\sigma_*$ relation for BHs with no recoil kicks (black circles) and with~\vk/\vesc~$=0.9$ (red triangles).  $M_{\rm BH}$ is in all cases the BH mass at the end of the simulation (\tmrg~$+\, 2.9$ Gyr) and $\sigma_*$ is the stellar velocity dispersion averaged over 100 random sight lines, where error bars give the range of sampled values.  All merger models with orbits a \& w - z are shown.  The solid black line is a least-square fit to the no-recoil data and the red dashed line is fit to the high-recoil data; the fit parameters are indicated on the plot.  Right panel: same data as left panel, but also including data for the six merger models in which we have varied~\vk/\vesc.  The blue diamonds show these results for~\vk/\vesc~$= 0.4 - 0.8$, and the blue dot-dashed line is a fit to these data.  The green triple-dot-dashed line is a fit to all data shown.\label{fig:msigma}}
\end{figure*}

We can conclude from Fig.~\ref{fig:tagn_vmin} that in most of our models, low-luminosity recoiling AGN may be distinguishable via spatial offsets $> 1$ kpc for kick speeds $\ga 0.5-0.7$~\vesc, but as kinematically-offset AGN with $\Delta v_{\rm min} = 800$~\kms, they may be distinguishable only for kick speeds $\ga$~\vesc.  These offset AGN generally have luminosities $\la 3\% L_{\rm Edd}$, and their luminosity owes mainly to accretion from the disk ejected with the BH.  The exceptions are equal-mass, gas-rich mergers (here, q1fg0.6a \& q1fg0.3a), in which recoiling BHs on bound trajectories may experience multiple phases as bright ($L > 10\% L_{\rm Edd}$), kinematically-offset AGN accreting from ambient gas during pericentric passages.  This scenario can only occur in merger remnants with large central escape speeds ($\ga 1100 - 1300$~\kms), such that BHs may receive kicks $\ga 800$~\kms and still undergo short-period oscillations.  In lower-$q$, lower-\fgas~mergers, the central gas density is lower, and the kick speeds for bound trajectories are smaller, so this bright, velocity-offset phase does not occur.  If kinematic offsets can be observed with a resolution of 500~\kms~instead of 800~\kms, velocity-offset-AGN lifetimes increase significantly, and measurable velocity offsets can be attained for a much wider range of kick speeds.

In the lowest $q$,~\fgas~merger for which we have varied~\vk~(the q0.5fg0.1a model), the offset AGN lifetimes are always $< 10$ Myr for $L > L_{\rm lim}$ and $< 2$ Myr for $L > 3\% L_{\rm Edd}$.  Our q0.5fg0.1a model has 10\% gas initially, but by the time of the recoil kick only $\sim 4\%$ of the baryonic mass is in gas.  We can conclude that merger remnants with gas fractions lower than this and with $q \la 0.5$ are unlikely to fuel bright offset AGN according to our criteria, at least for recoiling BHs with $M_{\rm BH} \sim 10^7$~\msun.  Larger BHs will have larger gas reservoirs and longer AGN lifetimes, and the reverse will be true of smaller BHs.   For example, in the q0.5fg0.1a model, a BH with~\vk/\vesc~$= 0.9$ has a velocity-offset ($\Delta v_{\rm min} = 800$~\kms) AGN lifetime $< 1$ Myr by all AGN criteria, while in the $10\times$ more massive q0.5fg0.1M10x model, the velocity-offset~\tagn~is 27 Myr (for $L > L_{\rm min}$).  In general, larger BHs will have longer offset-AGN lifetimes for fixed~\vk/\vesc.  However, larger BHs are found in larger galaxies, where a higher kick speed is required to reach an appreciable fraction of the escape speed, and such kicks have a lower probability of occurring.  We discuss recoil kick probabilities further in \S~\ref{ssec:traj_discuss}.

As a final caveat, we note that when Bondi accretion is dominant, the offset-AGN lifetimes are subject to the assumption that no significant delay occurs between when material becomes bound to the BH and when it is actually accreted.  The brightest AGN phases of this type occur during pericentric passages, when the velocity is high and the BH is subsequently traveling away from the galactic center.  Therefore, substantial delays in gas accretion would likely decrease lifetimes of velocity-offset AGN and increase lifetimes of spatially-offset AGN.

\subsection{M$_{\rm BH}$ - $\sigma_*$ Relation}
\label{ssec:msigma}

Several authors have suggested that GW recoil may contribute to scatter in the observed black hole - host galaxy bulge relations due to ejected BHs \citep{volont07} or bound recoiling BHs that leave the central galactic region \citep{bleloe08, sijack10}.  \citet{volont07} used semi-analytic models to estimate the effect of ejected BHs at high redshifts on the BH mass - bulge stellar velocity dispersion relation at $z=0$.  In this work, we have already demonstrated that, even when they remain bound to the host galaxy, recoiling BHs can undergo substantially different accretion histories than their stationary counterparts, and that the resulting AGN lightcurves may also vary significantly as a function of kick speed (see Figs.~\ref{fig:sfr_fgas_vks},~\ref{fig:tagn_compare_ejdisk},~\ref{fig:tagn_vmin}).  Here, we attempt to quantify the effect of individual GW recoil events on the M$_{\rm BH} - \sigma_*$ relation.  Because we have simulated all of our merger models with both with no recoil kick and with~\vk~$= 0.9$~\vesc, we are able to compare the M$_{\rm BH} - \sigma_*$ relations that result from each of these samples.  The final BH masses and LOS-averaged stellar velocity dispersions from both sets of simulations are plotted in Fig.~\ref{fig:msigma}; the right window also shows simulations with~\vk/\vesc~$0.4 - 0.8$.  We include only models with the fiducial merger orientation angles (orbits a \& w - z) in this plot to avoid heavily weighting the fit toward the ($q$,~\fgas) combinations for which we have varied the orbit.  

By examining the mass deficits of these recoiling BHs relative to the stationary BHs in our simulations, we can quantify the interruption to BH growth caused by GW recoil.  Note that we are probing only the effect of GW recoil on the growth of kicked BHs; the actual BH mass deficit resulting from BHs merging in a cosmological framework may differ from these values.  In particular, the~\vk~$= 0.9$~\vesc~simulations result in BH wandering times of at least several Gyr, so in reality these BHs will be lost to the galaxy and never contribute to the M$_{\rm BH} - \sigma_*$ relation.  If the galaxy experiences subsequent mergers, especially minor mergers, the kicked BH may be replaced with a smaller BH from the incoming galaxy, creating a larger mass deficit.  We also stress that the data plotted here are {\em not} meant to represent a random sample, as we have varied parameters in a systematic way.  Accordingly, the relations shown are not expected to reproduce the observed M$_{\rm BH} - \sigma_*$ relation.  In particular, the scatter in the~\vk~$ = 0$ relation (0.13 dex) is smaller than the observed scatter \citep[$\sim 0.25 - 0.3$ dex, e.g.,][]{tremai02,allric07,hopkin07b}, owing to a number of factors such as the single merger orbital configuration used in this plot, the constant feedback efficiency assumed for all simulations, and small range of total galaxy mass spanned by the majority of our models.  Furthermore, although scaling each kick to 0.9~\vesc~essentially allows us to capture the maximum effect of a single recoil event on BH growth, this does not reflect the fact that in galaxies with lower escape speeds, large~\vk/\vesc~recoil events will occur more frequently than in massive galaxies.  Such a trend with galaxy mass would steepen the slope of the M$_{\rm BH} - \sigma_*$ relation, though this would also depend on the distribution of other merger parameters, such as $q$ and~\fgas, with varying galaxy mass.  Despite these limitations in comparing our simulations to the observed M$_{\rm BH} - \sigma_*$ relation, our results provide substantial insight into how GW recoil affects the growth of kicked BHs and the inherent complexities of this process.  The relative differences in Fig.~\ref{fig:msigma} between the simulations with and without recoil are robust and can easily be understood in physical terms.  

It can be seen by visual inspection of Fig.~\ref{fig:msigma} that the effect of high-velocity recoils is an overall downward offset in normalization and an increase in scatter.  Because the LOS-averaged values of $\sigma_*$ do not change significantly in the presence of a recoil kick, we focus our discussion on variation in the final BH mass.

\begin{figure}
\resizebox{\hsize}{!}{\includegraphics{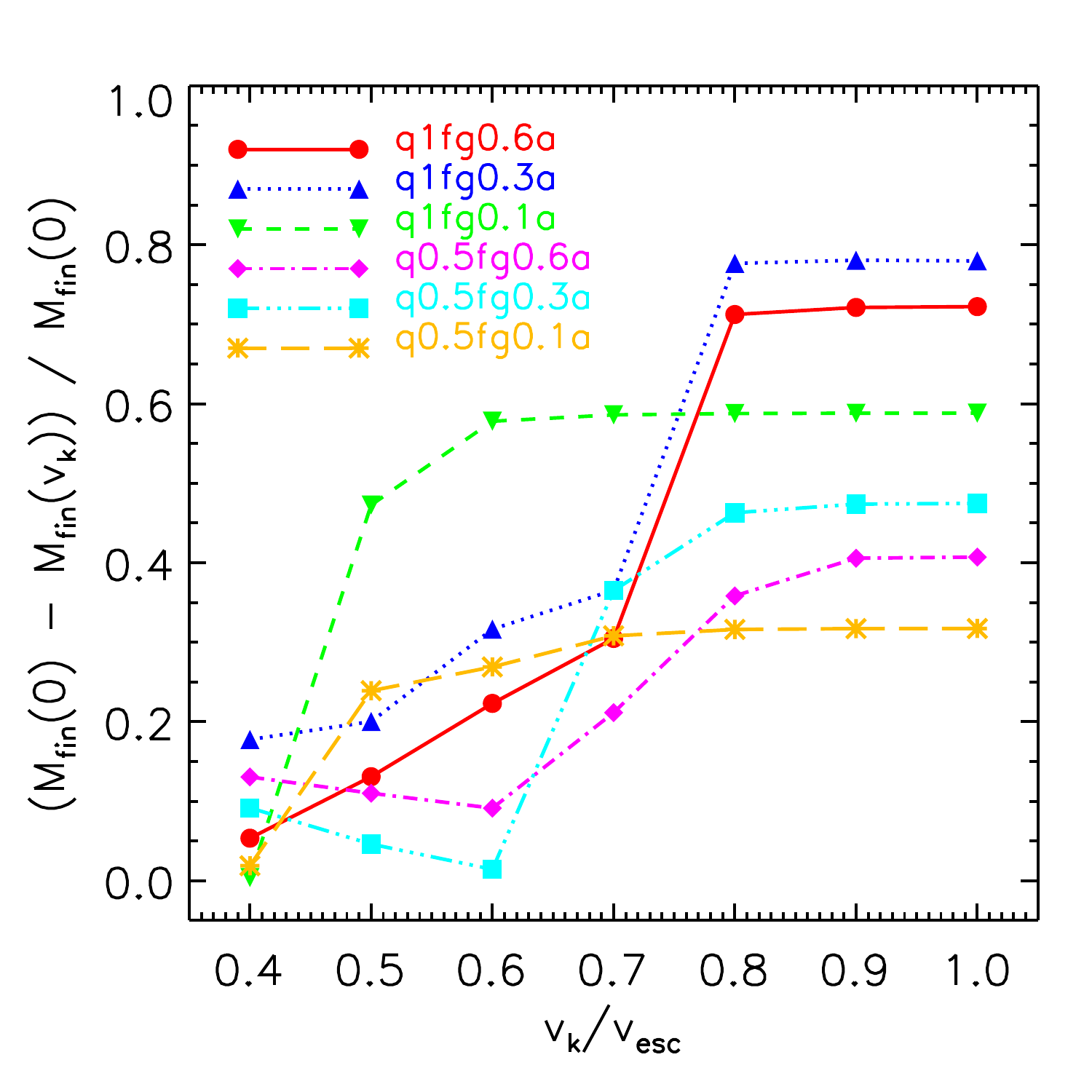}}
%\resizebox{\hsize}{!}{\includegraphics{all_mdef_vk.pdf}}
\caption[]{Fractional mass deficit of recoiling BHs versus stationary BHs as a function of~\vk/\vesc.  $M_{\rm fin}(0)$ and $M_{\rm fin}$(\vk) are the final BH masses (at~\tmrg~$+ \, 2.9$ Gyr) for stationary and recoiling BHs, respectively.  The plot legend lists the models shown. \label{fig:mdef_vk}}
\end{figure}

The downward shift in Fig.~\ref{fig:msigma} reflects the fact that GW recoil always reduces the final mass of a merged BH.  We quantify the ``mass deficit" of recoiling BHs relative to stationary BHs by calculating the fractional mass difference for each recoil simulation relative to its \vk~$= 0$ counterpart, ($M_{\rm fin}$(0) - $M_{\rm fin}$(\vk)) / $M_{\rm fin}$(0), where $M_{\rm fin}$ is the final BH mass in each case.  Fig.~\ref{fig:mdef_vk} shows the mass deficit plotted versus~\vk~for the six merger models discussed previously. The deficits generally increase with~\vk, though the details vary substantially between models.  In the q0.5fg0.1a model, recoiling BHs with~\vk~$=$~\vesc~have $\sim 1/3$ lower final mass than do stationary BHs, while in the q1fg0.3a model, a \vk~$=$~\vesc~recoil results in a BH almost five times smaller than its stationary counterpart.  Note also that the curves in Fig.~\ref{fig:mdef_vk} all level out at~\vk/\vesc~$= 0.7-0.9$; this marks the kick speed in each model above which recoiling BHs are unable to accrete more gas after their ejected disk has been exhausted.  The models with lower~\fgas~achieve this limiting value at lower~\vk/\vesc, owing to the lower gas drag and shallower potentials in these remnants.  The blue diamonds in Fig.~\ref{fig:msigma} mark the positions of these varying-\vk~simulations on the $M_{\rm BH}-\sigma_*$ relation. The points corresponding to lower~\vk/\vesc~lie close to the no-recoil (black) line; those with higher~\vk~have lower BH masses.  

In Fig.~\ref{fig:msigma}, the normalization offset between the~\vk~$=0$ BH population and the~\vk/\vesc~$=0.9$ population is 0.44 dex, and the offset between \vk~$=0$ and the total population is 0.22 dex.  However, it has been demonstrated that the normalization of the black hole - galaxy bulge correlations depends on the efficiency of AGN feedback \citep{dimatt05, robert06a, hopkin07a}.  In our simulations, we assume a constant fraction (5\%) of the BH's bolometric luminosity couples to the surrounding gas as thermal energy.  Stationary accreting BHs with lower feedback efficiencies would grow larger before heating the surrounding gas enough to slow or halt accretion.  Depending on how much accretion occurs prior to versus after the BH merger, higher feedback efficiencies could reduce the normalization offset between our recoiling and stationary BH populations.  In general, however, stationary BHs will accrete more after the BH merger than recoiling BHs, so there is likely to be a non-zero downward shift in $M_{\rm BH}-\sigma_*$ normalization caused by GW recoil.

As previously stated, the disproportionate effect of recoil events on galaxies with small~\vesc~could have some effect on the $M_{\rm BH}-\sigma_*$ slope.  In our simulations, where we have scaled kick speeds to~\vesc, there is no significant difference between the $M_{\rm BH}-\sigma_*$ slope for recoiling versus stationary BHs.  This is to be expected, because the depth of the central potential well scales more or less self-similarly with total galaxy mass.  We do, however, see a trend toward larger BH mass deficits for higher-$q$ mergers, which arises because gas is driven more efficiently to the galaxy centers during coalescence, and recoiling BHs ``miss" a larger accretion phase when they are kicked out of this central dense region.  

The variation in BH mass deficits between merger models results in larger intrinsic $M_{\rm BH}-\sigma_*$ scatter for BH populations that have undergone recoils.  The correlation for the no-recoil sample has a scatter of 0.13 dex, and the~\vk/\vesc~$= 0.9$ correlation has a scatter of 0.24 dex.  In other words, the high-recoil sample has almost twice as much intrinsic scatter as the no-recoil sample.  This increase in scatter is a universal consequence of GW recoil; recoil events essentially add an extra variable to the determination of final BH masses.  As opposed to stationary BHs, the final mass of recoiling BHs depends not just on how much gas is available to be accreted, but also {\em when} this accretion occurs relative to the time of the kick.  In mergers with no recoil, $M_{\rm BH}$ is a strong function of the amount of gas driven into the central region (i.e., the depth of the central potential well). The details and timing of the cold gas flow in galaxy mergers depend nontrivially on factors such as $q$,~\fgas, the galaxy orbits, and the star formation rate, so it is natural that larger intrinsic scatter is found in the final masses of recoiling BHs.  If the kick speed is lower, such that the BH can settle back to the center while an ample gas supply is still present, the BH mass deficit will be reduced, but the details of this late-phase accretion add yet another element of unpredictability to the final value of the BH mass.  

In addition to the larger intrinsic $M_{\rm BH} - \sigma_*$ scatter for a BH population with fixed~\vk/\vesc, we also anticipate an increase in overall scatter caused by the range of kick speeds that would occur in a realistic population of BHs and by BHs that have undergone multiple recoils through their formation history.  The contribution of GW recoil to the total $M_{\rm BH} - \sigma_*$ scatter is difficult to predict from our simulations; quantitative predictions would require following detailed BH formation histories in a cosmological framework, which is beyond the scope of this paper.  However, we at least gain a sense of how a range of kick speeds affects the $M_{\rm BH}-\sigma_*$ relation by plotting our varied-\vk~simulations along with the~\vk~$=0$ and~\vk/\vesc~$=0.9$ populations.  The right panel in Fig.~\ref{fig:msigma} shows the same data as the left panel but also includes data for the simulations in which we have varied~\vk/\vesc~(blue diamonds).  The blue line is a fit to these points, and the green line is a fit to all data shown.  The fit to all the data has a scatter of 0.25 dex, which as expected is larger than the scatter for the~\vk~$= 0$ population.  

The most robust of our conclusions from this analysis is the increase in intrinsic $M_{\rm BH}-\sigma_*$ scatter for BHs that have undergone a single recoil event, by a factor of $\sim 2$ in our simulations.  This finding suggests that GW recoil events may be a non-negligible contribution to the scatter in the observed $M_{\rm BH}-\sigma_*$ relation, especially because BHs at $z=0$ may have undergone multiple recoil events throughout their formation history.

\subsection{Evolution of Central Galactic Region}
\label{ssec:centralevol}

\begin{figure}
\resizebox{\hsize}{!}{\includegraphics{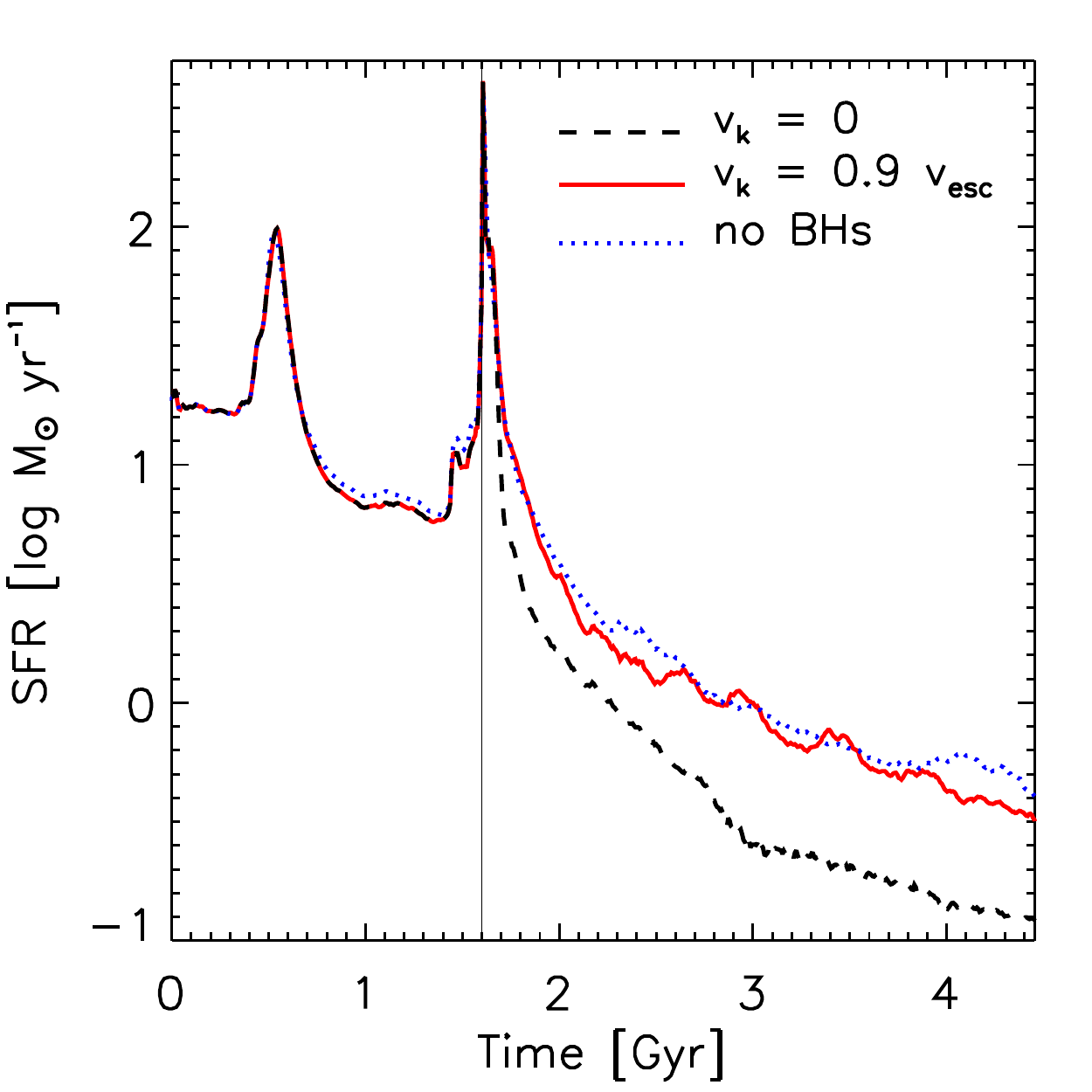}}
%\resizebox{\hsize}{!}{\includegraphics{q1_fg40_compare_v0v90_sfr.pdf}}
\caption[]{Total star formation rate (SFR) vs. time for the q1fg0.4a merger model.  The black dashed line represents the~\vk~$= 0$ simulation, the red solid line represents the~\vk/\vesc~$= 0.9$ simulation, and the blue dotted line represents a simulation with no BHs. The vertical line marks the time of BH merger in the simulations with BHs.\label{fig:comparesfr}}
\end{figure}

Fig.~\ref{fig:comparesfr} compares the SFRs for simulations with no GW recoil (black dashed curve), with~\vk/\vesc~$= 0.9$ (red solid curve), and with no BHs (blue dotted curve) in a gas-rich, equal-mass merger model (q1fg0.4a).  It is clear that after the merger, the simulation with a large recoil kick has a higher SFR than the simulation with~\vk~$=0$.  The post-merger SFRs are similar in the high-recoil case and in the simulation with no BH at all.  This same behavior can be seen in the high-recoil simulations of model q0.5fg0.3a shown in Fig.~\ref{fig:sfr_fgas_vks}, though to a lesser degree.  These enhanced SFRs occur because when the BH is removed from the center of the merger remnant, the AGN feedback is also displaced.  The lack of central energy input allows star formation to continue unhindered in this region for a longer time.  We can actually see the effect of the recoiling AGN feedback on the instantaneous SFR in Fig.~\ref{fig:comparesfr}.  The SFR curve for the~\vk/\vesc~$=0.9$ simulation has a vaguely periodic shape; in fact, each small trough in the SFR coincides with a pericentric passage of the BH through the galactic center, which imparts a small amount of feedback energy to the dense region.  This supports the connection between the recoiling BH dynamics and the galactic SFR, but the main consequence of the recoil event is that star formation is never as strongly quenched as when the BH does not recoil.  In the example shown in Fig.~\ref{fig:comparesfr}, the merger remnant with a stationary BH has almost twice the amount of gas at the end of the simulation as the merger with~\vk/\vesc~$=0.9$.  This corresponds to an increase in total star formation of $3.4\times10^9 M_{\odot}$ in the recoil simulation, or $\sim 3\%$ increase in total stellar mass, solely due to the displacement of the BH and its feedback.  

\begin{figure}
\resizebox{\hsize}{!}{\includegraphics{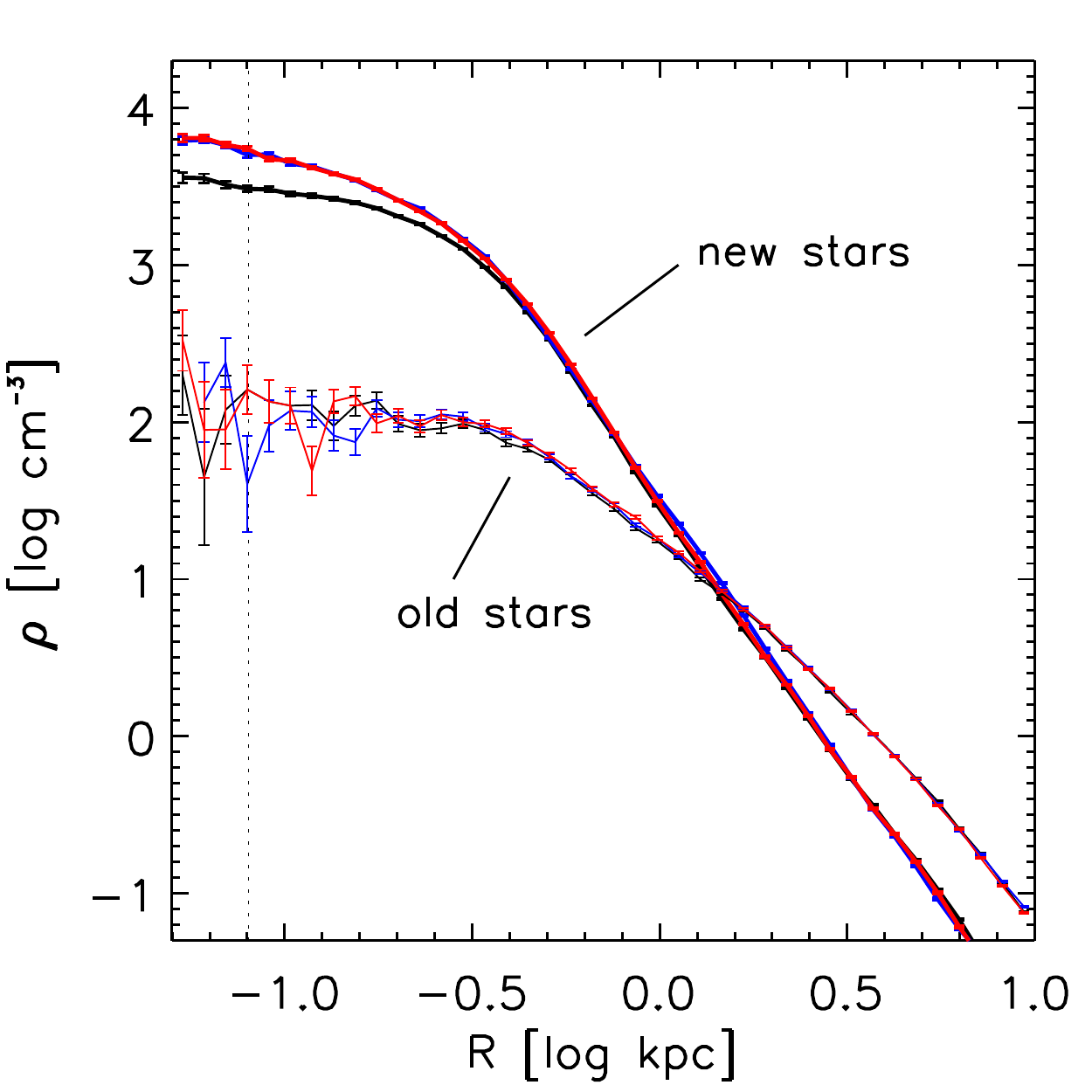}}
%\resizebox{\hsize}{!}{\includegraphics{compare_starsprof_446.pdf}}
\caption[]{Final stellar density profiles (2.9 Gyr after~\tmrg) for the same three simulations shown in Fig.~\ref{fig:comparesfr}, with the same color coding.  The thick curves are the ``new" stellar populations that formed during the merger and that dominate within the central $\sim$ kpc, and the thin curves are the ``old" stellar populations from the initial stellar disks of the progenitor galaxies.  In each case, the black curves are the profiles for the~\vk~$= 0$ simulation, the red curves represent the~\vk/\vesc~$= 0.9$ simulation, and the blue curves represent the simulation with no BHs.  Poisson error bars are shown on the densities in each radius bin.  The vertical dotted line denotes the gravitational softening length (80 pc) in these simulations. \label{fig:compare_profiles}}
\end{figure}

In Fig.~\ref{fig:compare_profiles}, we can clearly see the effect of this enhanced star formation on the remnant stellar density profile.   The figure shows the remnant stellar density profiles for the same three simulations shown in Fig.~\ref{fig:comparesfr}, with \vk~$ = 0$ (black curves),~\vk/\vesc~$= 0.9$ (red curves), and without BHs (blue curves).  The results are further broken down into the contribution from stars formed prior to the galaxy merger (``old'' stars, thin curves) and those formed during merger-driven starbursts (``new'' stars, thick curves).  In all cases, the ``new" stellar population dominates in the central kpc, but in the remnant with a recoiling BH the central stellar density is almost twice that of the remnant with a stationary BH.  In fact, this stellar profile very closely resembles that of a remnant without a BH.  Therefore, recoil events in gas-rich mergers can prolong starburst phases, creating denser and more massive stellar cusps.  We have also examined the (unattenuated) $U - B$ color evolution of the stellar component in these no-recoil, large-recoil, and no-BH simulations.  The colors in all three cases are similar; by the end of the simulation all have $U - B \sim 0$, i.e., these merger remnants are entering the ``green valley" in transition from blue to red.  However, in the~\vk/\vesc~$=0.9$ and no-BH simulations, which have prolonged star formation, the stars are slightly bluer than in the no-recoil simulation.  While this difference is small, $\sim 0.03$ mag, it further demonstrates that in merger remnants with rapidly-recoiling BHs, the (blue) starburst phase may be prolonged, and hence the transition to red elliptical slightly delayed, solely due to the effect of GW recoil.  In \S~\ref{ssec:ulirg_discuss}, we discuss the implications of these findings for observable properties of ultraluminous infrared galaxies (ULIRGs) and elliptical galaxies.

We note that this is the {\em opposite} effect of one that has been proposed for gas-poor mergers, where repeated BH oscillations through the central region of the merger remnant may scour out a stellar core \citep{boylan04,guamer08}.  Such core-scouring is expected to occur on scales of $\la$ 10s of pc, which are below our resolution limit.  However, we do comment that when even a small amount of gas is present in our merger simulations, it is efficiently driven to the center of the merger remnant.

\section{Discussion}
\label{sec:disc}

Using the SPH/N-body code {\footnotesize GADGET-3}, we have generated a suite of galaxy merger simulations both with and without a recoil kick applied to the central SMBH at the time of BH merger.   Owing to the large range of parameters sampled, we were able to analyze systematic trends in recoiling BH behavior with variation in galaxy mass ratio, total galaxy mass, initial gas fraction, orbital configuration, and BH merger time.  In addition, we have used a range of recoil kick velocities, which are scaled to the galactic central escape speed in each case.  We followed the trajectories and accretion history of the recoiling BHs, as well as evolution of host galaxy properties such as the star formation rate and depth of the central potential well.  For the BH accretion, the Bondi-Hoyle accretion rates were used, but we also included a time-dependent model for accretion from a gas disk carried along with the recoiling BH.  

\subsection{Recoil Trajectories \& Kick Probabilities}
\label{ssec:traj_discuss}

The recoiling BH trajectories in our simulations are characterized by low-angular-momentum orbits (i.e., ``oscillations" about the galactic center) that are also highly three-dimensional.  In other words, the baryonic component of the merger remnant dominates the BH orbits even if they extend well into the halo, but the unrelaxed nature of this remnant creates complicated BH trajectories.  We find that the amplitude and duration of these oscillations varies widely between different galaxy models and kick velocities; some BHs settle back to equilibrium at the galactic center within a few Myr, while others may remain on large orbits through the halo for a Hubble time.

The amplitude and duration of BH oscillations in a given recoil event clearly depend on the central escape speed, i.e., the depth of the central potential well.  However, by normalizing kick speeds in our simulations to~\vesc~at the time of the kick, we are able to compare different galaxy models regardless of their escape speed.  We find substantial variation in recoil trajectories between models even for fixed~\vk/\vesc.  Recoil oscillations are more readily suppressed in mergers with higher $q$ and~\fgas.  Such merger remnants are more centrally concentrated and gas-rich, resulting in steeper central potential wells and increased gas drag and dynamical friction.

In the models for which we have varied~\vk/\vesc, we find that, with the exception of our collisionless simulations, recoiling BHs with~\vk/\vesc~$= 0.4 - 0.7$ reach galactocentric distances $\la 1$ kpc and settle back to the center in $< 0.5$ Gyr.  The corresponding range of kick speeds is 310-880~\kms.

For large recoil velocities, GW recoil events can produce BH oscillations that persist through the end of our simulations, 2.9 Gyr after the kick.  The largest orbits can extend well into the galactic halo, with galactocentric distances up to $\sim 100$ kpc.  In our models these long-lived oscillations occur for kicks $\ga 0.6 - 0.9$~\vesc, again depending on the mass ratio and gas content of the merger. 

Furthermore, as illustrated in \S~\ref{ssec:tmrg}, in gas-rich mergers with $q \sim 1$, the central escape speed may increase rapidly around the time of BH binary coalescence.  In this case, the trajectory of the recoiling BH may also depend on the {\em time} of the BH merger relative to the formation time of the central density cusp.  Of all the factors influencing the BH trajectories, this is the most unpredictable, as it depends on the relative timing of two short-timescale processes.  This adds an element of uncertainty to any attempt, including the current study, to model populations of recoiling BHs in gas-rich galaxies.  Our assumption of rapid BH mergers (\tmrg~$=$~\tcoal) means that our BH oscillation amplitudes may be upper limits in mergers with strong~\vesc~evolution.  However, the effect of this uncertainty on the results presented here is limited, due to the relatively small number of merger models affected and our use of the normalization~\vk/\vesc. 

We have also tested whether recoil trajectories are affected by the direction of the kick.  Our generic merger remnants (those that are not formed from highly-aligned galaxy orbits) are highly disturbed at the time of the BH recoil and therefore do not have disk-like structures, i.e., there is no ``special" direction in which a recoiling BH would clearly experience much greater drag.  Even with the clumpy, irregular density profiles of our merger remnants, we find that the trajectories of recoiling BHs have little dependence on the kick direction.  Much more important in determining the amplitude and duration of recoiling BH oscillations is the {\em magnitude} of the kick velocity relative to the escape speed, along with the aforementioned global galaxy properties such as the initial gas fraction and mass ratio.  We note that there should indeed be a dependence of the recoil trajectory on the kick direction with respect to the plane of the accretion disk; recoils directed into the disk plane likely experience more gas drag and have shorter oscillation timescales.  In addition, \citep{rossi10} have shown that the ejected disk mass and AGN luminosity may depend strongly on the direction of the recoil kick with respect to the accretion disk plane.  We do not resolve this small-scale region in our simulations; however, recoil kicks $\ga 500$~\kms will {\em always} be nearly perpendicular to the orbital plane of the binary \citep{campan07a, campan07b, louzlo09}, which is unlikely to be perpendicular to the orbital plane of the accretion disk.  Therefore, for the recoil velocities we consider ($\ga 300$~\kms), we can assume that the BH is unlikely to be kicked directly into the accretion disk plane.  

We mention in \S~\ref{sec:simulations} that according to the results of NR simulations, the recoil kick speeds in our fiducial-mass merger models, which we have scaled to the escape speeds of $\sim 700 - 1250$~\kms, should occur with fairly high probabilities.  Even if the BH spin magnitudes are randomly distributed between 0 and 1, $41.6\%$ of major mergers should have kicks above 500~\kms, and $13.5\%$ should be above 1000~\kms.  Furthermore, \citet{lousto10a} have conducted statistical analysis of GW recoil spins and kick speeds using post-Newtonian approximations calibrated to numerical relativity results.  They provide kick probability distributions for several mass ratio bins, which we can use to estimate the relative probability of ``high" (\vk~$\sim$~\vesc) or ``moderate" (\vk~$\sim 0.6$~\vesc) kick speeds in various merger models.  Coincidentally, for mass ratios $0.3 \leq q \leq 1.0$ and $300 \la$~\vk~$\la 1000$~\kms, which represent a large fraction of our fiducial-mass simulations, the kick probability distribution is quite flat.  For these models, the relative probability of a given kick speed between models, as well as the relative probability of high versus moderate~\vk~in a given model, is roughly unity.  In our highest-mass (q1fg0.3M20x) simulation, the probability of a kick near~\vesc~$= 2500$~\kms~is about 10 times smaller than the probability of a kick with 0.6~\vesc~and is also $\sim 10$ times smaller than the kick probability for~\vk~$\sim$~\vesc~in the q1fg0.3a model.  For mergers with $q \la 0.25$, which we do not model in our study, the kick distribution becomes heavily skewed toward velocities $< 500$~\kms.

An important caveat to the interpretation of recoil kick probabilities is that the relevant velocity for observations is the line-of-sight (LOS) velocity, not the actual velocity.  For random lines of sight, the probability of observing a given kick~\vk~as a kinematic offset $> \Delta v_{\rm min}$ is therefore reduced by a factor $P(\Delta v_{\rm LOS} > \Delta v_{\rm min}) = 1 - \Delta v_{\rm min}/$\vk.  For example, about 20\% of recoils with~\vk~$= 1000$~\kms~will have $\Delta v_{\rm LOS} > 800$~\kms.  \citet{lousto10a} find that the overall probability distribution for $\Delta v_{\rm LOS}$ is lower than the~\vk~distribution by a factor of between $\sim 2$ (for kicks $> 500$~\kms) and $\sim 20$ (for kicks $> 2500$~\kms).  Similarly, the probability of observing a recoiling BH with spatial offset $\Delta R_{\rm LOS} > \Delta R_{\rm min}$ is $\sqrt{ 1 - \Delta R_{\rm min}^2 / \Delta R^2 }$; i.e., a majority of spatial offsets $\ga 1.2 \, R_{\rm min}$ will have LOS offsets exceeding the minimum observable offset.

\subsection{Recoiling AGN Lightcurves}
\label{ssec:agn_discuss}

The final coalescence of two merging, gas-rich galaxies typically triggers both a burst of star formation and a phase of rapid BH accretion.  Assuming the BHs themselves coalesce on a short timescale, the resulting recoil event coincides with a period of high activity in the central region of the merger remnant.  Thus, we expect recoiling BHs in gas-rich mergers to have substantially different accretion histories than their stationary counterparts.  This is indeed the behavior seen in our simulations, with greater disruption to the BH accretion history for higher kick velocities.

Recoil events that produce short-period BH oscillations (\vk/\vesc~$\sim 0.4-0.8$, depending on the merger model) cause the BH accretion rate to drop at the time of the kick, thereby reducing the peak luminosity of the BH.  However, the BH remains in the central few kpc of the merger remnant where gas densities are large enough that Bondi-Hoyle accretion is efficient.  Because the BH's feedback energy is deposited in a much larger volume than for stationary BHs, it is unable to heat the gas sufficiently to completely terminate the AGN phase.  The recoiling BH may therefore experience a longer active phase by a factor of a few than its stationary counterpart, although at lower luminosities.  This phase continues in some cases until the end of the simulation, 2.9 Gyr after the BH merger.  Owing to the lower luminosities of these extended AGN phases, we find that they do not increase the final BH mass relative to stationary AGN.  However, this type of accretion episode is distinct from other merger-triggered AGN phases in that it is not accompanied by a simultaneous starburst.  The copious amounts of dust produced by starbursts can easily obscure an active central source, and the starburst luminosity may overwhelm the AGN luminosity.  Recoiling AGN that continue accreting long after the starburst is complete therefore may be more easily observable.  

When recoil kicks are large enough to produce long-period ($\ga 100$ Myr) BH oscillations (\vk/\vesc~$\ga 0.6-0.9$ depending on the merger model), the Bondi-Hoyle accretion rate drops precipitously as the BH is rapidly ejected to a lower-density region.  Even though the BH remains on a low-angular-momentum orbit and returns several times to the dense central region, its velocity is generally too high during these pericentric passages to allow for substantial gas accretion, and the gas is depleted by star formation between subsequent passages.  

In such high-velocity recoil events, a portion of the inner accretion disk may remain bound to the BH.  We account for this by assigning to each recoiling BH a time-dependent, isolated accretion disk with a mass and initial accretion rate based on the BH mass, recoil velocity and the accretion rate at the time of the kick.  The mass of these ejected disks is only a few percent of the BH mass, so they do not contribute significantly to the final BH mass.  However, the accretion from these disks does greatly extend the active lifetime of rapidly-recoiling BHs in the regime where Bondi-Hoyle accretion is inefficient.  We find that the AGN lifetime in our models may increase from $\ll 1$ Myr to $\sim 20$ Myr when ejected-disk accretion is included, though these are generally low-luminosity AGN.  The peak lifetimes in the Bondi-Hoyle regime are generally longer and correspond to brighter AGN, but for a fairly narrow range of kick speeds. ~\tagn~in the Bondi-Hoyle regime may be up to $\sim 300$ Myr for low-luminosity AGN and between $\sim 10 - 100$ Myr for AGN accreting at $3 -10\% L_{\rm Edd}$.

\subsection{Offset AGN Lifetimes}
\label{ssec:offsetagn_discuss}

From an observational standpoint, our primary interest is not the total active lifetime, but rather the time for which the BH is active {\em and} can be distinguished as a recoiling BH.  We find that kinematically-offset AGN -- those that are luminous enough to be detected {\em and} have velocity offsets large enough to be spectrally resolved -- occur in two distinct physical scenarios.  In the first scenario, the recoiling AGN travel far from the galactic center and are powered by accretion from the ejected disk.  This occurs for recoil velocities near or above the central escape speed.  In our simulations, AGN with velocity offsets $> 800$~\kms~have lifetimes $\la 10$ Myr in most cases.  In most of our galaxy models, recoiling AGN can only meet this velocity criterion for~\vk~$>$~\vesc.  For reference, the SDSS quasar study conducted by~\citet{bonshi07}, which yielded a null result, had a spectral offset limit of 800~\kms. 

The other physical mechanism for producing kinematically-offset recoiling AGN is via multiple pericentric passages through a central dense region.  Note that marginally-bound recoiling BHs may also experience close passages, but these will be rapid and few in number; the cumulative probability of observing such an event is low.  If a recoiling BH is able to settle back to the galaxy center within $\sim 0.5-1$ Gyr, it will undergo many short-period oscillations through the central region.  In this case, the BH may be able to accrete efficiently from the ambient gas during high-velocity close passages, producing a kinematically-offset AGN.  This scenario favors massive, gas-rich galaxies with high central densities: the larger the central escape speed and the steeper the potential well, the higher the velocity with which a BH can be kicked and still remain tightly bound to the galaxy.  In our $q=1$, gas-rich mergers, offset AGN with $\Delta v > 800$~\kms can have lifetimes up to 100 Myr, and may have lifetimes $\ga 10$ Myr at bright ($> 10\%\, L_{\rm Edd}$) luminosities.  Especially for these AGN, an increase of spectral resolution to achieve a limit of 500~\kms~could make a substantial difference; in our models, the lifetimes increase by a factor of $\sim 5 - 10$ with this lower value of $\Delta v_{\rm min}$.  

The maximum spatial offset achieved by a recoiling AGN in our high-mass simulations is $\Delta R = 112$ kpc by our most generous AGN definition ($L > 1.3\times10^{43}$ erg s$^{-1}$) and 3.2 kpc by our strictest definition ($> 10\% L_{\rm Edd}$), but for our fiducial-mass simulations these maximum offsets are only 15 and 1.3 kpc, respectively.  To calculate spatially-offset AGN lifetimes, we require a minimum offset $\Delta R_{\rm min} = 1$ kpc.  Such an offset could be resolved by {\em HST} or {\em JWST} at $z \la 0.1$, or by {\em SDSS} or {\em Chandra} at $z \la 0.03$.  In our simulations, recoiling BHs achieve $R_{\rm max} > 1$ kpc for~\vk/\vesc~$\ga 0.5-0.7$.  AGN lifetimes may be quite long near this threshold kick speed ($\sim 100$ Myr), though only at low luminosities.  At higher kick speeds, lifetimes are generally $\sim 1-10$ Myr and slowly decrease with higher~\vk, as less gas bound to the recoiling BHs.

Based on our calculated AGN lifetimes, we conclude that for escaping BHs (\vk~$>$~\vesc), probabilities for observing recoils via spatial offsets and kinematic offsets are similar.  For recoiling BHs on bound trajectories, velocity-offset AGN with high luminosities are favored for massive, gas-rich galaxies ($q=1$,~\fgas~$=0.3-0.6$ in our models).  Spatially-offset AGN with lower luminosities have similar lifetimes for all models.  Recoiling BHs in merger remnants that are both gas poor (\fgas~$\la 0.1$ initially) and result from lower-$q$ mergers ($q \,\la \,1/3$) are not expected to produce long-lived offset AGN.

\subsection{$M_{\rm BH}-\sigma_*$ Scatter \& Offset}
\label{ssec:msigma_discuss}

We have demonstrated that recoiling BHs have lower final masses than their stationary counterparts, though the mass deficit varies considerably with kick speed and merger remnant properties.  This may affect the formation of the observed correlations between SMBHs and the bulges of their host galaxies.  In particular, we consider the possible effects of these mass deficits on the observed $M_{\rm BH}-\sigma_*$ relation by comparing the $M_{\rm BH}-\sigma_*$ correlations of our sets of~\vk~$=0$ and~\vk~$=0.9$~\vesc~simulations, as well as our simulations with varying~\vk.  We do not find any evidence for a change in $M_{\rm BH} - \sigma_*$ slope caused by recoils with fixed~\vk/\vesc, but in a realistic population of galaxy mergers, galaxies with smaller~\vesc~would be disproportionately affected by recoils with~\vk~near or above~\vesc.  This could contribute to some steepening of the $M_{\rm BH} - \sigma_*$ slope.  In our simulations, we find that GW recoil events may contribute to a downward shift in normalization, an increase in overall scatter, and an increase in intrinsic scatter of the $M_{\rm BH}-\sigma_*$ relation (see Fig.~\ref{fig:msigma}).  The magnitude of these effects, particularly the normalization offset and total scatter, are sensitive to the galaxy population considered and the detailed merger and recoil histories of the central BHs.  Therefore, we cannot make quantitative predictions about the contribution of GW recoil to the observed normalization and scatter of the $M_{\rm BH} - \sigma_*$ relation; this would in fact be difficult to predict with any method.  A more realistic approach involving BH and galaxy populations derived from a cosmological framework would still be plagued with uncertainties in the recoiling BH masses.  We have shown that these masses are sensitive numerous factors that do not lend themselves to semi-analytical modeling, including the (possibly evolving) central potential depth, the BH merger timescale, and the detailed BH accretion history. 

Despite these uncertainties, the existence of a recoiling-BH contribution to the $M_{\rm BH} - \sigma_*$ correlation and its scatter is a necessary consequence of GW recoil events.  In particular, we demonstrate that a universal effect of GW recoil is to increase the intrinsic scatter in BH masses.  The disruption to BH accretion caused by a recoil event introduces an extra variable into the determination of the final BH mass, namely, the timing of the recoil kick relative to the merger-induced burst of BH accretion (i.e. quasar phase).  In our recoil simulations, the intrinsic $M_{\rm BH} - \sigma_*$ scatter of our rapidly-recoiling BH population (\vk/\vesc~$= 0.9$) is almost $2 \times$ the scatter in our stationary-BH population.  This is the contribution to scatter caused solely by interruption of BH growth by a recoil kick; other factors not accounted for here could easily increase the contribution of GW recoil to $M_{\rm BH} - \sigma_*$ scatter.  While a more realistic population of BHs will have smaller average mass deficits resulting from a range of kick speeds, the BHs are also likely to have undergone numerous mergers and recoil events in the past, which would increase the overall scatter.  Subsequent merger events, in which an incoming BH may replace an ejected BH, could also increase scatter.  The relatively small scatter of the empirical relation, 0.25 - 0.3 dex, places constraints on the contribution of GW recoil and, hence, on the frequency of large~\vk/\vesc~recoil events.  Similarly, a non-zero contribution of GW recoil to the observed scatter places constraints on the contribution from other sources, such as redshift evolution of the $M_{\rm BH} - \sigma_*$ relation \citep[e.g.,][]{haekau00, shield03,robert06a}.

\subsection{Broader Implications: Starbursts, ULIRGs, and Elliptical Galaxies}
\label{ssec:ulirg_discuss}

Previously, a model for galaxy evolution has been outlined in which mergers play a central role \citep[e.g.,][]{hopkin06a,hopkin08c,somerv08}.  In this picture, galaxies form originally as disks, and the population of ellipticals is built up through time from mergers.  Gas-rich mergers pass through intermediate phases in which they host significant star formation and black hole activity, first as ULIRGs, then as quasars, and finally as dormant elliptical galaxies.

There is a great deal of observational support for this scenario.  Nearby ULIRGs are invariably associated with gas-rich mergers \citep{sander88a,sanmir96} and contain large quantities of molecular gas in their centers \citep{sander91}, driving nuclear starbursts.  Many display evidence for central AGN through their ``warm'' spectra \citep{sander88b}, leading to the interpretation that quasars are the descendants of ULIRGs.  As the starbursts and quasar activity fade, the merger remnant evolves into a passive elliptical \citep{toomre72,toomre77}, making the transition from blue to red.  The relics of this process can be seen in the central light concentrations in elliptical galaxies that were put into place during the starburst phase of their creation \citep[e.g.,][]{mihher94b,hopkin08d}, and correlations between supermassive black holes and e.g. the central potentials of their hosts \citep{hopkin07b,allric07}.

These fossil signatures, and the timing of the events that led to their formation, are sensitive to the interplay between star formation and black hole growth and the feedback processes that accompany each of these.  In principle, black hole recoil could impact all of these
phenomena to varying degrees, providing constraints on the modeling we have presented here as well on the interpretation of the observations more generally.

For example, some ULIRGs display evidence for significant AGN activity in the central starburst region.  This is strong evidence that the BHs in these systems did not undergo rapid recoil events, and that any recoil motion was quickly damped out.  However, merger-triggered central AGN, particularly those in ULIRGs, may be dust-obscured or dwarfed by the luminous starburst.  AGN that are instead kicked out of the central region by a large recoil may be more easily detected, provided a large enough region of the galaxy is observed.  Additionally, we have shown that in some cases, bound recoiling AGN may have longer lifetimes than stationary AGN.  Such AGN may continue shining after the starburst is complete, which may also make them easier to observe.  

Moreover, in the simulations, at least, feedback from black hole growth plays a significant role in quenching star formation \citep{dimatt05,spring05c}, allowing the remnant to evolve from a blue, actively star-forming galaxy, to a red, quiescent object.  During this transition, the remnant can, for some period of time, be classified as a K+A or E+A galaxy \citep{snyder10}.  If the central BH were ejected, star formation would linger at higher rates than for a stationary BH, extending the duration of the blue to red transition, modifying the stellar populations, and ultimately yielding an elliptical with denser, more massive central stellar cusps.  If the BH did not eventually return to the nucleus, subsequent gas-poor mergers would be less strongly affected by BH scouring, obscuring the relationship between core and cusp ellipticals of the same mass \citep{hopher10b}.

Fig.~\ref{fig:compare_profiles} shows that if the BHs recoil at sufficiently high speed, the remnant relaxes to a state similar to what would have happened had the galaxies not had black holes.  Compared to the case of no BH recoil, the remnants with recoiling BHs (or no BHs) have more massive central stellar cusps.  These more massive stellar cusps reflect a more extended period of active star formation following final coalescence of the progenitor galaxies, lengthening the period of time to make the blue to red transition.  Furthermore, the reduced efficiency of AGN feedback means that the central regions would remain more heavily dust-obscured than if there were no BH recoil, making the remnant less likely to be visible as a K+A galaxy \citep{snyder10} and potentially significantly reducing the number of K+A galaxies expected to be observable as merger remnants.

Using our suite of merger models, we can make some qualitative predictions about the morphologies of recoiling BH hosts.  Fig.~\ref{fig:3proj} shows that our merger remnants have some common morphological features.  In general, at the time of BH merger (and recoil) the merging galaxies have undergone final coalescence (i.e., they no longer have two distinct cores), but are still highly disturbed and have prominent tidal features.  This is true even when the progenitor galaxies have low initial gas fractions (4\%) or relatively small ratios (0.25).  For minor mergers with $q < 0.25$, the disruption to the primary galaxy would be less pronounced, but such mergers would also yield small GW recoil kicks.  Highly-aligned (i.e. coplanar) galaxy merger orbits may result in remnants with regular disk structures (see Fig.~\ref{fig:3proj}), but these orbits should comprise only a small fraction of major galaxy mergers.  Therefore, we conclude that the majority of galaxy merger remnants that produce large GW recoil events have single cores but are still tidally disturbed at the time of the recoil kick.  Furthermore, we have shown that the offset-AGN lifetimes for recoiling BHs in our simulations are $< 100$ Myr and are in most cases $\sim 1 - 10$ Myr.  Tidal structures in these major merger remnants generally persist for at least a few hundred Myr, so it follows that most observable recoiling AGN should be found in unrelaxed remnants that may still have visible rings or tidal tails.

\subsection{Comparison to Recent Work}
\label{ssec:compare_discuss}

 Shortly before this paper was completed, \citet{guedes10} and \citet{sijack10} completed independent studies of recoiling BHs in gaseous mergers.  However, both of these papers are exploratory in nature, in that only a few examples of galaxy mergers are studied.  While there is agreement between these studies on some fundamental conclusions, our large parameter study allows us to expand upon these findings, connect them into a more coherent picture, and present some entirely new results.  Here we give a brief comparison of this work to the work of \citet{guedes10} \& \citet{sijack10}. 

\citet{guedes10} use one equal-mass and two minor merger remnants as initial conditions for their simulations.  The mass and spatial resolution used are comparable to ours, though in their equal-mass merger, higher resolution is obtained at late stages via particle splitting.  The merger orbits were coplanar and prograde, a configuration that results in highly disk-like remnants (see our Fig.~\ref{fig:3proj}).  Recoils with a range of kick speeds were applied to the BH, and after a brief initial phase, the BH trajectories were followed semi-analytically.  We note that they use~\vk/\vesc~as low as 0.14, while we use only~\vk~$> 0.4$~\vesc, such that the BHs do not spend most of their time at radii $\sim R_{\rm soft}$.

\citet{sijack10} use isolated galaxy models as well as one galaxy merger model for their recoiling BH simulations.  Like our simulations, theirs were performed using {\footnotesize GADGET-3}, with similar spatial resolution.  However, they use only one merger model, which is massive and gas-rich, resulting in a remnant with a very high central escape speed (3500~\kms) and steep central potential.  This model is therefore fairly extreme, in that recoils with~\vk~$\sim$~\vesc~are required to eject the BH beyond the central few kpc.   

Despite the substantially different approaches of these two studies and our own, a common finding is that gas inflow during galaxy mergers can create centrally-dense remnants that impart significant gas drag to recoiling BHs, thereby reducing the amplitude and return time of their trajectories.  That this effect remains important even for modest gas fractions ($\la 0.1$) illustrates the limitations of modeling GW recoil dynamics in purely collisionless systems.

\citet{guedes10} find that in their minor mergers, recoiling BHs exceed 600~\kms~only for~\vk~$>$~\vesc, and they do not calculate AGN lifetimes for escaping BHs.  Kinematically-offset AGN have lifetimes up to a few Myr in their $q=1$ merger.  \citet{sijack10} note that velocity offsets up to 500~\kms~may occur during the bright AGN phase  in their isolated disk model, but do not discuss offset AGN in their merger model.

Although they use similar values of $L_{\rm min}$ and $\Delta R_{\rm min}$, \citet{guedes10} calculate longer spatially-offset AGN lifetimes than we find in our models.  The disparity in offset AGN lifetimes may arise partly from the different dynamical treatments; \citet{guedes10} calculate the recoil trajectories semi-analytically, as opposed to our N-body approach.  Further, they assume $\dot M_{\rm BH} = \dot M_{\rm Edd}$ for the ejected-disk accretion rather than a time-dependent, isolated-disk model.  They argue that spatially-offset AGN are naturally longer-lived than kinematically-offset AGN because the former occur near apocenter and the latter near pericenter.  However, we show that in some cases, kinematically-offset AGN may have longer lifetimes, because the BH is much more likely to be actively accreting during passages though the central dense region. 

As we have found in our recoil simulations, \citet{sijack10} show that recoil events interrupt a phase of rapid BH accretion in their gas-rich merger, causing a BH mass deficit relative to a stationary BH.  They note that BH mass deficits may introduce scatter into BH mass - host galaxy relationships, which we demonstrate robustly in \S~\ref{ssec:msigma}. 
 
Furthermore, both our results and those of \citet{sijack10} show that the central SFR is higher after a large recoil, due to the displacement of AGN feedback.  \citet{sijack10} note that this further depletes the gas supply available to the BH once it returns to the center.  We demonstrate that SFR enhancement due to recoil is strongest in gas-rich, nearly-equal-mass mergers, though it is present in all of our simulations to some degree, and we have outlined several possible observational consequences in \S~\ref{ssec:ulirg_discuss}.

In addition to connecting and expanding upon the results of \citet{guedes10} and \citet{sijack10}, many of our results are entirely new to this work.  As we have discussed these previously, we list them here only briefly.  First, because we allow the BH merger time to be a free parameter, we are able to demonstrate that recoil trajectories may depend quite sensitively on the BH merger time in $q \sim 1$, gas-rich mergers.  We eliminate much of this uncertainty in our simulations by scaling~\vk~to~\vesc(\tmrg).  This approach also enables a much clearer comparison between models.  Also unique to our work is that we test a range of recoil kick directions in generic merger remnants, demonstrating that in general, the kick direction is less important for recoil trajectories than the kick magnitude.  

Another novel feature of our models is the inclusion of a time-dependent accretion disk around the recoiling BH.  We also calculate recoiling AGN lifetimes using three different luminosity limits, which allows us to differentiate between bright and faint AGN.  We learn from this analysis that some recoiling BHs may actually have longer AGN lifetimes than stationary BHs, though at low luminosities.  We also show that kinematically-offset AGN occur via two different mechanisms: repeated pericentric passages in massive, gas-rich remnants, and ejected-disk accretion in high-velocity recoils (\vk~$>$~\vesc~in our models). We demonstrate that spatially-offset AGN can occur for a wide range of kick speeds and merger models.

Finally, we use our suite of simulations to construct $M_{\rm BH}-\sigma_*$ relations for stationary and rapidly-recoiling BH populations.  We find that GW recoil introduces a normalization offset and larger overall and intrinsic scatter into this correlation.

\vspace{14pt}

We have shown that recoiling BHs may be observable as offset AGN; discoveries of these objects would both inform models of hierarchical galaxy formation and constrain event rates for future gravitational-wave observatories.  In the meantime, the richly varied effects of GW recoil on SMBHs and their host galaxies are not to be discounted as important components of merger-driven galaxy evolution.

\section*{acknowledgements}
{\rev  This work was supported in part by NSF grant AST0907890 and NASA grants NNX08AL43G and NNA09DB30A (for A. L.)  T. C. thanks the Keck Foundation for their support.}

\bibliography{refs_recoil_sims}

\end{document}